\def\glsp{$\wtil g$-LSP}
\def\lsim{\mathrel{\raise.3ex\hbox{$<$\kern-.75em\lower1ex\hbox{$\sim$}}}}
\def\gsim{\mathrel{\raise.3ex\hbox{$>$\kern-.75em\lower1ex\hbox{$\sim$}}}}
\def\ifmath#1{\relax\ifmmode #1\else $#1$\fi}
\def\half{\ifmath{{\textstyle{1 \over 2}}}}

\def\ejet{E_{\rm jet}}
\def\thetamuid{\theta(\mu\mbox{id})}
\def\mrecoil{M_{\rm recoil}}
\def\sigp{\sigma_{\rm P}^{\rm ann}}
\def\signp{\sigma_{\rm NP}^{\rm ann}}
\def\alsp{\alpha_s^{\rm P}}
\def\alsnp{\alpha_s^{\rm NP}}
\def\mpi{m_{\pi}}
\def\sigann{\sigma^{\rm ann}}

\def\vev#1{\langle #1 \rangle}
\def\lam{\lambda}

\def\gtino{\wt G}
\def\mgtino{m_{\gtino}}

\def\sq{\wt q}

\def\msq{m_{\sq}}

\def\delgs{\delta_{GS}}

\def\ptmiss{/ \hskip-8pt p_T}

\def\mhalf{m_{1/2}}
\def\gl{\wt g}
\def\mgl{m_{\gl}}

\def\etc{{\it etc.}}

\def\etc{{\em etc.}}

\def\eg{{\it e.g.}}
\def\etal{{\it et al.}}
\def\mhalf{m_{1/2}}

\def\etc{{\it etc.}}

\def\rpm{R^{\pm}}

\def\rzero{R^0}
\def\mrzero{m_{\rzero}}
\def\etc{{\em etc.}}

\def\eg{{\it e.g.}}
\def\etal{{\it et al.}}
\def\mhalf{m_{1/2}}
\def\gl{\wt g}
\def\mgl{m_{\gl}}

\def\tanb{\tan\beta}

\def\mz{m_Z}

\def\mgut{M_U}

\def\cnone{\wt\chi^0_1}

\def\cntwo{\wt\chi^0_2}

\def\wt{\widetilde}

\def\cpmone{\wt \chi^{\pm}_1}

\def\MPL #1 #2 #3 {{\sl Mod.~Phys.~Lett.}~{\bf#1} (#3) #2}
\def\NPB #1 #2 #3 {{\sl Nucl.~Phys.}~{\bf #1} (#3) #2}
\def\PLB #1 #2 #3 {{\sl Phys.~Lett.}~{\bf #1} (#3) #2}
\def\PR #1 #2 #3 {{\sl Phys.~Rep.}~{\bf#1} (#3) #2}
\def\PRD #1 #2 #3 {{\sl Phys.~Rev.}~{\bf #1} (#3) #2}
\def\PRL #1 #2 #3 {{\sl Phys.~Rev.~Lett.}~{\bf#1} (#3) #2}
\def\RMP #1 #2 #3 {{\sl Rev.~Mod.~Phys.}~{\bf#1} (#3) #2}
\def\ZPC #1 #2 #3 {{\sl Z.~Phys.}~{\bf #1} (#3) #2}
\def\IJMP #1 #2 #3 {{\sl Int.~J.~Mod.~Phys.}~{\bf#1} (#3) #2}
\def\NIM #1 #2 #3 {{\sl Nucl.~Inst.~and~Meth.}~{\bf#1} {#3} #2}

\def\calm{{\cal M}}
\def\wtil{\widetilde}

\def\lam{\lambda}
\def\br{BR}

\def\gam{\gamma}

\def\etal{{\it et al.}}
\def\etc{{\it etc.}}

\def\anti{\overline}
\def\epem{e^+e^-}

\def\rts{\sqrt s}
\def\ie{{\it i.e.}}
\def\eg{{\it e.g.}}
\def\eps{\epsilon}
\def\anti{\overline}

\def\mz{m_Z}

\def\tanb{\tan\beta}

\def\fbi{~{\rm fb}^{-1}}
\def\fb{~{\rm fb}}
\def\pbi{~{\rm pb}^{-1}}
\def\pb{~{\rm pb}}

\def\gev{~{\rm GeV}}
\def\tev{~{\rm TeV}}

\newcommand{\nc}{\newcommand}
\nc{\beq}{\begin{equation}}   \nc{\eeq}{\end{equation}}
\nc{\bea}{\begin{eqnarray}}   \nc{\eea}{\end{eqnarray}}
\nc{\baa}{\begin{array}}      \nc{\eaa}{\end{array}}
\nc{\bit}{\begin{itemize}}    \nc{\eit}{\end{itemize}}
\nc{\ben}{\begin{enumerate}}  \nc{\een}{\end{enumerate}}
\nc{\bce}{\begin{center}}     \nc{\ece}{\end{center}}
\def\beqa{\begin{eqnarray}}
\def\eeqa{\end{eqnarray}}

\def\tanb{\tan\beta}

\documentstyle[12pt,equations,epsf]{article}
\def\ie{{\it i.e.}}
\def\etal{{\it et al.}}
\def\9{\phantom 0}      
\renewcommand\linebreak{\unskip\break} 
\begin{document}
\newlength{\captsize} \let\captsize=\small 
\newlength{\captwidth}                     

%
\font\fortssbx=cmssbx10 scaled \magstep2
\hbox to \hsize{
$\vcenter{
\hbox{\fortssbx University of California - Davis}\medskip
}$
\hfill
$\vcenter{
\hbox{\bf UCD-98-8} 
\hbox{\bf FSU-HEP-980612}
\hbox{\bf hep-ph/9806361}
\hbox{June, 1998}
\hbox{Revised: September, 1998}
}$
}

%
\medskip
\begin{center}
\bf
A HEAVY GLUINO AS THE LIGHTEST SUPERSYMMETRIC PARTICLE
\\
\rm
\vskip1pc
{\bf Howard Baer$^{1,2}$, Kingman Cheung$^1$ and John F. Gunion$^1$}\\
\medskip
{\it $^1$Davis Institute for High Energy Physics}\\
{\it University of California, Davis, CA 95616}\\
{\it $^2$Department of Physics, Florida State University}\\ 
{\it Tallahassee, FL 32306}\\
\end{center}

\begin{abstract}
We consider the possibility that the lightest supersymmetric particle
is a heavy gluino. After discussing models in which this
is the case, we demonstrate that the \glsp\ could evade cosmological
and other constraints by virtue of having a very small relic density.
We then consider how neutral and charged hadrons containing a
gluino will behave in a detector, demonstrating that
there is generally substantial apparent missing momentum associated with
a produced \glsp. We next investigate
limits on the \glsp\ deriving from LEP, LEP2 and
RunI Tevatron experimental searches for excess events
in the jets plus missing momentum channel and for stable heavily-ionizing
charged particles. The range of $\mgl$ that can be excluded depends
upon the path length of the $\gl$ in the detector, the amount
of energy it deposits in each hadronic collision, and the probability
for the $\gl$ to fragment to a pseudo-stable charged hadron after
a given hadronic collision. We explore how the range of excluded $\mgl$
depends upon these ingredients, concluding that for non-extreme cases
the range $3\gev\lsim\mgl\lsim 130-150\gev$
can be excluded at 95\% CL based on currently available OPAL and CDF analyses.
We find that RunII at the Tevatron can extend the excluded
region (or discover the $\gl$) up to $\mgl\sim 160-180\gev$.
For completeness, we also analyze the case
where the $\gl$ is the NLSP (as possible in gauge-mediated
supersymmetry breaking) decaying via $\gl\to g+\mbox{gravitino}$.  
We find that the Tevatron RunI data excludes $\mgl\leq 240\gev$.
Finally, we discuss application of 
the procedures developed for the heavy \glsp\
to searches for other stable strongly interacting
particles, such as a stable heavy quark.

\end{abstract}

\section{Introduction}

In the conventional minimal supergravity (mSUGRA)
and minimal gauge-mediated (mGMSB) supersymmetry models, the
gaugino masses $M_i$ at low energy are proportional to the
corresponding $\alpha_i$ and are in the ratio 
\beq
M_3:M_2:M_1\sim \alpha_3:\alpha_2:\alpha_1\,,
\label{standard}
\eeq
as would, for example, apply if the $M_i$ evolve to a common
value $\mhalf$ at the GUT scale $\mgut$ in the SUGRA model context.  
However, well-motivated
models exist in which the $M_i$ do not obey Eq.~(\ref{standard}).
In particular, the focus of this paper will be on models
in which $M_3$ is the smallest of the gaugino masses, implying
that the gluino will be the lightest supersymmetric particle (LSP).
(We note that
we explicitly do not consider $\gl$ masses as low as those
appropriate in the light gluino scenario \cite{glennys},
which some \cite{nogluinos} would claim has now been ruled out.)

One such model is the O-II string model 
in the limit where supersymmetry breaking is
dominated by the universal `size' modulus \cite{nonuniv,guniondrees2}
(as opposed to the dilaton). Indeed, 
the O-II model is unique among the models considered in \cite{nonuniv}
in that it is the only string model in which the limit
of zero dilaton supersymmetry breaking is consistent
with the absence of charge/color breaking.
In the absence of dilaton supersymmetry breaking, the gaugino masses
arise at one-loop and are therefore
determined by the standard renormalization group equation coefficients
and by the Green Schwartz parameter $\delgs$. The O-II model
in this limit results in the ratios
\beq
M_3:M_2:M_1\stackrel{\scriptstyle O-II}{\sim}-(3+\delgs):(1-\delgs):
({33\over 5}-\delgs)\,,
\label{oiibc}
\eeq
and a heavy gluino is the LSP when $\delgs\sim -3$ (a preferred
range for the model). 

In the GMSB context, the possibility of a 
heavy \glsp\ has been stressed in Ref.~\cite{raby}. There,
the $\gl$ is the LSP as a result of mixing between the 
Higgs fields and the messenger fields, both of which
belong to $5$ and $\anti 5$ representations of SU(5), which
are, in turn, contained in $10$'s of SO(10). The basic idea
is as follows. First, one implements
the standard mechanism for splitting the color-triplet
members of the Higgs from their SU(2)-doublet partners
in the $5,\anti 5$ representations using an `auxiliary' $10$. 
As a result of this splitting, 
the Higgs color triplets mix with the color triplet members of the
auxiliary $10$, both acquiring mass of order the unification
scale, $\mgut$.
If one now identifies the fields in the auxiliary $10$ 
with the messenger sector $10$ fields, it is the messenger sector
fields that supply the standard doublet-triplet Higgs splitting
and whose color triplet members acquire mass $\sim \mgut$.
As a result, the color-triplet messenger fields
naturally become much heavier than their SU(2)-doublet counterparts.
Since the masses of the gauginos arise in GMSB via loop graphs
containing the messenger fields of appropriate quantum numbers,
the result is that the (colored) gluino mass is suppressed by $(M/\mgut)^2$
compared to the other gaugino masses, where $M$ is the typical mass
of a doublet messenger field. One requires that
$M/\mgut\lsim 0.1$ in order to adequately 
suppress baryon number violating interactions
mediated by the Higgs triplets 
(which are controlled by an effective mass of order $\mgut^2/M$).

Early outlines of the phenomenological
constraints and possibilities for a heavy \glsp\ appear in
\cite{starkman,steigman,wolfram,mohnus,raby,gunbarcelona}. 
Here, we attempt to refine
these phenomenological discussions.  For our phenomenological
studies, we will make the assumption
that all supersymmetric particles are substantially heavier than
the \glsp.\footnote{This is natural for the sfermions
in the O-II model, since the $m_0$ SUSY-breaking
scalar mass parameter is automatically much larger than $\mhalf$.}
This is a conservative assumption in that discovery of supersymmetry will be
easier in scenarios in which some of the other superparticles are 
not much heavier than the gluino.

The outline of the rest of the paper is as follows. In section 2, we
demonstrate the sensitivity of the relic gluino density 
to assumptions regarding the non-perturbative physics associated
with gluino and gluino-bound-state annihilation.  In section 3,
we examine how energetic massive gluinos produced at an accelerator
will be manifested in a typical detector. In section 4, we consider
the constraints from LEP and LEP2 data on a massive gluino produced
in $\epem\to q\anti q\gl\gl$.
In section 5, we examine constraints on a massive \glsp\
from the existing RunI data in the jets plus missing momentum
channel and explore the prospects for improvements at RunII.
In both sections 4 and 5, we discuss how the constraints/limits
depend on the manner in which a $\gl$ is manifested in a detector.
We consider limits on a heavy \glsp\ that 
arise from searches for heavy stable charged particles at OPAL and CDF
in sections 6 and 7, respectively.
In section 8, we present Tevatron limits on a gluino that is the NLSP
of a gauge-mediated supersymmetry breaking model, decaying
via $\gl\to g+\mbox{gravitino}$.
In section 9, we outline possible applications of the procedures
developed for the heavy gluino to other new particle searches,
in particular searches for a stable heavy quark.
Section 10 presents our conclusions. The reader is encouraged
to begin by scanning the concluding section 10 so as to get an overview
of our results and the issues upon which he/she should focus
while working through each section.

\section{The Relic Gluino Density}

Before embarking on our discussion of direct accelerator limits,
it is important to determine if a massive gluino LSP can 
have a relic density that is sufficiently small to be consistent
with all constraints. In particular, 
as discussed in \cite{starkman,steigman,wolfram,mohnus,raby}, 
its relic density must
be sufficiently small that it cannot constitute a significant fraction
of the dark matter halo density.  Otherwise, it would almost certainly
have been seen in anomalous matter searches, underground detector
experiments and so forth. We will show that non-perturbative physics
can lead to large enhancements in the relevant annihilation cross sections,
with the result that the relic density could be very small.

We begin with a very brief review of the standard approach for
computing a relic density. First, one determines the freeze-out
temperature $T_F$, which is roughly the
temperature at which the annihilation rate 
for two gluinos falls below the rate at which the universe is expanding.
The standard form of the freeze-out condition is \cite{gondolo}
\beq
\ln\left\{{\vev{\sigann v}\over 4\pi^3} \sqrt{{45 \over 2 g^*(T_F) G_N}}\,
\mgl g_{\gl} x_F^{-1/2}\right\}=x_F\,.
\label{tfform}
\eeq
Here, $G_N$ is Newton's constant,
$x\equiv\mgl/T$, $g_{\gl}=2\times 8$ is the number of gluino degrees
of freedom, and $g^*(T)$ is the density degree-of-freedom counting factor.
In all our computations, we employ the exact formula of
Ref.~\cite{gondolo} for $\vev{\sigann v}$:
\beq
\vev{\sigann v}={1\over 8\mgl^4
TK_2^2(\mgl/T)}\int_{4\mgl^2}^\infty\sigann(s)s^{3/2}\beta^2K_1(\sqrt s/T)ds\,,
\label{sigvform}
\eeq
where $\beta=\sqrt{1-4\mgl^2/s}$ is the velocity of the
$\gl$'s in the initial state center-of-mass frame; $\vev{\sigann v}$
is computed numerically.
The above $\vev{\sigann v}$ form assumes only that the $\gl$'s 
(or $\rzero$'s, see below) remain in
kinetic equilibrium for all temperatures (as seems highly
likely given that they re-scatter strongly on either
quarks/gluons or hadrons, respectively, even after freeze-out). 
We then numerically integrate the Boltzmann equation.
Defining as usual $Y=n_{\gl}/s$ (where $s$ is the entropy density
and $n_{\gl}$ is the gluino number density),
the standard result is 
\beq
{1\over Y_0}-{1\over Y_F}=\left[{45G_N\over\pi}\right]^{-1/2}
\int_{x_F}^{x_0} {h^*(T)\over \sqrt{g^*(T)}}{\mgl\over x^2}\vev{\sigann v}
dx\,,
\label{yform}
\eeq
where the subscript 0 ($F$) refers to current (freeze-out) temperature and
$h^*(T)$ is the entropy 
degree-of-freedom counting factor.\footnote{Note
that only standard model particles are counted 
in computing $g^*$ and $h^*$ since all
supersymmetric particle are presumed to be heavier than the $\gl$.}
As usual, $1/Y_F\ll1/Y_0$ and can be neglected. Finally, we compute
the current gluino mass density as
\beq
\rho_0=\mgl n_0=\mgl s_0Y_0=\mgl h^*(T_0){2\pi^2\over 45}T_0^3Y_0\,,
\label{rho0form}
\eeq
and
\beq
\Omega h^2={\rho_0 h^2\over \rho_c}=
{\rho_0\over8.0992\times 10^{-47}\gev^2}\,.
\label{omegahsq}
\eeq

The estimates in the literature \cite{steigman,wolfram,mohnus,raby}
for the relic density of a massive gluino 
differ very substantially, at least in part
due to different assumptions regarding
the size of the annihilation cross section. Perturbatively,
the annihilation cross section is $\sigp=\sigma(\gl\gl\to
gg)+\sum_q\sigma(\gl\gl\to q\anti q)$ with:
\bea
\sigma(\gl\gl\to gg)&=&{3\pi\alpha_s^2\over 16\beta^2
s}\left\{\log{1+\beta\over
1-\beta}\left[21-6\beta^2-3\beta^4\right]-33\beta+17\beta^3\right\}\,,
\label{sigglgltogg}\\
\sigma(\gl\gl\to q\anti q)&=&{\pi\alpha_s^2\anti\beta\over 16\beta s}
(3-\beta^2)(3-\anti\beta^2)\,.
\label{sigglgltoqq}
\eea
[In Eq.~(\ref{sigglgltoqq}), $\anti\beta=\sqrt{1-4m_q^2/s}$, $m_q$ 
being the quark mass.] We observe that 
as $\beta\to 0$, $\beta\sigp$ approaches
a constant unless the $\alpha_s$ employed is allowed to increase
in a non-perturbative manner.  (Note that this
is in sharp contrast to the $\beta\sigann\propto \beta^2$
$p$-wave behavior for the $\cnone\cnone$ annihilation cross sections; 
since the $\gl\gl g$ vertex does not contain a $\gamma_5$, 
$\gl\gl$ annihilation can occur in an $s$-wave
and is much stronger at low $\beta$.) For our
perturbative computations we employ $\alsp(Q)$
evaluated at $Q=\sqrt s$, where $\alsp(Q)$
is the usual moving coupling, $\propto 1/\log(Q^2/\Lambda^2)$
at one loop. (When employed at small $Q$, see below, $\alsp(Q)=1$
will be the maximum value allowed.)

However, near the
threshold, $\sqrt s\sim 2\mgl$, non-perturbative effects can be
expected to enter. There are many possibilities. Consider first
multiple gluon exchanges between interacting $\gl$'s. These will
give rise to a Sommerfeld enhancement factor \cite{sommer,appol,bgs,bhsz},
which we will denote by $E$, as well as logarithmic enhancements
due to soft radiation \cite{bhsz}.  Here, we retain only $E$,  which
takes the form\footnote{The Sommerfeld enhancement factor takes the form
$1+C\pi\alpha_s/(2\beta)$
for small $C\pi\alpha_s/(2\beta)$. We extend this to the region of large
$C\pi\alpha_s/(2\beta)$ by using the standard exponentiated form given.}
\beq
E={C \pi \alpha_s\over \beta}
\left[1-\exp\left\{-{C \pi \alpha_s\over \beta}\right\}\right]^{-1}\,,
\label{esom}
\eeq
with $C$ being a process-dependent constant. The $E_{gg}$ ($E_{q\anti q}$)
for $\gl\gl\to gg$ ($\gl\gl\to q\anti q$)
is given by taking $C=1/2$ ($C=3/2$).
If one examines the derivation of $E$, then one finds that
the typical momentum transfer of the soft gluon exchanges responsible
for $E$ is $Q\sim \beta\mgl$. Thus, we choose to evaluate $E$
using $\alpha_s(\beta\mgl)$.\footnote{In
the perturbative next-to-leading order results of
\cite{bhsz}, $E_{\gl\gl}$ and $E_{q\anti q}$ are evaluated
at the factorization scale $\mu$. In the perturbative
expansion approach, a next-to-next-to-leading order
calculation is required to determine the appropriate effective
scale at which to evaluate the next-to-leading Sommerfeld factor.}
The $C$ values quoted above are those appropriate to color averaging
in the initial $\gl\gl$ state. Color averaging is relevant since
the high scattering rate of gluinos (off gluons \etc),
continually changes the color state of any given gluino,
and, in particular,
does not allow for the long time scales needed for the Sommerfeld
enhancement to distort \cite{bgs} the momenta 
of the relic gluinos so that they become organized into 
color-singlet pairs with low relative velocity.  
In what follows,
we will employ the shorthand notation $E\sigp\equiv E_{gg}\sigma(\gl\gl\to
gg)+E_{q\anti q}\sigma(\gl\gl\to q\anti q)$.

As an aside, we note
that multiple soft-gluon interactions between the final state
$q$ and $\anti q$ in $\gl\gl\to q\anti q$ result in 
a repulsive Sommerfeld factor at small $\anti\beta$
(since the $q\anti q$ are in a color octet
state). However, this is not an important effect since 
the $\gl\gl\to q\anti q$ cross section vanishes as $\anti\beta\to 0$
anyway. We do not include this final-state Sommerfeld factor 
in our calculations.

We will consider two possibilities for computing $E\sigp$. In the first case,
$\sigp$ is computed using $\alsp(\rts)$ and $E$ is computed using
$\alsp(\beta\mgl)$, with the result that $\beta E\sigp\propto 1/\beta$
as $\beta\to 0$, recalling that $\alsp(\beta\mgl)$ is not
allowed to exceed 1.
In the second case, we employ a `non-perturbative' form for $\alpha_s$,
denoted $\alsnp$,
defined by replacing $1/\log(Q^2/\Lambda^2)$ in the $\alsp$
form by $1/\log(1+Q^2/\Lambda^2)$. (This
form is that which corresponds to a roughly linear potential
a large distance, and was first discussed in Ref.~\cite{richardson}
with regard to the charmonium bound state spectrum.)
$\sigp$ and $E$ are evaluated using
$\alsnp(\rts)$ and $\alsnp(\beta\mgl)$, respectively.
The result is that $\beta E\sigp\propto 1/\beta^3$
at small $\beta$. In both cases,
the growth of $E\sigp$ will be cutoff by requiring
that $E\sigp$ not exceed $E\sigp=\beta^{-1}/\mpi^2$, the largest annihilation
cross section that we wish to consider.

Of course, as is well-known from the charmonium analogue \cite{appol},
the Sommerfeld enhancement at best provides an
average (in the dual sense) over the resonance structure that is likely to
be present.  Further, just as in charmonium,
the Sommerfeld enhancement is a precursor to the formation
of $\gl\gl$ bound states that will occur once the temperature
falls below the typical binding energy. This binding energy
would be of order $\sim \alpha_s^2\mgl$ to the extent that
short range Coulomb-like color attraction is most important,
but terms in the potential between the two gluinos
(that possibly rise linearly with the separation)
can also play an important role. Thus, it is difficult to
be precise about the temperature at which this transition occurs,
but it is almost certainly above the temperature of the quark-gluon
deconfinement transition. If $\gl\gl$ bound state formation were
to be complete,
the annihilation rate $n_{\gl}\sigma^{\rm ann} v$ 
(where $n_{\gl}$ is the number density of
gluinos per unit volume) would be replaced by the decay
rate for the $\gl\gl$ bound state. In the charmonium analogy, this decay
rate is proportional to $|\calm|^2 |\Psi(0)|^2$, where $\calm$
is the matrix element for the decay, $\propto \alpha_s^2/\mgl^2$, and
$|\Psi(0)|$ is the magnitude of
the wave function at the origin, $\propto [\alpha_s\mgl]^{3/2}$.
The result is a decay rate proportional to $\alpha_s^5\mgl$. 
The important feature of
this result is that the bound state draws the two gluinos
together (as represented in $|\Psi(0)|$) so as to overcome
the perturbative behavior of the annihilation $|\calm|^2$.
A full treatment would have to implement a coupled-channel treatment in which
the $\gl\gl$ bound state formation would be treated in analogy
with the standard approach to $e^-p$ recombination in the early universe.
Those $\gl$'s that are not absorbed by $\gl\gl$ bound state formation prior to
the temperature falling below the deconfinement transition temperature
would end up inside bound states containing one $\gl$
and one or more gluons or light quarks; most likely the $\rzero=\gl g$
bound state would be dominant. The rate of annihilation of
the $\rzero$'s is far from certain (as discussed below).
Although we \cite{kiskis}
are exploring the possibility of implementing this
full scenario, there are clearly many uncertain ingredients.  
We presume that the resulting relic density will be bracketed 
on the high side by the Sommerfeld
enhancement result and on the low side by 
the limit where very few $\gl\gl$ bound states
form before the confinement transition, below which 
strong $\rzero\rzero$ annihilation takes over.

In this latter extreme non-perturbative scenario, we imagine
that at small $\beta$ there will be
a transition where the $\gl$'s condense into color singlet
bound states containing one $\gl$ and light quarks and/or gluon(s); 
as noted above, 
we shall assume here that the lightest is the $\rzero=\gl g$.
(An electrically charged LSP bound state has 
much stronger cosmological constraints
and is easier to see at accelerators.)
For $\beta$ above the transition point, we will employ $\sigp$
without any enhancement factor $E$. For $\beta$ below the transition,
the appropriate annihilation cross section will be that
for $\rzero\rzero\to  \pi's$. It is often assumed 
(see, \eg, \cite{steigman,wolfram,mohnus,raby}) that 
the non-perturbative $\signp$
will be $\signp= A\beta^{-1}/\mpi^2$, 
where the $\beta^{-1}$ factor is the standard result for $s$-wave annihilation
of spin-0 particles and $A$ is
an uncertain constant not too different from unity. We will consider
this possibility even though
we regard such a large annihilation cross section as being unlikely
since annihilation must remove the gluino quanta, implying,
in a parton picture, gluino exchange in the $t$-channel.\footnote{In, 
for example, the model of Ref.~\cite{gunsop} for strong scattering, $A$ would 
scale as $1/\mgl^2$ for annihilation, in sharp contrast
to the $\rzero\rzero\to\rzero\rzero$
scattering cross section which would scale with the inverse size squared
of the $\gl g$ bound state (which would have comparable size to a pion
or proton bound state).} Note that if $A$ scales as $1/\mgl^2$, we would obtain
$\signp\sim \sigp$ (both behaving as $1/\beta$ as $\beta\to 0$
and having similar normalization);
the result would be
a relatively smooth transition as the temperature crosses the deconfinement
boundary, yielding a result not very different from our perturbative
case (with no Sommerfeld enhancement factor).

In our numerical work, the choice of $\signp=A\beta^{-1}/\mpi^2$ with
$A=1$ is labelled as I. As an alternative, we also consider
a second choice (II): $\signp=1/\mpi^2$, such that $\beta\signp$
vanishes (like $\sigp$) as $\beta\to 0$. Although II has no particular
model motivation (other than representing a kind of average
of $s$-wave and $p$-wave behavior), it allows us to assess the importance
of the small $\beta$ behavior of $\signp$. We will see that it leads
to significantly larger relic densities than I. 
For a given choice of $\signp$,
the exact point of transition between $\sigp$ and $\signp$ and
its smoothness are also crucial ingredients in determining the relic density.
\bit
\item
For the transition point we consider two choices: (a)
the total $\gl\gl$ kinetic energy
in the center-of-mass
falling below a given limit $L$, with $L\sim 0.2-1\gev$
(we employ $L=1\gev$ in our numerical results --- the relic
density increases with decreasing $L$);
(b) twice the $\gl$ momentum falling below $L$.
We note that the transition occurs roughly at $\beta\sim \sqrt{L/\mgl}$
and $\beta\sim L/\mgl$ in cases (a) and (b), respectively.
To the extent that the condensation
of $\gl$'s into bound states is controlled by the typical temperature,
the KE criterion is the most natural. It is because it leads
to large increases in the relic density that we have considered
the more moderate (b) possibility.
\item
For the smoothness of the transition we also consider two options: (i)
use $\sigp$ for larger $\beta$ with
an abrupt transition to the non-perturbative annihilation form
for $\beta$ below the appropriate limit; (ii) a smooth transition
in which $\sigp$ is evaluated using $\alsnp(Q)$
and $Q$ is taken to be
the net kinetic energy, $\sqrt s - 2\mgl$, or $2p_{\rm cm}^{\gl}$ 
in cases (a) and (b) above, respectively.
The modified $\sigp$ is employed until it exceeds $\signp$,
after which point the latter is employed. A smooth transition
will lead to larger relic density than the sudden transition choice.
\eit

Altogether, we shall consider eleven cases. The first three are:
(1) $\sigp$ ($E=1$); (2) $E\sigp$ with $E$ as given in Eq.~(\ref{esom})
evaluated using $\alsp(Q=\beta\mgl)$;
and (3) $E\sigp$ with $E$ computed using $\alsnp(Q=\beta\mgl)$; in (2) and (3)
$E\sigp$ is not allowed to exceed $E\sigp=\beta^{-1}/\mpi^2$.
The remaining eight cases are specified by various $\signp$
scenarios: (4) (I,a,i); (5) (II,a,i); (6) (I,b,i); (7) (II,b,i);
(8) (I,a,ii); (9) (II,a,ii); (10) (I,b,ii); (11) (II,b,ii).

\begin{figure}[ht]
\leavevmode
\begin{center}
\epsfxsize=4.25in
\hspace{0in}\epsffile{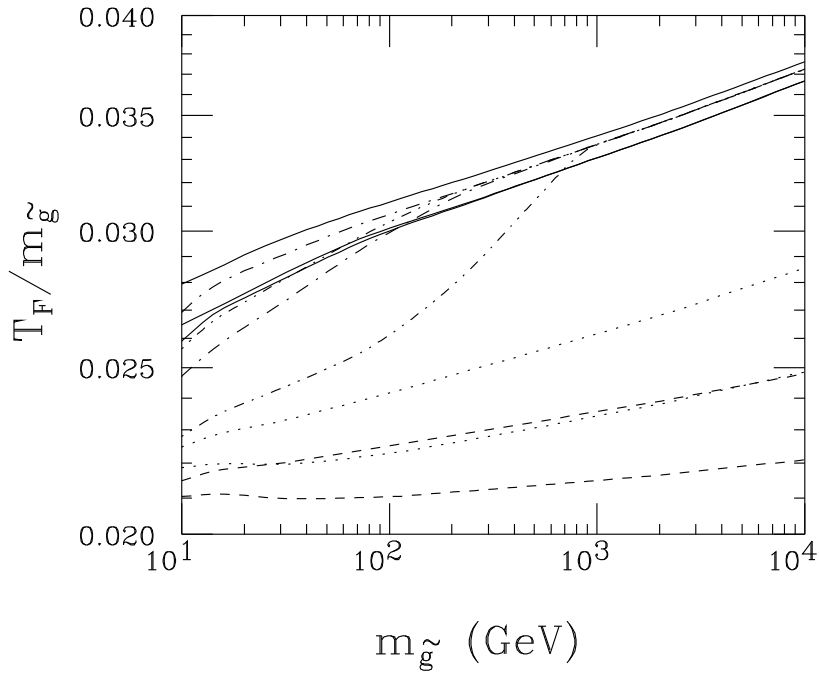}
\end{center}
\caption[]{$T_F/\mgl$ as a function of $\mgl$ for the
11 cases described in the text. The solid lines
correspond to results for cases (1), (2) and (3),
respectively, in order of decreasing $T_F$. 
Results for cases (4) (I,a,i) (5) (II,a,i) (6) (I,b,i) 
(7) (II,b,i) are the lower dashed, dotted, dot-dot-dash and dash-dot
lines, respectively. Results for cases
(8) (I,a,ii) (9) (II,a,ii) (10) (I,b,ii) and
(11) (II,b,ii) are the upper dashed, dotted, dot-dot-dash and dash-dot
lines, respectively. This figure assumes $L=1\gev$, see text.}
\label{relicxf} 
\end{figure} 

\begin{figure}[ht]
\leavevmode
\begin{center}
\epsfxsize=4.25in
\hspace{0in}\epsffile{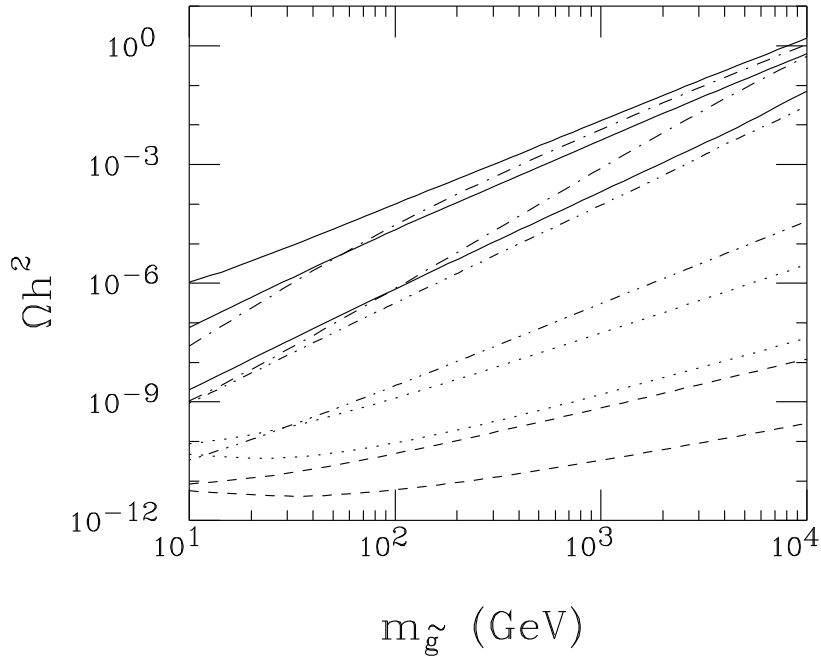}
\end{center}
\caption[]{$\Omega h^2$ as a function of $\mgl$ for the
11 cases described in the text. Line notation as in
Fig.~\protect\ref{relicxf},
with solid lines for cases (1), (2) and (3) in order
of decreasing $\Omega h^2$.}
\label{relicd} 
\end{figure} 

Results for the freeze-out temperature and the
relic gluino density for the 
eleven cases detailed above are shown in Figs.~\ref{relicxf}
and \ref{relicd}, respectively. As expected, the freeze-out
temperature for a relic gluino (relative to the mass $\mgl$ 
of the gluino relic) is lower (by roughly a factor of two)
than in the case of a weakly-interacting relic particle. The ordering of the
curves for the eleven different cases can be easily understood on
the basis of the strength of the annihilation cross section for each
case as a function of $\beta$.

After freeze-out takes place, annihilation remains substantial
(especially in cases where $\sigann$ jumps to a large value at small $\beta$)
and the relic-density continues to decline. The current relic density
is thus very strongly dependent upon the model employed.
Fig.~\ref{relicd} shows that $\Omega h^2$ can be substantial (even
corresponding to an over-closed universe for $\mgl\gsim 10\tev$)
if a purely perturbative approach
is followed, or it can be extremely small out to very large $\mgl$,
as in case (I,a,i) where $\signp=\beta^{-1}/\mpi^2$ and an abrupt transition
from $\sigp$ to $\signp$ based on the KE criterion is employed.\footnote{This
and the other related $\signp$ cases evade the upper bound 
on the mass of the dark matter particle of Ref.~\cite{grkam},
based on $s$-wave dominance of the cross section and partial wave unitarity,
by virtue of the fact that $\signp\sim
\beta^{-1}/\mpi^2\gg\pi\beta^{-2}/\mgl^2$ (the latter being the $s$-wave
unitarity limit) can arise
from, for example, the coherent contribution of many partial waves.}
Almost any result in between is also possible.
Further, the second sub-electroweak scale inflation discussed by some
(see, for example, Ref.~\cite{murayamaetal})
would dilute even the purely perturbative relic densities to an unobservable
level. Until the non-perturbative physics issues can be clarified, 
and late time second inflation can be ruled out, we must assume that
the relic $\gl$ (or more properly $\rzero$) density is small enough
that constraints from anomalous nuclei in seawater, 
signals associated with annihilation in the core
of the sun, interactions in underground detectors \etc\ are not 
significant. In the following sections,
we discuss the extent to which accelerator
experiments can place definitive constraints on the heavy \glsp\ scenario.

\section{How a heavy gluino LSP is manifested in detectors}

Before turning to accelerator constraints on the \glsp\ scenario,
we must determine how a stable gluino will manifest itself inside a detector.
This is a rather complicated subject. The important question is how
much momentum will be assigned to the jet created by the gluino as it
traverses a given detector. This depends on many ingredients,
including, in particular, the probability $P$ that the gluino fragments
to a charged $R$-hadron, $\rpm$, vs. a neutral $R$-hadron, $\rzero$.
It is useful to keep in mind the following two extremes.
\bit
\item
Very little energy would be assigned to the $\gl$
if it always fragments into an $\rzero$
which interacts only a few times in the detector and deposits little
energy at each interaction.
\item
Large energy would be assigned to the $\gl$ 
if it undergoes many hadronic interactions
as it passes through the detector, with large energy deposit at each
interaction, and/or if it fragments often to a $\rpm$ following a hadronic
collision. In particular, when the $\gl$ moves with low velocity
through the detector while contained within an $\rpm$,
it will deposit a substantial fraction of its
energy in the form of ionization as it passes through the calorimeters.
Further, for non-compensating calorimeters
this ionization energy is overestimated when the calorimeter
is calibrated to give correct energies for electrons and pions.

In addition, in the OPAL analysis to be considered later,
if the gluino $R$-hadron is charged in the tracker
and at appropriate further out
points in the detector, it will pass cuts that cause it 
to be identified as a muon, in which case the momentum 
as measured in the tracker is added to the energy measurement
from the calorimeter
and a (much too small) minimal ionization energy deposit
is subtracted from the calorimeter response. In this case,
the energy assigned to the $\gl$ `jet' can actually exceed its true momentum.
\eit
In all our discussions, it should be kept in mind that in current analysis
procedures jets or jets containing a muon are always assumed
to have a small mass, so that the momentum of a jet is presumed
to be nearly equal to its measured energy.

\begin{figure}[ht]
\leavevmode
\begin{center}
\epsfxsize=4.25in
\hspace{0in}\epsffile{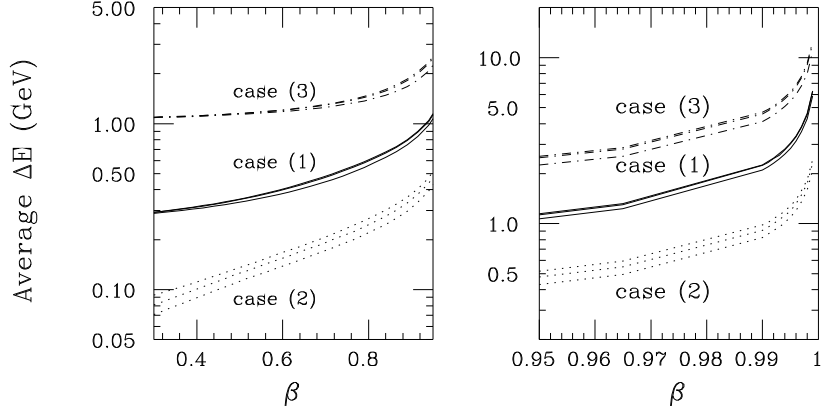}
\end{center}
\caption[]{Average energy loss, $\vev{\Delta E}$, in a collision
as a function of $\beta$ for the three cases
described in the text. 
Results are shown for $\mrzero=5$, $25$ and $140\gev$.
At high $\beta$, curves are ordered according to increasing $\mrzero$.}
\label{deltae} 
\end{figure} 

\subsection{Hadronic energy losses: the $\gl\to \rzero$ case}

In this subsection, we explore the energy loss 
experienced by a heavy $\gl$ passing through a detector as
a result of hadronic collisions.
An early discussion of the issues appears in Ref.~\cite{dreestata}.
These would be the only energy losses if
the $\gl$ almost always moves through the detector
as part of an $\rzero$ state. (This would be the case if
charge-exchange reactions are significantly suppressed because
the charged $\gl$ bound states are substantially heavier than the $\rzero$
or if the $\rpm$ states undergo rapid decay to an $\rzero$ state.)
The first question is how much energy will the $\rzero$ lose in
each hadronic collision as a function of its current $\beta$ value.
As a function of $|t|$ and $m_X^2$ (where 
$t$ is the usual momentum transfer invariant for the $\rzero$
and $m_X$ is the mass
of the system produced in the $\rzero N\to \rzero X$ collision)
the energy loss is given by
\beq
\Delta E={m_X^2-m_N^2+|t|\over 2 m_N}\,,
\label{deform}
\eeq
where we have assumed that the appropriate target is a single
nucleon $N$ rather than the nucleus as a whole or a parton (both
of which are estimated to be irrelevant in \cite{dreestata}).
To estimate the average $\Delta E$ per collision, we must assume
a form for ${d\sigma\over d|t|dm_X}$. We have examined three
different possibilities:
\begin{description}
\item{(1)}
${d\sigma\over d|t|dm_X}\propto 1$ for $|t|\leq 1\gev^2$ and zero for
$|t|>1\gev^2$.
\item{(2)}
${d\sigma\over d|t|dm_X^2}$ given by a triple-Pomeron form \cite{moshe}:
\beq
{d\sigma\over d|t|dm_X^2}\propto {1\over m_X^2}\beta^2(|t|)
\left({s\over
m_X^2}\right)^{2(\alpha_{P}(|t|)-1)}\left[m_X^2\right]^{\alpha_{P}(0)-1}
\,,
\label{ddform}
\eeq
where $\alpha_P(|t|)=1-0.3|t|$ and $\beta(|t|)=1/(1+|t|/0.5\gev^2)^2$
is a typical parameterization. 
For the parameterization of Eq.~(\ref{ddform}), the result for
the average energy loss $\vev{\Delta E}$ is independent
of the maximum value (if $\gsim 0.5\gev^2$) allowed for $|t|$.
\item{(3)}
${d\sigma\over d|t|dm_X}\propto 1$ for $|t|\leq 4\gev^2$ and zero for
$|t|>4\gev^2$.
\end{description}
We compute the average value of $\Delta E$ as a function of 
the $\beta$ of the $\rzero$ in the rest frame of the target nucleon:
\beq
\vev{\Delta E}= {\int_{m_N}^{\rts-\mrzero} dm_X
\int_{|t|_{\rm min}(m_X)}^{|t|_{\rm max}(m_X)} d|t| 
\,\Delta E{d\sigma\over d|t|dm_X}
\over 
\int_{m_N}^{\rts-\mrzero} dm_X
\int_{|t|_{\rm min}(m_X)}^{|t|_{\rm max}(m_X)} d|t| {d\sigma\over d|t|dm_X}}\,,
\label{veve}
\eeq
where 
$|t|_{\rm min,max}(m_X)=2[E(m_N)E(m_X)\mp p(m_N)p(m_X)-\mrzero^2]$
with $E(m)=(s+\mrzero^2-m^2)/(2\rts)$ and
$p(m)=\lam^{1/2}(s,\mrzero^2,m^2)/(2\rts)$
[with $\lam(a,b,c)=a^2+b^2+c^2-2(ab+ac+bc)$], where $s=\mrzero^2+m_N^2+2\gam
\mrzero m_N$ [with $\gam=(1-\beta^2)^{-1/2}$].
In integrating down to $m_X=m_N$ in Eq.~(\ref{veve}), 
we include both elastic and inelastic scattering (using the same cross section
form).\footnote{For large $\beta\gsim 0.95$, the purely elastic scattering
component gives smaller $\vev{\Delta E}$ than the inelastic scattering
component. This should be incorporated in a more complete treatment.}
We note that the above kinematic limits for $|t|$ as a function of $m_X$
must be carefully incorporated in order to get correct results 
for $\vev{\Delta E}$; in particular, $|t|_{\rm min}\to|t|_{\rm max}$ as
$m_X\to\rts-\mrzero$.

The results for $\vev{\Delta E}$ obtained from Eq.~(\ref{veve}) in the above
three cross section cases are plotted in Fig.~\ref{deltae}
for three masses that will later prove to be of interest:
$\mrzero=5$, $25$ and $140\gev$. 
We note that $\vev{\Delta E}$ as a function of $\beta$ is almost
independent of the $\rzero$ mass so long as $\mrzero\geq 5\gev$. 
In what follows we will use the $\mrzero=25\gev$ results
for $\vev{\Delta E}$ for all $\mrzero$. 

\begin{figure}[h]
\leavevmode
\begin{center}
\epsfxsize=4.25in
\hspace{0in}\epsffile{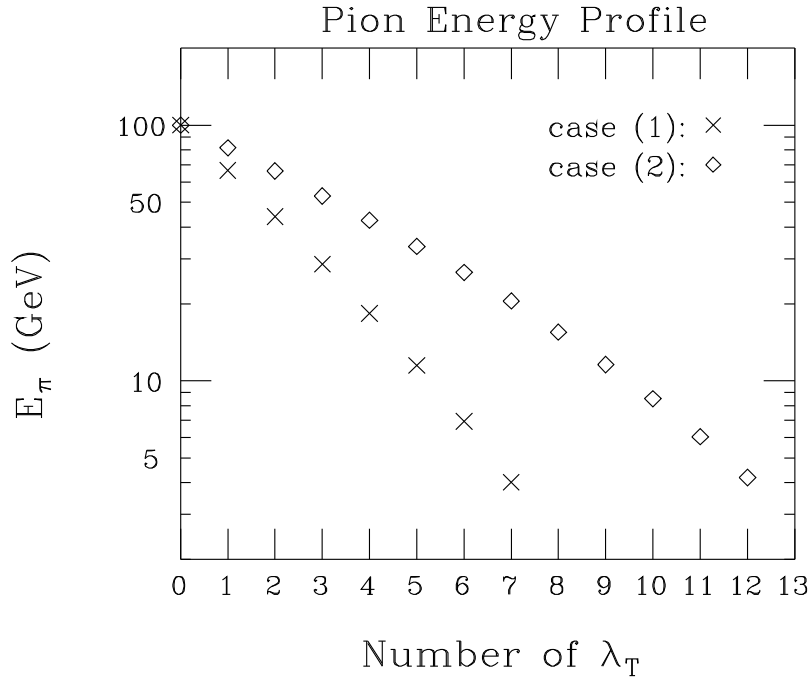}
\end{center}
\caption[]{We plot the energy of an incident 100 GeV pion
after a certain number of hadronic collisions 
for the case (1) and (2) cross section models.}
\label{picascade} 
\end{figure} 

\begin{figure}[h]
\leavevmode
\begin{center}
\epsfxsize=4.25in
\hspace{0in}\epsffile{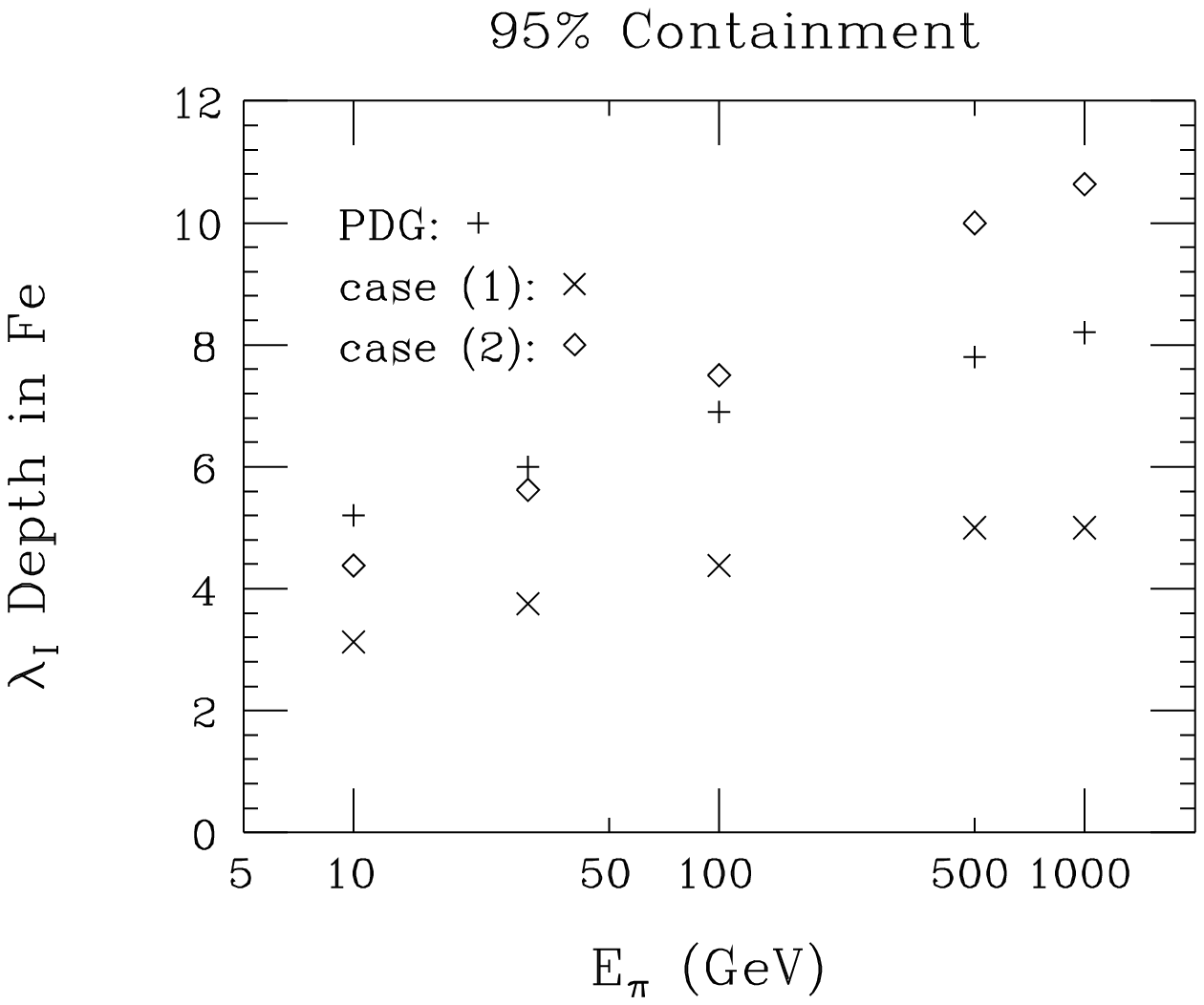}
\end{center}
\caption[]{We plot the number of $\lam_I=17$ cm (\ie\ in iron)
path lengths required for 95\% containment of the energy of
a pion. Experimental results from the PDG, Fig.~24.2 of Ref.~\cite{pdg},
are compared to predictions based on Eq.~(\ref{veve})
for the case (1) and (2) cross section models.}
\label{pieloss} 
\end{figure}

In order to understand whether any of the three models
for ${d\sigma\over d|t|dm_X}$ is reasonable, and, if so, which
is the most reasonable, we examined the results given by our procedure in the
case where the $\rzero$ is replaced by a pion. In so doing,
the pion is viewed as retaining its identity (aside from possible
charge exchange) as it traverses the detector, slowing down
after each hadronic collision by an amount determined
by the $\vev{\Delta E}$ for the then current $\beta$ of the pion.
In our approach, since the elastic cross section is effectively included
in our cross section parameterizations, the average distance
between hadronic interactions of the pion is characterized by 
its path length $\lam_T$ (in the notation of Ref.~\cite{pdg}) in iron (Fe)
as determined by the total cross section.
(We will also need to refer to
the inelastic collision length, denoted by $\lam_I$.)
In Fig.~\ref{picascade}, we show how the energy of a 100 GeV pion
deteriorates to below 5\% of its initial energy
as it undergoes successive hadronic collisions separated
by $\lam_T$, using cross section models (1) and (2).\footnote{Note that 
the $\vev{\Delta E}$ values in Fig.~\ref{deltae} are not correct for a light 
hadron; we employ Eq.~(\ref{veve}) computed numerically
for the current $\beta$ value just prior to a given collision.}
In Fig.~24.2 of Ref.~\cite{pdg}, results for the number of $\lam_I=17$ cm
interaction lengths in iron required for 95\% of the kinetic energy
of a pion to be deposited as a result of hadronic collisions
are given as a function of initial energy. We have computed this number
for the $\vev{\Delta E}$ predictions of our three cross section models;
note that in our approach, hadronic interactions occur every $\lam_T=11$ cm.
The results\footnote{Results are
independent of whether the pion is assumed to be charged or not;
\ie\ $dE/dx$ losses are not important.} for cross
section models (1) and (2) are given in Fig.~\ref{pieloss}
along with the results from Fig.~24.2 of Ref.~\cite{pdg}.
For moderate energies, Fig.~\ref{pieloss} shows that 
the triple-Pomeron case (2) yields rough agreement,
but at higher energies predicts that 95\% containment requires
more $\lam_I$ than experimentally measured.  
The case (1) cross section
predicts 95\% containment for fewer $\lam_I$ than actually measured
for all initial energies.
(Case (3) would predict that even fewer $\lam_I$ would be required
for 95\% containment.) 

As we shall see, the main issue for detecting a \glsp\ signal is
the amount of kinetic energy of the $\gl$'s $R$-hadron that is not deposited
in the calorimeter. Deposited energy has many critical impacts
in the context of the experimental analyses that we will later employ.
We mention two here. First, for an event that is accepted by other cuts,
larger missing kinetic energy implies a stronger 
missing momentum signal. This is the dominant effect for
a $\gl$-jet that propagates primarily as part of a neutral
$R$-hadron bound state. For the OPAL and CDF jets + missing momentum signals,
considered in later sections,
case (1) would then be conservative in that it leads to smaller missing
momentum. Second, for larger missing kinetic energy
a $\gl$-jet that is propagating as a charged $R$-hadron will be
more frequently identified as being a muon. In the CDF jets + missing
momentum analysis, muonic jets are discarded. As a result, case (2) will
weaken this CDF signal for a charged $R$-hadron
(but not the jets + missing momentum OPAL signal, 
for which muonic jets are retained). 
In later sections, we will use case (1) as part
of our normal scenario-1, or ``SC1'', choices.
Clearly, it will be important
to explore sensitivity to the $\vev{\Delta E}$ case choice.
Of course, the net amount of energy deposited by a $\gl$-jet is
also influenced by the path length, $\lam_T$, of the $\gl$.
As discussed below, a simple model suggests that $\lam_T$ for the $\gl$
is longer than $\lam_T$ for a pion. For the graphs of this section, we will 
use the value of $\lam_T=19$~cm derived from this model (see below).
In later sections, however,
we will discuss sensitivity to doubling and halving $\lam_T$ relative
to this ``SC1'' value.

\begin{figure}[ht]
\leavevmode
\begin{center}
\epsfxsize=4.25in
\hspace{0in}\epsffile{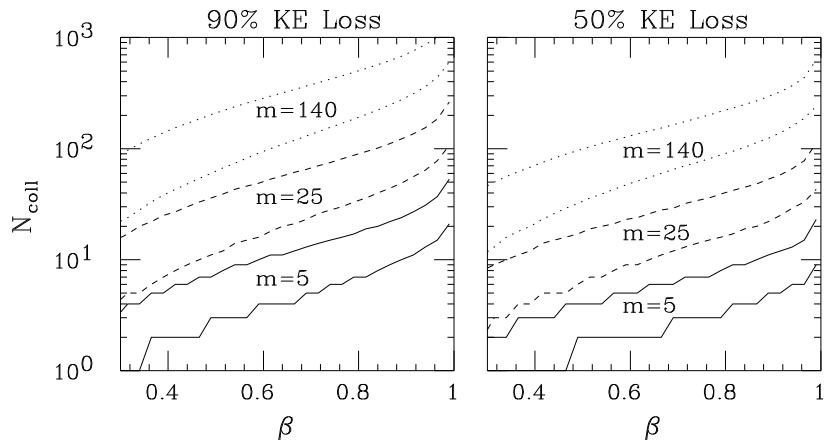}
\end{center}
\caption[]{Number of collisions, $N_{\rm coll}$, required for
an $\rzero$ of the indicated mass (in GeV units)
to deposit 90\% or 50\% of its kinetic energy given the initial $\beta$
plotted on the $x$ axis. The upper and lower lines of a given
type are for $\vev{\Delta E}$ cases (2) and (1), respectively.
The last $\beta$ point plotted is $\beta=0.99$.}
\label{nstop} 
\end{figure} 

Turning now to the $\rzero$,
we compute the number of collisions $N_{\rm coll}$ required to 
deplete a certain percentage of the $\rzero$'s initial kinetic energy.
We carry out this computation by starting the $\rzero$ out
with a given $\beta$ and stepwise reducing its kinetic energy
according to the $\vev{\Delta E}$ given in Fig.~\ref{deltae}.
Results for $\vev{\Delta E}$ cases (1) and (2) are plotted in 
Fig.~\ref{nstop} for $\mrzero=5$, $25$ and $140\gev$. 
It is clear from this figure that what is important is how
the initial $\beta$ correlates with $\mrzero$ in the experimental
situations of interest. 
The initial $\beta$'s that will be of relevance for these masses
(which will prove to be of particular interest) are:
$\beta\sim 0.95-0.99$ for $\mrzero\sim 5\gev$ at LEP 
and $\mrzero\sim 25\gev$ at 
the Tevatron; and $\beta\sim 0.5-0.8$ for $\mrzero\sim 25\gev$ at LEP
and $\mrzero\sim 140\gev$ at the Tevatron. In all cases, we see that
a substantial number of collisions are required in order that 
the $\rzero$ deposit a large fraction of its kinetic energy
as a result of hadronic collisions.

To interpret the above results it is necessary to know
the number of hadronic collisions that the $\rzero$ is likely to experience
as it passes through the detector. Further,
it is important to know how much of the energy deposited
in a given hadronic collision will be measured as 
visible energy and, therefore, used in determining the energy of 
the associated `jet'. 
In assessing the latter, we employ the following approximations.
\bit
\item
For a neutral $\rzero$ (which interacts
strongly only --- no ionization), we presume that the energy deposited in
both elastic and inelastic hadronic
collisions in the calorimeters will contribute to `visible' energy
in much the same way as do energy losses by a pion. In this case,
the calorimeter (which is calibrated using pion beams) will correctly
register the amount of energy deposited by the $\rzero$. 
This should probably be more thoroughly studied in the case of elastic
collisions for which all the energy deposited resides in recoiling nucleons
which could have a somewhat different probability for escaping the absorbing
material and creating visible energy in the scintillating material.
\item
We assume that the energy deposited in uninstrumented iron,
such as that which separates the calorimeters
from the muon detection system in the CDF and D0 detectors,
is not visible.
\eit

For our cross section models,
the number of hadronic collisions of the $\rzero$
as it passes through the detector is determined by the
total (and not just the inelastic) cross section for $\rzero$ scattering
on the detector material.  This is normally rephrased in terms
of the interaction length $\lam_T$ in iron (Fe). 
The average number of collisions is then given by the number of equivalent
Fe $\lam_T$ interaction lengths that characterizes the detector.  
(However, it is conventional for detectors to be
characterized in terms of their thickness expressed
in terms of the number of inelastic collision lengths, $\lam_I$, in Fe.)
For the pion (which we take to be representative of a
typical light hadron), we have already noted that 
$\lam_T(\pi)\sim 11~\mbox{cm}$
and $\lam_I(\pi)\sim 17~\mbox{cm}$ \cite{pdg}.
The equivalent CDF and D0 detector `thicknesses' are specified
in terms of the number of $\lam_I(\pi)$. For all but a small angular
region, the D0 detector thickness ranges from $13-19$ $\lam_I(\pi)$,
depending upon the angle (or rapidity) (the smallest number
applying at $\eta=0$ and the larger number at $\eta\sim 1.5$).
However, of this, a large fraction is in the CF or EF toroid magnets
and is uninstrumented. The instrumented thickness in which energy
deposits are recorded ranges from $\sim 7\lam_I(\pi)$ at $\eta=0$ to 
$\sim 9\lam_I(\pi)$ at $\eta\sim 1.5)$. The CDF detector thickness
at $\eta=0$ consists of about $4.7\lam_I(\pi)$ of instrumented calorimetry
and $\sim 2.9\lam_I(\pi)$ of uninstrumented steel in front of the outer muon
chamber. The instrumented portion of the muon detection system
is fairly thin and will lead to little energy deposit.
The LEP detectors have similar thickness for 
the instrumented category. In particular, at $\eta=0$ 
OPAL has about $2\lam_I(\pi)$
of electromagnetic calorimetry and about $4.7\lam_I(\pi)$ in the
instrumented iron return-yoke hadron calorimeter.
Further, no additional uninstrumented iron is placed between the
magnet return yoke and the muon detectors (which are drift chambers).
To summarize, instrumented thicknesses at $\eta=0$ are
$\sim 5\lam_I(\pi)$ for CDF, $\sim 6.5\lam_I(\pi)$
for OPAL and $\sim 7\lam_I(\pi)$ for D0.
At $\eta=1.5$ the thickness is 
perhaps as large as $9\lam_I(\pi)$ at D0. 
For $\eta\lsim 1$, uninstrumented
sections add about $3\lam_I(\pi)$ for CDF and $6\lam_I(\pi)$ for D0
in front of the muon chambers.
To get the number of $\lam_T(\pi)$
that corresponds to a given number of $\lam_I(\pi)$, multiply the latter
by $\sim 1.6$. Thus, the 5 (CDF), 6.5 (OPAL) and 7 (D0)
$\lam_I(\pi)$ for small $\eta$ 
convert to roughly 8 (CDF), 10 (OPAL) and 11 (D0) $\lam_T(\pi)$. 
At $\eta\sim 1.5$ add about 3 $\lam_T(\pi)$ to the CDF and D0 numbers
and perhaps 2 $\lam_T(\pi)$ to the OPAL result.
Uninstrumented thicknesses
for $\eta<1$ are $\sim 5\lam_T(\pi)$ (CDF) and $\sim 10\lam_T(\pi)$ (D0).
OPAL has no additional uninstrumented iron prior to its muon chamber.

We must now correct these thicknesses 
for the relative size of $\sigma_{\rzero N}$ as compared to $\sigma_{\pi N}$,
using the fact that $\lam_T(\pi)\propto 1/\sigma^T_{\pi N}$.
To estimate $\sigma^T_{\rzero N}$, we employ the two-gluon exchange model for
the total cross section developed in detail in Ref.~\cite{gunsop}.
Compared to the $\pi N$ cross section, the $\rzero N$ cross section
must be increased by the ratio of $C_A/C_F=9/4$ to account for the
color octet nature of the $\rzero$ constituents, and it must be
multiplied by $\vev{r^2_{\rzero}}/\vev{r^2_{\pi}}$, where $\vev{r^2}$
is the (transverse) size-squared of the particle.  In the simplest
approach, which has substantial phenomenological support, $\vev{r^2}$
is inversely proportional to the square of the reduced constituent 
mass of the bound
state constituents: $\vev{r^2_{\pi}}\propto 4/m^2_q$ vs.
$\vev{r^2_{\rzero}}\propto 1/m^2_g$ (for $\mgl\gg m_g$), 
where $m_q$ and $m_g$ are constituent
light quark and gluon masses, respectively.  Taking them to be similar in
size, we find $\sigma^T_{\rzero N}\sim (9/16)\sigma^T_{\pi N}$,
yielding $\lam_T(\rzero)\sim (16/9)\lam_T(\pi)\sim 19~\mbox{cm}$.
Using the factor of 9/16, and rounding up,
the 8 (CDF), 10 (OPAL) and  11 (D0) $\lam_T(\pi)$ 
instrumented thicknesses at small $\eta$ convert to 5 (CDF), 
6 (OPAL) and 7 (D0) $\lam_T(\rzero)$. 
About 2 $\lam_T(\rzero)$ should be added for $\eta\sim
1.5$. For $\eta<1$, about 3 (CDF) or 6 (D0) $\lam_T(\rzero)$
uninstrumented interactions occur before the $\rzero$ reaches the outer muon
detection chambers. Below, we present results for 6, 7 and 8
instrumented hadronic interactions, as appropriate for 
the average measured energy deposit of $\rzero$'s in the $\eta<1.5$
region at CDF, OPAL and D0, respectively.
For later reference, it is important to note that the 8 hadronic
interaction results are also appropriate for
the total energy lost (even though not all is measured)
due to hadronic collisions
before reaching the outer (central) muon chambers at CDF.

Obviously, a refined analysis by the
detector collaborations to improve on the above will be quite worthwhile.
More important, however, is understanding the extent to which
the $\mgl$ region that can be excluded experimentally
is sensitive to $\lam_T(\rzero)$.
This will be examined when we consider exclusion limits based
on OPAL and CDF analyses.

\begin{figure}[ht]
\leavevmode
\begin{center}
\epsfxsize=4.25in
\hspace{0in}\epsffile{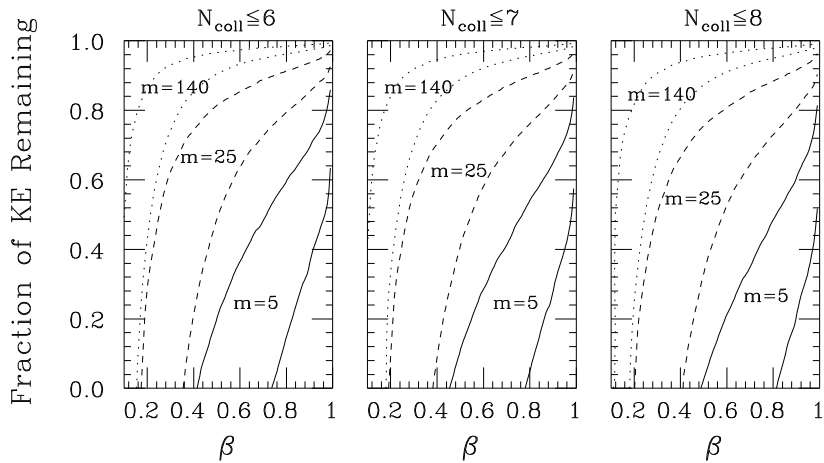}
\end{center}
\caption[]{The fraction of kinetic energy retained by
the $\rzero$ is plotted as a function of its initial $\beta$
for the cases of $N_{\rm coll}\leq 6$, 7 and 8 for $\mrzero=5$, $25$
and $140\gev$. Upper and lower curves for a given mass are
for $\vev{\Delta E}$ cases (2) and (1), respectively.}
\label{eloss} 
\end{figure}

Our results for the fraction of the $\rzero$ kinetic energy
that is not deposited in the calorimeter (which will be the same
as one minus the fraction included in the visible $\gl$-jet energy/momentum)
after $N_{\rm coll}=6$, 7 and 8 hadronic collisions are presented
in Fig.~\ref{eloss} as a function of the initial $\beta$ of the $\rzero$.
Below, we make several observations that will be useful for understanding
borderline cases that will arise in subsequent sections.

For OPAL at LEP (recalling
that the number of hadronic collisions of the $\rzero$
in the OPAL detector is close to 7):
\bit
\item
For a 5 GeV $\rzero$ with large $\beta\sim 0.98$, 
the triple-Pomeron [case (2)]
$\vev{\Delta E}$ implies that 7 interactions will deposit
only about 20\% of the $\rzero$ kinetic energy. The constant cross section
case (1) $\vev{\Delta E}$ implies that about 45\% of the KE
would be deposited in 7 interactions. 
\item
For $\mrzero=25\gev$, $N_{\rm coll}=7$ and initial $\beta\gsim 0.5$, 
the case (2) [(1)] cross section form would predict
that no more than 20\% [60\%], respectively,
of the $\rzero$ kinetic energy would be deposited in the calorimeter.
\eit
For our CDF Tevatron analysis:
\bit
\item
For $\mrzero=25\gev$ and initial $\beta\gsim 0.95$, less than
8\% of the KE would be deposited in 6 interactions
for the case (2) triple-Pomeron parameterization and less than 15\% for
the case (1) constant cross section choice.
\item
For $\mrzero=140\gev$ and initial $\beta\gsim 0.5$, no more than 5\% [10\%]
of the $\rzero$'s KE would be deposited in case (2) [(1)] and contribute to
visible energy in the detector.
\eit
The key overall observation is that, in all cases,
a large fraction of the gluino's
kinetic energy will not contribute to visible energy in the detector.

We now specify how events containing a stable $\rzero$ must be treated 
at the parton level in the
standard OPAL and CDF analyses of the jets plus missing momentum channel
that will be of special interest in what follows. The procedure
given below assumes that the calorimeter calibration is such
that energy deposited in the calorimeter by hadronic
interactions is correctly measured. (This should be the case
given that calorimeter calibration is established using a pion
beam of known energy.)
\bit
\item
As usual, in each event 
the visible three-momentum for a $q$, $\anti q$ or $g$ jet
is taken equal to its full three-momentum and its energy 
is taken equal to the magnitude of its three-momentum.
\item
The visible energy of a $\gl$ (as measured by the calorimeter)
is taken equal to the total energy deposited in the instrumented
calorimeter due to the $\gl$'s hadronic collisions. 
\item
The magnitude of the three-momentum assigned to a $\gl$
is taken equal to its visible energy (\ie\ as if the visible
$\gl$-jet were massless) and the direction of the three-momentum
is given by the direction of the $\gl$.
\item
The invisible or missing momentum three-vector is computed as minus
the vector sum
of all the final-state three-momenta as defined above. 
Only transverse missing momentum
is relevant for the experimental analyses.
\item 
As usual, the absolute magnitude of the missing transverse momentum
is termed the invisible or missing transverse energy.
\eit
An alternative way of thinking about this is that for each $\gl$-jet
one computes the missing momentum as the difference
\beq
|\vec p_{\rm true}|-|\vec p_{\rm
apparent}|=\mgl(\beta\gamma-X(\gamma-1))\,,
\label{etform}
\eeq
where $X$ is the fraction of the $\gl$'s kinetic energy deposited
and measured in the calorimeters of the detector:
$|\vec p_{\rm apparent}|= X\times KE=X\mgl(\gamma-1)$.
The direction of a given $\gl$'s contribution to missing momentum
is the direction of the $\gl$. Note that even if $X=1$, \ie\
all the kinetic energy is seen by the detector, we find
missing momentum associated with the $\gl$-jet
of magnitude $\mgl\left(1-\sqrt{(1-\beta)/(1+\beta)}\right)$,
which is substantial for large $\mgl$ unless $\beta$ is small.

In the LEP and Tevatron analyses it will be important to note
that since $\gl$'s are produced in pairs and in association
with other jets with significant transverse momentum,
the net missing momentum from combining the missing momenta of 
the two $\gl$'s will not generally point
in the direction of either of the $\gl$-jets.
Thus, $\gl$-pair events will normally pass cuts requiring an azimuthal
or other separation between the direction of the missing momentum
in the event and the directions of the various jets.

\subsection{Ionization energy deposits and the $\gl\to \rpm$ possibility}

We must now consider the possibility that the $\gl$ does not
fragment just to an $\rzero$ that propagates through the detector
without charge exchange.  It might also have a significant
probability for fragmenting to a (pseudo-stable) charged state, $\rpm$,
when initially produced and after each subsequent hadronic interaction
in the detector. (An example of an $R^+$ state would be a $\gl u \anti d$ bound
state.) We will assume that the initial and subsequent 
fragmentation probabilities
are all the same. (We denote the common probability by $P$.)
This would be the case if each time the $R$-hadron containing the $\gl$
undergoes a hadronic interaction in the detector
the light quarks and/or gluon(s)
are stripped away and the $\gl$ then fragments independently of
the previous $R$-hadron state.
A simple model for estimating $P$ is the following.  
First, assume that the $\gl$ is 
more likely to pick up a quark-antiquark pair to form a mesonic $R$-hadron
than three quarks to form a baryonic $R$-hadron.  If $u,d$ ($u,d,s$) 
quark and antiquark types are equally probable, then 
of the 4 (9) possible quark-antiquark pairs only 2 (3) are charged
and $P=1/2$ (1/3) {\it if the probability for fragmentation to $\gl g$ is
zero.}
Of course, if the $\rzero=\gl g$ bound state is the lightest $R$-hadron
or is at least very close in mass to the $\gl q\anti q$ $R$-hadrons, 
we expect that this latter probability is actually quite
significant. If we assign the $g$ 
a probability equivalent to all the quark-antiquark
pair combinations included above, then $P=1/4$ (1/6) in the $u,d$
($u,d,s$) cases, respectively. 
Thus, it would seem that $P<1/2$ is quite likely.
In considering the $\rpm$ states and the various neutral $R$-hadron
states on a similar footing, we are implicitly assuming that 
all are stable against decay as they traverse the detector, \ie\
that their lifetime is longer than $\sim 10^{-7}~\mbox{sec}$.
This will not be the case unless all the mass differences 
between the various states are smaller
than $\mpi$. Current estimates for the mass differences
are too uncertain to reliably ascertain whether or not this is the case
\cite{sharpe}.

\begin{figure}[ht]
\leavevmode
\begin{center}
\epsfxsize=4.25in
\hspace{0in}\epsffile{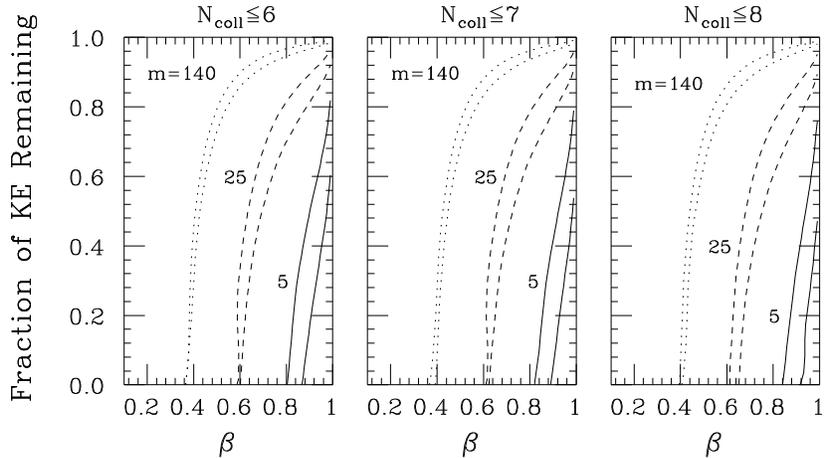}
\end{center}
\caption[]{The fraction of kinetic energy retained 
(\ie\ that is not deposited) by
a singly-charged $\gl$ bound state is 
plotted as a function of its initial $\beta$
for the cases of $N_{\rm coll}\leq 6$, 7 and 8 for $m=5$, $25$
and $140\gev$. Upper and lower curves for a given mass are
for $\vev{\Delta E}$ cases (2) and (1), respectively.}
\label{elossion} 
\end{figure} 

It is useful to consider first the extreme where $P=1$
and compute the total amount of energy deposited, including both hadronic
interactions and ionization. 
The hadronic energy losses are presumed to be the
same as already discussed for the $\rzero$. For the ionization energy
losses we employ the standard 
result for $dE/dx$ from Ref.~\cite{pdg}. As before,
we will parameterize the detector in terms of its equivalent size
as if entirely made of Fe. Our procedure will be to integrate the 
ionization energy
loss up to the point of the first hadronic collision at distance $\lam_T$.
The hadronic energy loss at this first collision will be computed for the
then current $\beta$ following our earlier procedures. 
We then integrate $dE/dx$ starting from the $\beta$ value retained
by the $R^{\pm}$ after this first collision over a second $\lam_T$ of distance,
compute the energy loss for this 2nd hadronic collision
using the new current $\beta$, and so forth. We will consider, as before,
a certain number of hadronic collisions, $N_{\rm coll}=6$, 7 or 8.
The $\lam_T$ employed will be 19 cm, as discussed above. Ionization
energy loss will be computed for $N_{\rm coll}$ segments of length $\lam_T$.
The results corresponding to our earlier Fig.~\ref{eloss} are presented
in Fig.~\ref{elossion}. There we plot, as a function
of initial $\beta$, and for $N_{\rm coll}=6$, 7 and 8, the fraction of
kinetic energy of a singly-charged gluino bound state
that is not deposited, after allowing for energy
losses both from hadronic collisions and from ionization.

\begin{figure}[ht]
\leavevmode
\begin{center}
\epsfxsize=4.25in
\hspace{0in}\epsffile{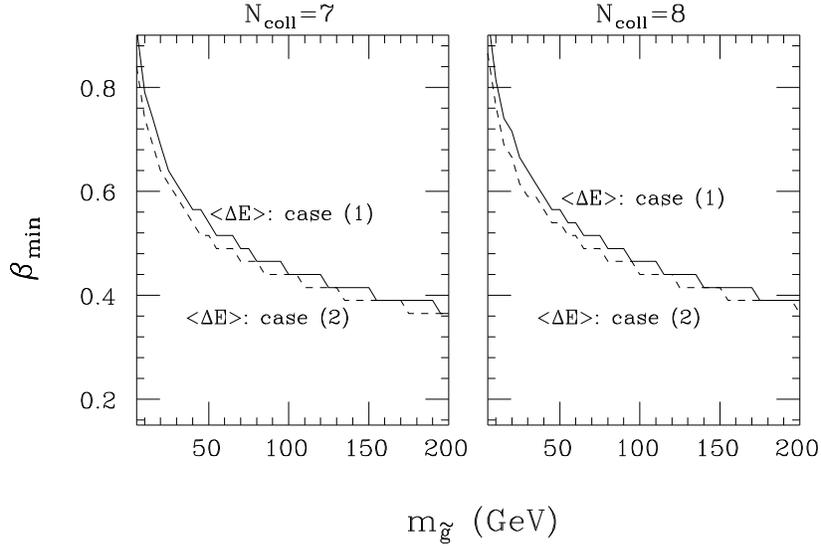}
\end{center}
\caption[]{The minimum velocity $\beta_{\rm min}$
required for a singly-charged $\gl$ bound state to retain
non-zero kinetic energy after $N_{\rm coll}=7$ or 8. The former (latter)
is a rough
estimate of what is required to penetrate to the OPAL (CDF) muon
chambers. Results are plotted for $\vev{\Delta E}$ cases (1) and (2).}
\label{betamin} 
\end{figure}

From Fig.~\ref{elossion} we see that for low enough $\beta$ the $\rpm$
will be stopped in the detector. (For smaller initial $\beta$,
the ionization energy losses are larger and the velocity decreases 
rapidly.) This will be important when
considering limits on a \glsp\ coming from searches for 
a stable charged particle that is heavily-ionizing.
For example, CDF has placed strong constraints on such
a stable charged object if its $\beta$
is small enough for the particle to be
at least twice minimal-ionizing (as measured soon after leaving the
interaction vertex) but large enough that it will penetrate to
the outer muon chamber \cite{stable}. 
For a singly-charged state, twice minimal-ionizing
requires $\beta\gam\lsim 0.85$ or $\beta\lsim 0.75$.
At CDF, roughly $N_{\rm coll}=8$ collisions
are experienced by the charged hadron containing the gluino before
reaching the outer central muon detector system.
Fig.~\ref{elossion} shows that for $\mgl\sim 140\gev$ ($\mgl\sim 25\gev$)
$\beta\gsim 0.4$ ($\gsim 0.6$), respectively,
is required in order that the $\gl$ not
lose all its kinetic energy before reaching the outer muon chamber.
A plot as a function of $\mgl$ 
of the minimum initial $\beta$, $\beta_{\rm min}$,
needed in order that the $\gl$ retain non-zero KE after 7 (8) collisions,
and, therefore, penetrate to the OPAL (CDF) outer muon chambers, 
respectively, is presented in Fig.~\ref{betamin}. Results are given for both
the energy loss case (1) and case (2) models. 
We will later employ the lower limits for $N_{\rm coll}=8$
and case (1) in assessing
our ability to observe a charged gluino bound state as 
a penetrating heavily-ionizing particle in the Tevatron CDF experiment.

Of course, if the $\gl$ fragments part of the time to a neutral
hadronic state and part of the time to a charged state and/or if charge
exchange occurs as a result of hadronic interactions, \ie\
if $P<1$ in the model discussed earlier, the 
results for energy loss and $\beta_{\rm min}$ will be intermediate
between the neutral and purely charged cases discussed above. However,
in obtaining the accelerator limits based on heavily-ionizing
tracks, to be discussed later, 
the reduced value of $\beta_{\rm min}$ that would
apply for $P<1$ is not important since the typical $\beta$
for the produced gluinos is substantially above $\beta_{\rm min}$
for the cases of interest.

\subsection{The momentum experimentally 
assigned to the $\gl$-jet: general $\gl\to\rzero,\rpm$ case}

Let us now return to the visible energy associated with $P>0$
probability for $\gl$ appearance as an $\rpm$.
In the case of a $\gl$ traversing the detector
and sometimes (or always) appearing as an $\rpm$,
the procedure for determining this visible energy is analysis-
and detector-dependent.

First, we must note that both the OPAL and CDF hadronic calorimeters
are constructed out of iron layers.  These are intrinsically non-compensating
in that purely ionization energy losses contribute more to the output
energy measured by the calorimeter than do hadronic collision
losses. For example, the CDF calorimeter is calibrated so that a 50 GeV
pion beam is measured to have energy of 50 GeV.  Using this same
calibration, a 50 GeV muon beam is measured  \cite{cdfscint}
to deposit 2 GeV of energy
whereas its actual energy loss as computed using the standard $dE/dx$
of a muon in iron is only $\sim 1.3\gev$. We define
the ratio of calorimeter response to actual $dE/dx$
loss from ionization as $r$. From the above, $r=1.6$ for iron.
The ionization energy deposited by an $\rpm$ as it moves
through the iron will be converted into $r$ times as much
measured calorimeter energy (which will be included in the
visible energy/momentum of the $\gl$-jet). 
The net energy deposited in the calorimeter
after one complete interaction length will be measured to be
$E_{\rm calorimeter}=rE_{\rm ionization}+E_{\rm hadronic}$, after
including the hadronic energy deposit at the end. 

The next important consideration
is whether there is a track, associated with the $\gl$-jet,
that is identified as a muon. 
\bit
\item
In the CDF jets + missing
energy analysis discussed later, the $\gl$-jet  would be 
declared to be ``muonic'' if:\footnote{We thank H. Frisch and J. Hauser
for clarifying this procedure for us.}
a) the $\gl$ emerges from the interaction in an $\rpm$
whose track is seen in the central tracker and if the $\gl$ is
also in an $\rpm$ state either in the inner muon chamber
or in the outer muon chamber (it is not required that the track be seen
in both);
b) the momentum of the $\rpm$ track in the tracker is measured
to be $>10\gev$. 
c) the energies measured (in an appropriate cone surrounding the charged track)
by the hadronic calorimeter and electromagnetic
calorimeter are less than 6 GeV and 2 GeV, respectively (both conditions
are required to be satisfied, but only the first is relevant
for a $\gl$-jet).

If an event contains a muonic jet, then the event is discarded in the CDF
analysis we later employ. Otherwise, the energy of 
every jet is simply taken equal to the energy as measured by the calorimeters.
\item
At OPAL\footnote{We thank R. Van Kooten for clarifying
the OPAL procedures for us.}
the final magnet yoke acts both as the hadron calorimeter
and the final iron prior to the muon detector. 
A jet is said to contain a muon if there is a charged track
in the central tracker, 
an associated charged track in one of the scintillation
layers of the hadronic calorimeter and a track in the muon chamber.
For a $\gl$-jet, we have approximated
their procedure by requiring that the $\gl$ be
in an $\rpm$ state: a) in the tracker;
b) as it enters the hadronic calorimeter; and c) as
it exits the hadronic calorimeter.

OPAL does not discard events when one or more of
the jets contains a muon identified in
the above way. Rather, the jet energy is corrected assuming
that the charged track identified as a muon is, indeed, a muon.
The procedure for computing the jet energy is as follows.
\bit
\item
Four-momentum vectors are formed for each track and calorimeter cluster
to be included in the jet,
and then summed. The three-momentum employed for a given track is directly
measured in the tracker and the energy component
for the track is computed by assigning it
the pion mass, unless it is identified as an electron or muon. 
(For our purposes, we can neglect the $e,\mu,\pi$ masses.)
Calorimeter clusters are treated as massless particles; the magnitude
of the three-momentum is taken equal to the energy of
the cluster as measured by the calorimeter.
\item To reduce double counting, four-vectors based on the average expected
energy deposition in the calorimeter of each charged track are then subtracted.
\eit
For a $\gl$-jet that has $\rpm$ tracks in the tracker and muon chamber 
that are identified as belonging to a muon, 
this means that the energy and momentum vector magnitude
assigned to the $\gl$-jet will be given by adding the
$\rpm$ track momentum as measured in the tracker to the
total calorimeter response, and then subtracting $2\gev$
to account for the energy deposit of the supposed
minimal-ionizing muon.  If an $\rpm$
track in the tracker does not have an associated penetrating
track in the muon system (according to the above-stated criterion), 
the track is assumed to be that of
a charged pion (it would not be identified as an electron), 
in which case the energy subtracted
will be taken to be that of a pion with the same momentum 
as measured for the $\rpm$ in the tracker. Neglecting the pion
mass, this subtraction is equal to the measured momentum, with
the result that the energy assigned to the $\gl$-jet will equal that
measured by the calorimeter.
Algebraically, we can represent these alternatives by writing
\beq
\ejet =p_{\rm jet}= 
E^{\rm tot}_{\rm calorimeter}+
\thetamuid(\mgl\beta\gam-\mbox{2 GeV})\,,
\label{muonjet}
\eeq
where $\thetamuid=1$ or 0 according to whether there is or is not,
respectively, an $\rpm$  
track identified as a muon associated with the $\gl$-jet.
Note that it is always presumed that the $\gl$-jet is massless so
that $\ejet=p_{\rm jet}$ is presumed to apply.
In the OPAL analyses, $\ejet=p_{\rm jet}$ 
will be defined by this experimental procedure and
will not be the true jet energy or momentum.

\item
A possibly tricky case arises when the $R$ hadron is neutral
and undergoes a hadronic interaction 
in the iron of the hadronic calorimeter (or in the uninstrumented iron
preceding the outer muon chamber at CDF) at a location that is less than
(roughly) a pion interaction length away from a muon chamber.
This could result in a charged track or, even more probably, a 
``shower'' of particles entering
the muon chamber from the outer edge of the iron. The result would
be an anomalous muon signal in the muon chamber. 
In addition, for a track or shower from a hadronic interaction
at the edge of the hadronic calorimeter, the full energy 
loss of the $R$-hadron from this interaction would not be measured
by the calorimeter. These effects fall
outside the simplified treatment that we shall employ, described above,
which assumes that the shower
from a hadronic interaction is completely contained in the iron.
They will be discussed at the end of this section.
For now, we present results obtained assuming complete containment.

\eit

In order to assess the implications of the OPAL
and CDF procedures, we have computed the average result for the energy
(= momentum), $\ejet$, assigned to a gluino jet for 1000 $\gl$'s produced
with a given initial $\beta$, following the OPAL and CDF procedures.
Since the missing momentum for a given $\gl$-jet 
is the difference between the experimental measurement, $\ejet$,
and the true initial momentum of the $\gl$, our focus will
be on expectations for the ratio $\ejet/p_{\rm true}$.
All results for $\ejet$, here and in future sections,
will assume that the shower from a hadronic interaction
occurring in the iron of the hadronic calorimeter is fully contained.
As discussed just above, we believe that the effects of incomplete
shower containment are small.

Consider first the CDF detector configuration.
We assume $N_{\rm coll}=6$ interactions in instrumented iron
and $N_{\rm coll}=2$ uninstrumented interactions between the 
inner muon chamber (which is just outside the hadronic
calorimeter) and the outer muon chamber. When the gluino is initially
produced, and after each subsequent hadronic interaction,
it is assigned charge $|Q|=1$ with probability $P$ and
$Q=0$ with probability $1-P$. Ionization energy losses are incorporated
for any path segment between hadronic interactions for which $|Q|=1$.
Ionization energy losses are multiplied by $r=1.6$ when computing
the calorimeter response. At each hadronic interaction the $\vev{\Delta E}$
of Fig.~\ref{deltae} is assumed to be deposited in the calorimeter
and included in the calorimeter response (with coefficient 1).
If the $\gl$ is charged in the first track segment, charged after
6 interactions and/or also charged after 8 interactions, 
(and has non-zero kinetic energy where it is seen to be charged),
and the earlier described momentum and energy deposit requirements
are satisfied, then 
we presume it will be identified as a muon and the $\gl$-jet is discarded.
If it is not identified as a muon then the $\gl$-jet is retained and
the jet energy is set equal to the energy as measured by the calorimeter. 

\begin{figure}[h]
\leavevmode
\begin{center}
\epsfxsize=4.25in
\hspace{0in}\epsffile{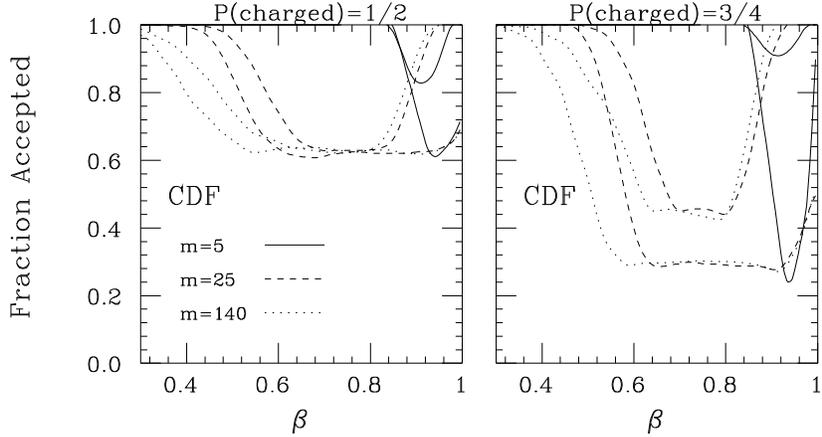}
\end{center}
\caption[]{For $P=1/2$ and $3/4$, we plot, vs. the gluino's initial $\beta$, 
the average fraction of gluino jets that is retained 
when the CDF procedure is followed.
Results are given for $\gl$ masses of 
$m=5$, $25$ and $140\gev$, taking $r=1.6$.
The two curves for a given mass are for $\vev{\Delta E}$ cases (1) and (2), 
the lower curve corresponding to case (2).}
\label{nacceptcdf} 
\end{figure} 

The first important issue with regard to the CDF procedure
is the fraction of $\gl$-jets that are discarded as a result of
the $\gl$-jet being declared to be ``muonic'' (according
to the earlier-stated criteria).
In Fig.~\ref{nacceptcdf}, we plot the average fraction of $\gl$-jets retained
as a function of the gluino's initial $\beta$, for $P=1/2$ and  $3/4$.
Results are given for $\mgl=5$, $25$ and $140\gev$.
This figure shows that there is an intermediate $\mgl$-dependent
range of $\beta$ for which the $\gl$-jet is ``muonic'' 
a significant fraction of the time.
This occurs as a result of the fact that the energy 
(from electromagnetic and hadronic energy deposits) measured
by the hadronic calorimeter drops below 6 GeV at intermediate $\beta$.
(This happens because, when present, 
the $\rpm$ is not sufficiently heavily-ionizing at intermediate $\beta$,
and hadronic energy deposits typically only become large at large $\beta$.)
Note that Fig.~\ref{nacceptcdf} shows that
events are discarded over a larger range of $\beta$ 
for $\vev{\Delta E}$ case (2) as compared to case (1), 
in agreement with expectations following from the fact that case (2)
yields smaller hadronic energy deposits.
For $P=0$, all $\gl$-jets are, of course, non-muonic and are retained.
For $P=1/4$, the fraction of retained $\gl$-jets is above 0.87 for
all $\beta$ values for all masses and both $\vev{\Delta E}$ cases.
$P=1$ is a bit of a special case, as we now describe.

For $P=1$, there are no charge fluctuations
and, for a given $\beta$ and $\vev{\Delta E}$ case,
all $\gl$-jets are either retained or discarded.
For $\vev{\Delta E}$ case (1), we find that the $\gl$-jets are retained
for all values of $\beta$ for all three $\mgl$ values because
the hadronic calorimeter energy deposits (including both ionization
and hadronic collision energy deposits) are large enough to fail 
the $\leq 6\gev$ criterion for a muonic jet.
For $\vev{\Delta E}$ case (2), there is an intermediate range of $\beta$
(dependent upon the value of $\mgl$) for which 
the hadronic calorimeter energy deposits are small enough
to satisfy the $\leq 6\gev$ criterion and the $\gl$-jets are discarded
as being muonic. These intermediate ranges appear as gaps in the
$\vev{\Delta E}$ case (2) curves for $P=1$ in Fig.~\ref{pvisiblecdf} below.
As a result, it turns out that there is a very large difference
in the ability of the jets + missing energy CDF analysis 
to exclude a heavy \glsp\ in case (1), which yields good sensitivity,
as compared to case (2), which yields poor sensitivity.
This is clearly an artifact of the published CDF analysis procedures.
To avoid this sudden change in efficiency, we recommend that CDF
re-analyze their data without discarding muonic jets.

\begin{figure}[p]
\leavevmode
\begin{center}
\epsfxsize=4.25in
\hspace{0in}\epsffile{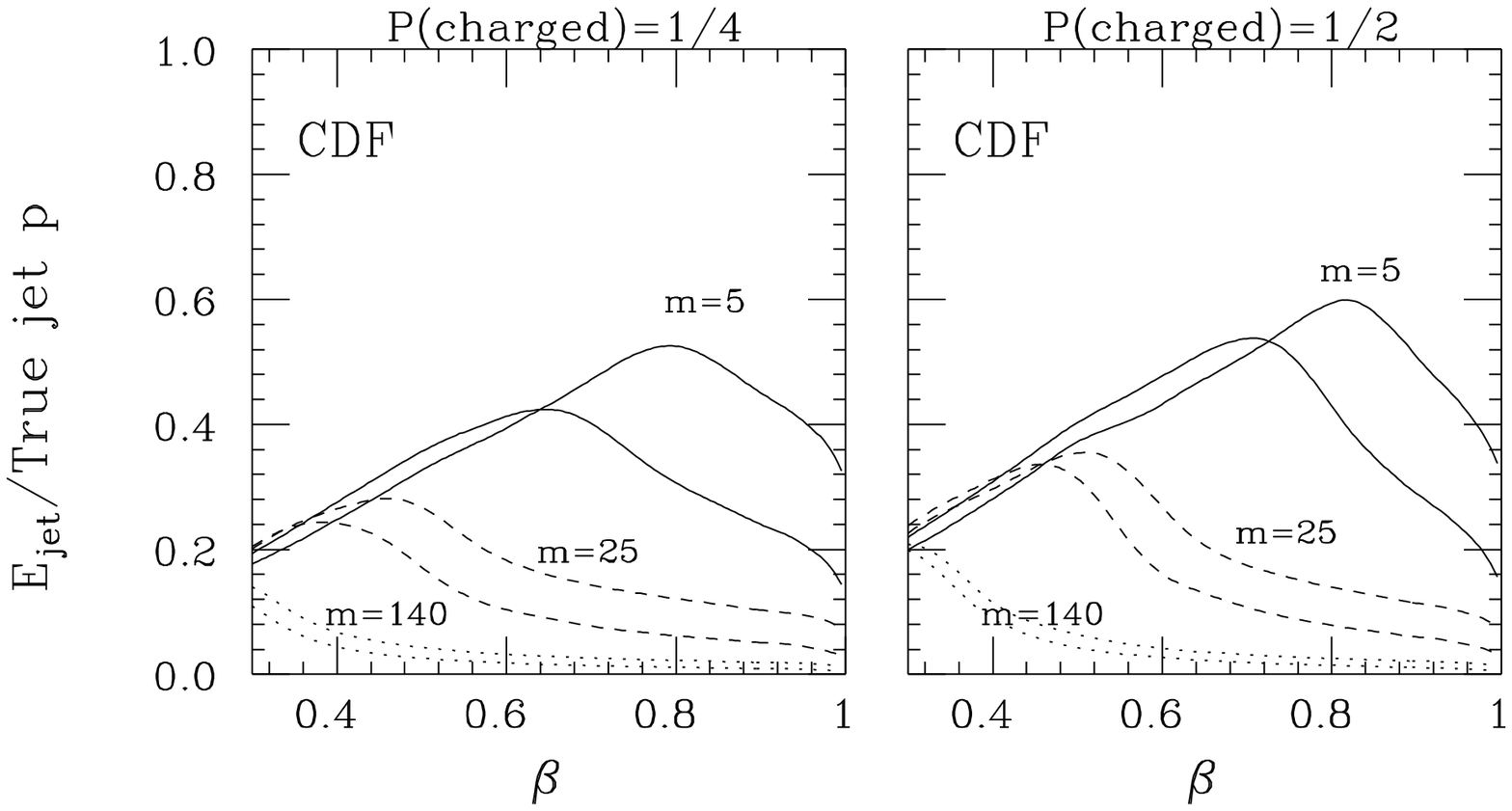}
\smallskip
\epsfxsize=4.250in
\hspace{0in}\epsffile{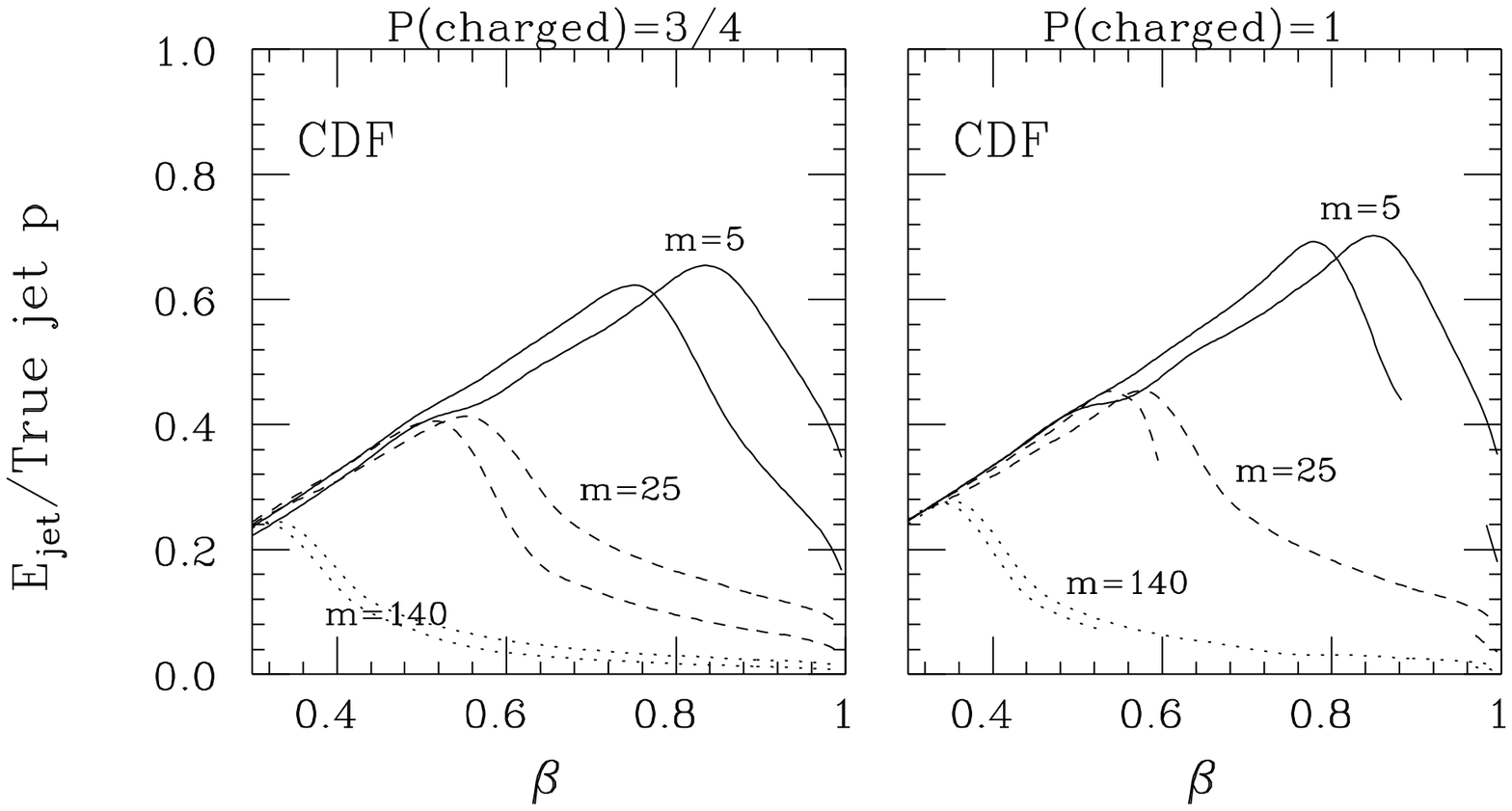}
\end{center}
\caption[]{For $P=1/4$, $1/2$, $3/4$ and 1, 
we plot, vs. the gluino's initial $\beta$, 
the average measured jet energy $\ejet $ 
as a fraction of the gluino's initial momentum
for $\gl$-jets that are not declared to be muonic
[using the CDF procedures].
Results are given for $m=5$, $25$ and $140\gev$, taking $r=1.6$.
The two curves for a given mass are
for $\vev{\Delta E}$ cases (1) and (2).
Raggedness in the numerical results, reflecting the fact that 
in our approximation the hadronic interactions
only occur at precise intervals of 19 cm whereas ionization losses
occur continuously, has been smoothed out in the plots.
Gaps in the case (2), $P=1$ curves are where the $\gl$-jet is
declared to be muonic.}
\label{pvisiblecdf} 
\end{figure}

The second important issue is the measured energy of the retained $\gl$-jets.
In Fig.~\ref{pvisiblecdf}
we plot the average (over 1000 produced $\gl$'s) energy assigned to the
accepted $\gl$-jets 
divided by their actual initial momentum for $P=1/4$, $1/2$. $3/4$ and $1$. 
Remarks relevant to borderline cases that will be important
in the CDF jets + missing momentum analysis are the following.
\bit
\item
For $\mgl=25\gev$ and initial $\beta\gsim 0.95$, 
the fraction $X$ of the $\gl$'s actual momentum that is 
included in the measured $\ejet$
is in the range $X\leq 0.15$ for all $P$ values
and both $\vev{\Delta E}$ cases.
\item
For $\mgl=140\gev$ and initial $\beta\gsim 0.6$, one
finds $X\leq 0.1$ for all $P$ values and both $\vev{\Delta E}$ cases.
\eit
The only exception to these generalities 
occurs when $P=1$ and for $\vev{\Delta E}$
case (2), for which $\gl$-jets with the above masses and $\beta$ values
are discarded as being muonic.
Aside from this, we can anticipate that $\gl\gl$ production at CDF will
result in an event with large missing momentum.

\begin{figure}[p]
\leavevmode
\begin{center}
\epsfxsize=4.25in
\hspace{0in}\epsffile{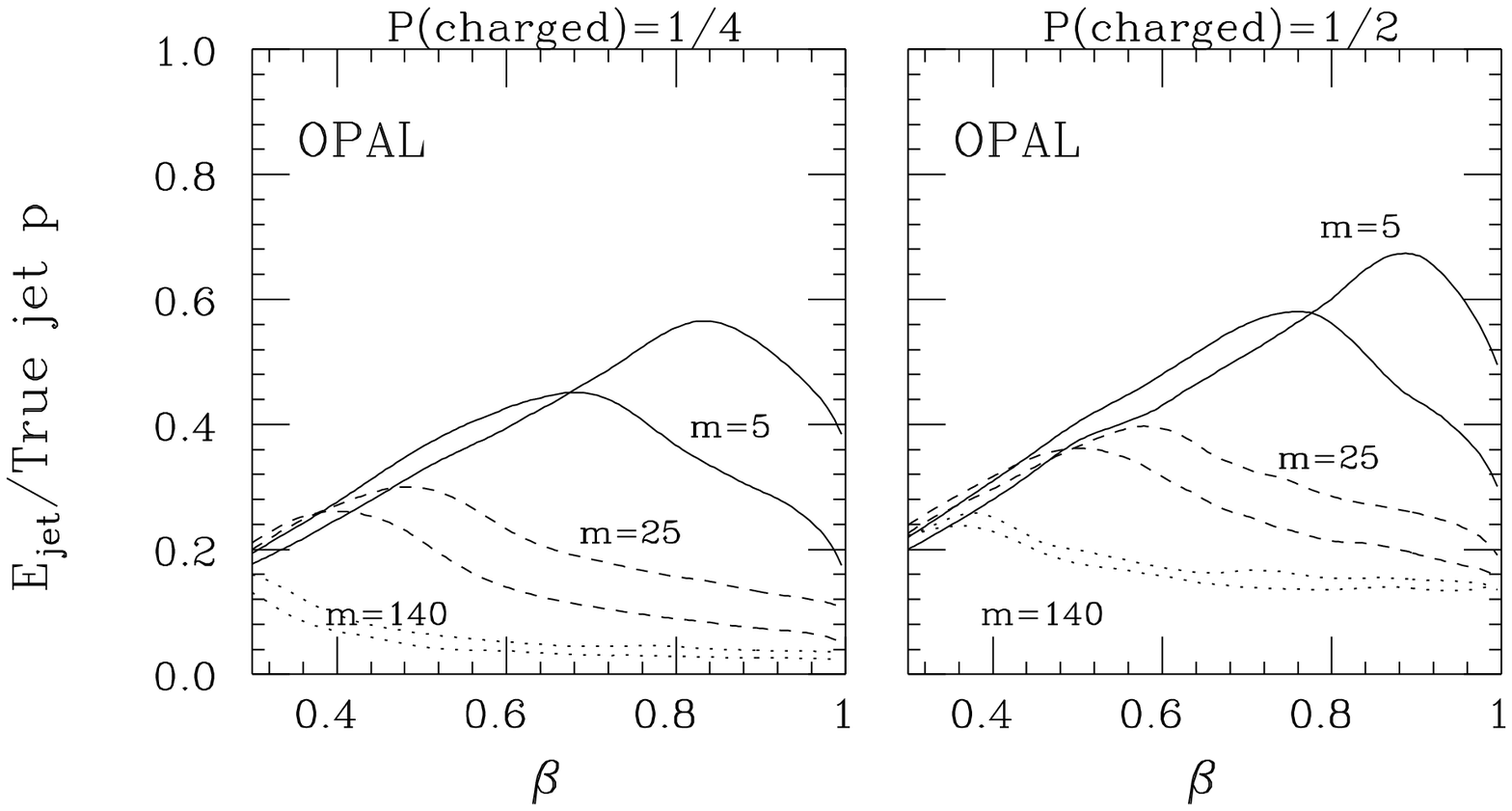}
\smallskip
\epsfxsize=4.250in
\hspace{0in}\epsffile{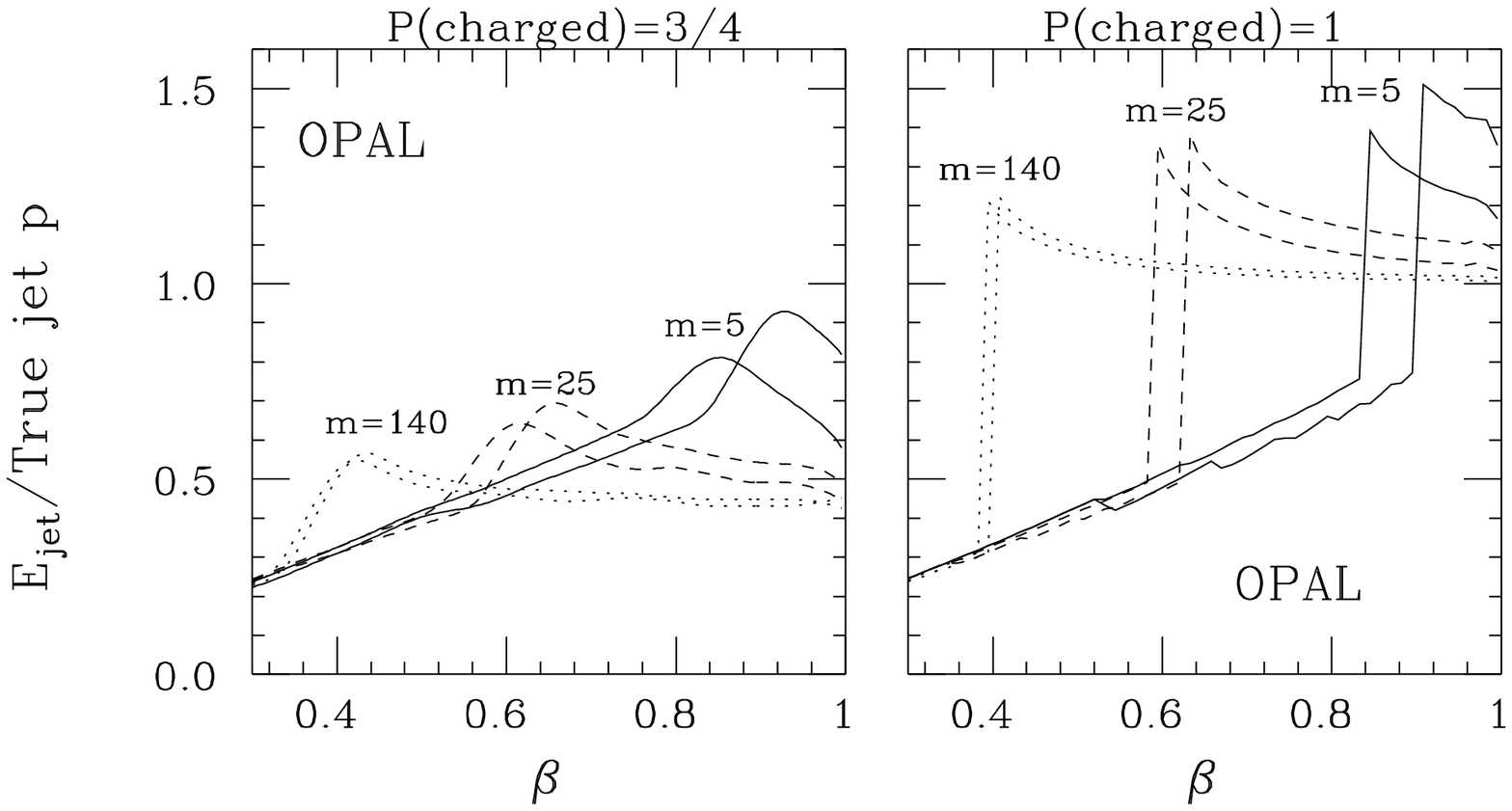}
\end{center}
\caption[]{For $P=1/4$, $1/2$, $3/4$ and 1, 
we plot, vs. the gluino's initial $\beta$, 
the average jet energy $\ejet $ [computed using the OPAL procedures,
cf. Eq.~(\ref{muonjet})] 
as a fraction of the gluino's initial momentum.
Results are given for $m=5$, $25$ and $140\gev$, taking $r=1.6$.
The two curves for a given mass are
for $\vev{\Delta E}$ cases (1) and (2).
Raggedness in the numerical results, reflecting the fact that 
in our approximation the hadronic interactions
only occur at precise intervals of 19 cm whereas ionization losses
occur continuously, has been smoothed out in the plots.}
\label{pvisibleopal} 
\end{figure}

In the case of OPAL, if the $\gl$-jet has $|Q|=1$ in the tracker and
if it emerges into the muon chamber
with $|Q|=1$ and positive kinetic energy after $N_{\rm coll}=7$ 
interactions then it is assumed that the track in the tracker
will be identified as a muon and that the jet energy correction of
Eq.~(\ref{muonjet}) will be applied.
If there is no track identified as a muon
then the jet energy is set equal to the 
energy as measured by the calorimeter. In Fig.~\ref{pvisibleopal},
we plot the average (over 1000 produced $\gl$'s) energy assigned to the
$\gl$-jet divided by its initial momentum for 
$P=1/4$, $1/2$. $3/4$ and $1$. 
For $P\leq 1/2$, the $\beta$ ranges of importance at LEP 
will be those where $\ejet$ is only a fraction of the full initial
momentum of the $\gl$. This is not unlike the CDF result. However,
for large $P$ there are very substantial differences as compared to CDF.
For example, when $P=1$ 
most of the $\rpm$ kinetic energy is deposited in the form of
ionization energy losses. If its $\beta$ is too small for
penetration to the muon detector, then the calorimeter
response gives $\ejet$ close to $r=1.6$ times the $\gl$ kinetic
energy. Once the $\beta$ is large enough for penetration to
the muon chamber and the $\rpm$ tracker track is identified as
a muon, $\ejet$, as determined 
from Eq.~(\ref{muonjet}), jumps to a level that
reflects the addition of the $\gl$ momentum as measured for the charged track
in the tracker. For $P=3/4$ one is in transition
from the typical low $P$ situation to $P=1$. 
To interpret $\ejet/p_{\rm true}>1$ it is important to 
recall that it is $|\ejet-p_{\rm true}|$
that determines whether the $\gl$-jet will result in missing momentum.
Values of $\ejet/p_{\rm true}$ significantly different
from 1 (whether larger or smaller) will lead to missing momentum.
Thus, at OPAL, events containing $\gl$'s will generally have some missing
momentum even when $P$ is large. 

With regard to values of $\mgl$ and associated
typical $\beta$'s that will be interesting borderline
cases for the OPAL jets + missing momentum analysis, we note the following.
\bit
\item
Consider first $\mgl=5$ and $\beta\sim 0.98$.
Fig.~\ref{pvisibleopal} shows that if $P$ is not large,
then the measured jet energy is small and there will be large missing
momentum associated with a $\gl$-jet. If $P\sim 1$,
$\ejet/p_{\rm true}$ is somewhat bigger than 1,
which as noted above will lead to some missing momentum,
but not as much as is typical at lower $P$.
\item
For $\mgl=25$ and $0.5\lsim\beta\lsim 0.8$, Fig.~\ref{pvisibleopal}
shows that the measured jet energy is typically a significant
fraction of the true momentum once $P>1/2$. For
$P=1$, $\ejet/p_{\rm true}$ is not far
from 1 for this $\beta$ range.
\eit
Thus, we can anticipate that $P=1$ will yield the weakest OPAL signal
at both ends of the mass range of interest.

Hopefully, the discussion of this subsection has provided intuition
as to the characteristics of $\gl$-jets as measured in the CDF
and OPAL detectors. We have presented results for what
we believe to be the most resonable choice of the interaction
length $\lam_T$ of the gluino. However, it will be important
to assess sensitivity to changes in $\lam_T$. Smaller $\lam_T$
(larger total cross section) yields more hadronic collisions
and, therefore, more hadronic energy deposit
and more slowing down of the $\gl$; larger
$\lam_T$, the reverse. We have found that the greatest sensitivity
to $\lam_T$ arises in the case of the CDF jets + missing
momentum analysis where larger $\lam_T$
implies that the smaller hadronic energy deposits and
smaller ionization energy deposits (due to less rapid slowing down
of the $\gl$) result in many $\gl$-jets being declared to be muonic when $P$
is large, implying a loss of sensitivity for the published
analysis procedures.  In order to provide a representative
sample of possibilities for both $\vev{\Delta E}$ and $\lam_T$,
we will consider three scenarios (denoted SC) in the jets + missing momentum
analyses that follow:
\bit
\item SC1: $\lam_T=19$~cm (as employed in the discussion 
and graphs given earlier in this section) and $\vev{\Delta E}$
case (1).
\item SC2: $\lam_T=9.5$~cm and $\vev{\Delta E}$ case (1),
implying twice as many hadronic interactions, 
and, therefore, larger measured energy for a given $\gl$-jet as
compared to the SC1 case.
\item SC3: $\lam_T=38$~cm and $\vev{\Delta E}$ case (2),
implying only half as many hadronic interactions
and small energy deposit per hadronic collision, leading
to much smaller measured energy for a given $\gl$-jet
as compared to the SC1 case.
\eit

In the OPAL and CDF analyses of the next sections, 
our procedure will be to generate
events containing a pair of gluinos, and then let each gluino
propagate through the detector allowing for charge changes
according to a given choice of the probability $P$ at each
hadronic interaction. The frequency of hadronic
interactions is determined by the choice of $\lam_T$,
and the amount of energy deposit at each interaction 
is determined by the $\vev{\Delta E}$ case.
The characteristics of each event are then computed, including
overall missing momentum, jet kinematics, \etc\ The relevant
cuts are then applied. Only this type
of Monte Carlo event-by-event procedure allows
for all the different types of fluctuations in charge, velocity
and so forth that take place if gluino-LSP's are being produced.

\subsection{Effects of incompletely contained hadronic interaction showers}

Finally, let us now return to the effects that arise if there is a hadronic
shower at the outer edge of the hadronic calorimeter and, in the case
of CDF, at the outer edge of the iron shield between the inner and outer muon
chamber. This mainly affects the jets + missing momentum
analyses of OPAL and CDF and the heavily-ionizing track
analysis of CDF. The details of these analyses will
be discussed in later sections, but we find it convenient
to summarize the influence of edge-showers here.
We have studied the effects on the analyses in the following very extreme
approximation. We assume: a) that 
the last hadronic interaction in the calorimeter
is completely uncontained and therefore does not contribute
to measured $\gl$-jet energy; and b) that the last hadronic interaction
in the hadronic calorimeter, and, for CDF, also the
last interaction in the iron shield, yields a
charged track in the subsequent muon chamber. We find the following results.
\bit
\item Small $P$:
In the OPAL and CDF jets + missing momentum analyses, 
the jet is declared to contain a muon
only if a charged track is also seen in the tracker.
For small $P$, this probability is small.
The main effect would then be that the energy
of the hadronic interaction shower 
at the edge of the calorimeter would not be deposited in the calorimeter,
thereby leading to a decrease in the measured jet energy.
We find that the resulting increase in missing momentum
would be modest ($\lsim 10-15\%$), even in our extreme approximation.
This would yield some enhancement in
the efficiency for the jets + missing momentum 
signal in the OPAL and CDF analyses, but not enough to significantly
alter the limits on $\mgl$ that are obtained.

The heavily-ionizing track
signature is not relevant for small $P$ since there is low probability
for a charged track in the tracker.

\item Large $P$:
For large $P$ values, in the jets + missing momentum
OPAL analysis, the $\gl$-jet will be declared to contain
a muon regardless of whether
there is an extra muon-chamber track or shower. Also, since
most of the $R$-hadron energy losses are in the form of ionization rather 
than from hadronic interactions, we find that
the measured $\gl$-jet energy only decreases
slightly. Thus, the OPAL jets + missing momentum 
results would be little affected. 

Turning to the CDF jets + missing momentum analysis, 
we again note that, when $P$ is large,
most of the measured energy is from ionization energy deposits and 
earlier hadronic interactions, and the incomplete
containment of the tracks/shower of a last hadronic interaction
in the hadronic calorimeter generally has little affect,
{\it provided} the $\gl$-jet is declared not to be muonic. 
(Note that if the incompletely contained shower originates in the 
outer edge of the iron between
the inner and outer muon chambers it would not have been instrumented,
\ie\ would not contribute to measured energy anyway.)
Unless one is right on a borderline, the
small decrease in measured energy due to losing the
shower from the last hadronic interaction in the calorimeter will not cause a 
$\gl$-jet that would otherwise be declared to be non-muonic
to fall into the muonic category. However, we have already seen
in Fig.~\ref{pvisiblecdf} that for $P=1$ we are right on such a borderline,
with case (2) $\vev{\Delta E}$ giving rise to large gaps (in $\beta$)
for which the $\gl$-jet is declared to be muonic whereas for our SC1
case (1) choice the $\gl$-jet is never declared to be muonic. 
We find that failure
to capture any of the energy of the last shower also pushes us past this
borderline. Thus, in our extreme approximation, the loss
of the shower results in much the same phenomenology for CDF
as the SC3 case defined earlier; one finds that 
a very substantial weakening
of the jets + missing energy signal occurs. Of course, as already
noted earlier, the way around this is to re-analyze the CDF
data without throwing away muonic jets, perhaps using something like the OPAL
procedure.

\item Moderate $P$:
For moderate $P$ values, the penetration of a hadronic interaction
shower to the muon chamber would tend
to increase the number of $\gl$-jets that are declared to contain
a muon in the OPAL analysis. The momentum computed for the
extra muon-jets via Eq.~(\ref{muonjet}) will be substantially larger
than otherwise.  On average this 
increase in momentum is only partially offset
by the decrease in the measured calorimeter energy deposit from the jet
due to non-containment of the final shower 
in the hadronic calorimeter. The net result is a modest
decrease in the efficiency
for the jets + missing energy signal. However, the $\mgl$
limit borderline is so sharp at moderate $P$ (see later OPAL results)
that there would be little change in the limits that can
be extracted from the OPAL analysis.

In the CDF analysis, there are two effects.
The extra muon-chamber signal will tend to decrease the number of non-muonic
events because a) there are more events with
tracks in the muon chambers and b) because the
energy deposit measured by the hadronic calorimeter decreases
as a result of incomplete containment of the tracks of the final shower.
However, a sizeable fraction (roughly, 50\% for $\vev{\Delta E}$ case
(1) and $P=1/4$, 1/2, and 3/4, in the $\beta$ regions of relevance) 
of the events that are retained at moderate $P$
(see Fig.~\ref{nacceptcdf}) are non-muonic because of the
absence of a charged track in the tracker. The retention of these
events would be unaffected by the presence of an anomalous
muon-chamber signal. Overall, we find that the decrease in the number
of accepted $\gl$-jets is typically of order 30\%. However,
this decrease is compensated by the fact that
the decrease in measured calorimeter energy
due to incomplete shower containment increases the missing
momentum and, therefore, the efficiency for non-muonic events that contain
such a shower. (Recall that, once accepted, the $\gl$-jet momenta are
computed in the CDF analysis without including any muon correction.) 
Changes in the extracted $\mgl$ limits would not be large.

\item For moderate or large $P$:
The heavily-ionizing track (HIT) searches that can be used to
eliminate a span of $\mgl$ values when $P\geq 1/2$ will be completely
unaffected by an anomalous muon-chamber signal in the case of OPAL
(since the OPAL HIT analysis, described later in section 6,
essentially only uses tracker information)
and will be enhanced in the case of CDF (since the CDF HIT
analysis, discussed in section 7, requires a track
in the inner and/or outer muon chamber in addition to
a HIT in the inner tracker).

\eit

Thus, we think that the effects upon our analyses
of a hadronic collision that leads to an anomalous muon-chamber 
track or shower are small, except in the case of large $P$
in the jets + missing momentum CDF analysis where one is
very sensitive to just how much of the energy
in the final hadronic calorimeter shower escapes into the muon chamber.
We repeat our expectation that this sensitivity could be eliminated by removing
the ``non-muonic'' jet requirement in the CDF analysis.
A study of the effects of incomplete shower containment
is probably best left to the detector groups themselves. 

Finally, we note that events having a shower entering the muon chamber would 
actually appear to provide a potentially spectacular signal for a \glsp\ ---
one that should be specifically searched for. This
signal would appear to be especially promising if $P$ is small
and one focuses on events in which there is no charged track
in the tracker associated with the jet pointing to the muon chamber shower.

\section{Constraints from LEP and LEP2}

\begin{figure}[ht]
\leavevmode
\begin{center}
\epsfxsize=4.25in
\hspace{0in}\epsffile{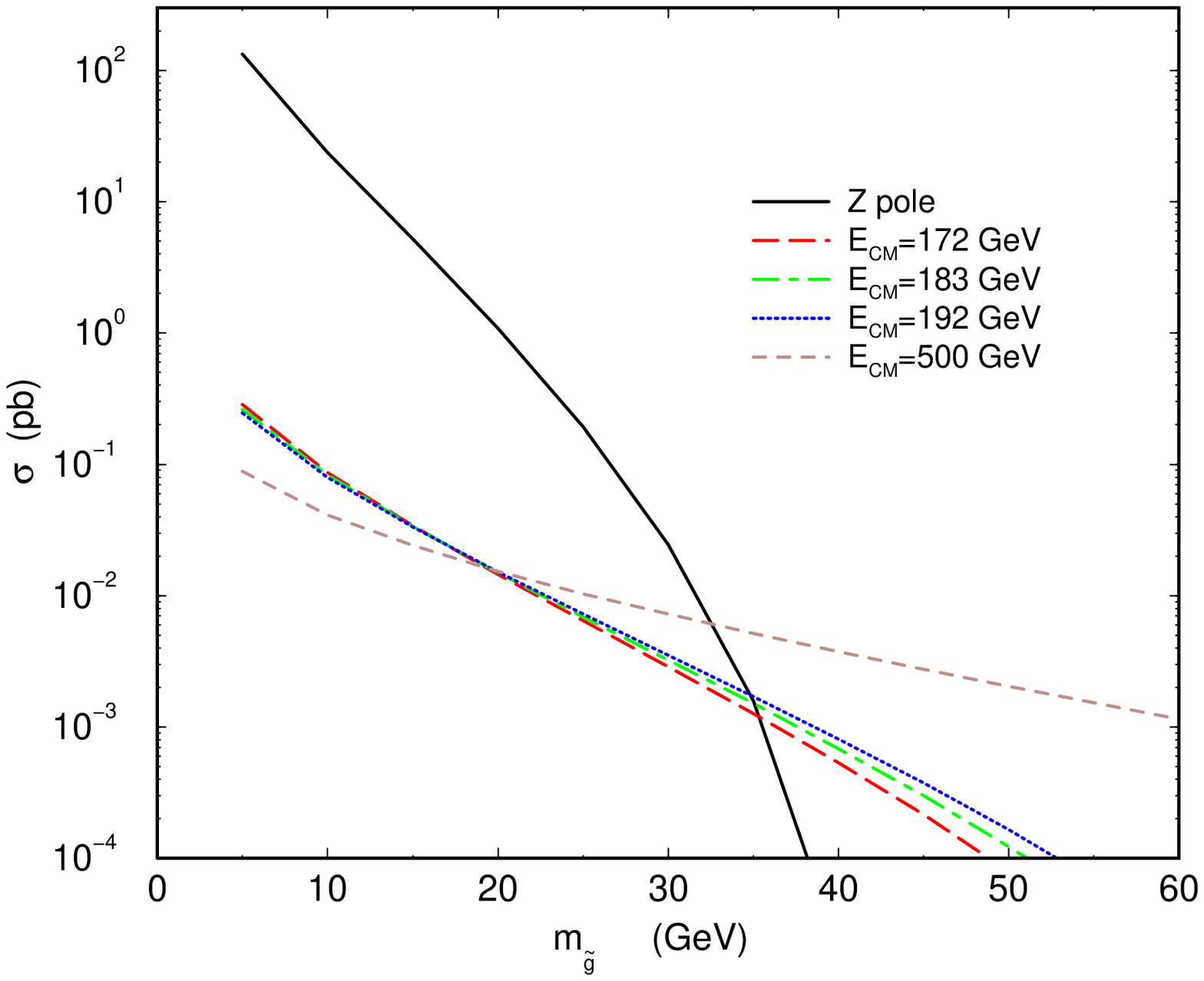}
\end{center}
\caption[]{$\sigma(\epem\to q\anti q \gl\gl)$ as a function
of $\mgl$ for $\rts=\mz$ (solid), 172 GeV (dashes), 183 GeV
(dot-dash),  192 GeV (dots) and $500\gev$. No cuts.}
\label{lepqqglgl} 
\end{figure}

At LEP and LEP2, we assume that 
all other SUSY particles are beyond the kinematic reach
of the machine. The only possible signal for SUSY is then
pair production of two gluinos. Gluinos can only be produced via two processes:
$\epem\to q\anti q\gl\gl$ \cite{cer,css,mts}, 
which can take place at tree-level,
and $\epem \to \gl\gl$ \cite{nper,krol,css}, which takes place via loop
diagrams (involving squarks and quarks).
As discussed later, the latter process
is very model dependent and can be highly suppressed.  Thus, we begin by
focusing on the $q\anti q\gl\gl$ final state. 
We consider both the LEP $Z$-pole data
and higher energy running at LEP2.
The (uncut) $q\anti q\gl\gl$ cross section\footnote{We
have employed a numerical helicity amplitude computation for $\epem\to q\anti q
\gl\gl$ valid for arbitrary $\mgl$; the program is available upon request.
A crossed version of the squared matrix element can also be found
in Ref.~\cite{carlson}.}
is plotted in Fig.~\ref{lepqqglgl}
as a function of $\mgl$ for $\rts=\mz$, $172\gev$, $183\gev$ and $192\gev$. 
Given that the total $\epem\to Z$ cross section is 
$\sim 6\times 10^4\pb$, Fig.~\ref{lepqqglgl} implies that $\br(Z\to q\anti
q\gl\gl)> {\rm few}\times 10^{-6}$ for $\mgl\lsim 25\gev$. Since
millions of $Z$'s have been produced at LEP, we can demonstrate that 
$\gl$'s lighter than this and heavier
than about $5\gev$ can be ruled out. In contrast, Fig.~\ref{lepqqglgl}
makes it clear that very substantial
luminosity at higher LEP2 energies will be required for constraints from
LEP2 data to be competitive. For example, $L=500\pbi$ at
$\rts=192\gev$ will yield only about 4 $\epem\to q\anti q\gl\gl$
events (before cuts) at $\mgl=25\gev$. Also shown in Fig.~\ref{lepqqglgl}
is the uncut $\epem\to q\anti q\gl\gl$ cross section at $\rts=500\gev$,
a possible choice for the next linear collider (NLC). 
One finds $\sigma(q\anti q\gl\gl)<1\fb$ for $\mgl\geq 60\gev$, which
would correspond to 50 events for $L=50\fbi$.  Even for $L=500\fbi$
one finds fewer than 5 events [$\sigma(q\anti q\gl\gl)<.01\fb$]
for $\mgl\geq 140\gev$, which will turn out to be close to the
lower limit that can already be set by using Tevatron data.

\begin{figure}[htp]
\leavevmode
\begin{center}
\epsfysize=2.4in
\hspace{0in}\epsffile{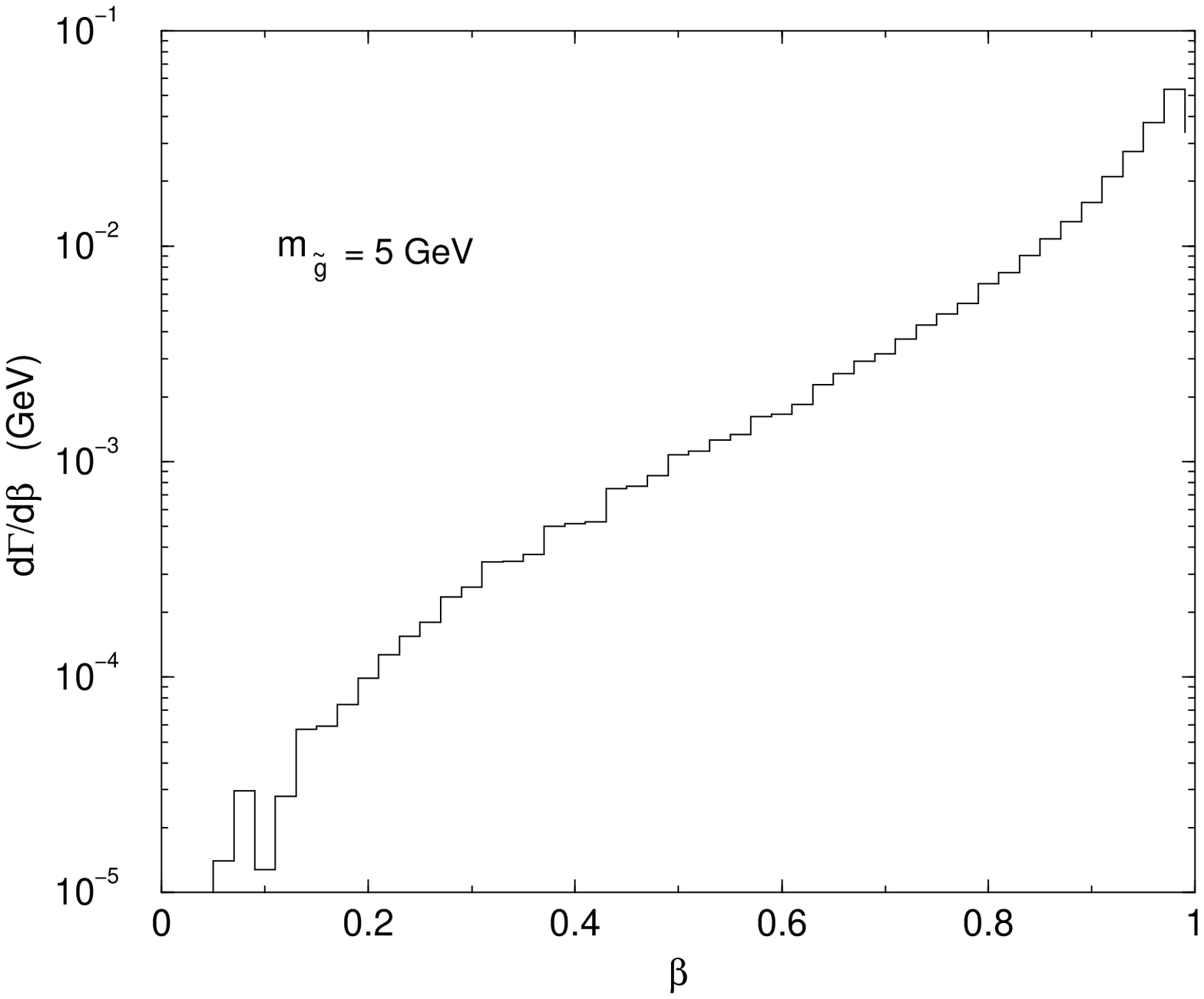}
\epsfysize=2.4in
\hspace{0in}\epsffile{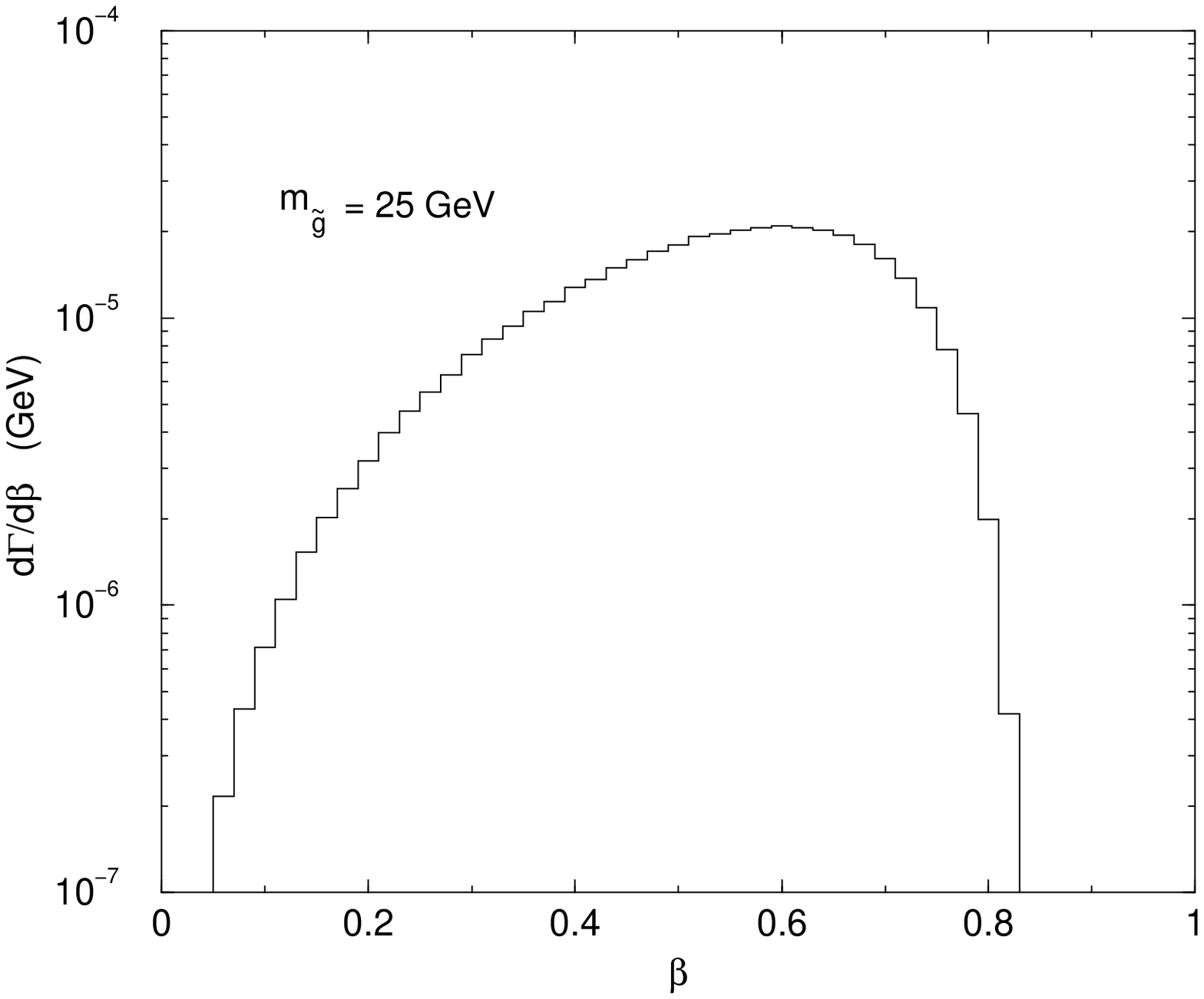}
\end{center}
\caption[]{Distributions of the number of $\gl$-jets
as a function of $\beta$ 
at LEP ($\protect\rts=\mz$) for $\mgl=5$ and $25\gev$. No cuts are imposed.}
\label{lepbetadist} 
\end{figure} 

Thus, we focus on $\rts=\mz$. 
The procedures for employing LEP $Z$-pole data to place constraints
on the \glsp\ scenario depend upon the manner in which the $\gl$-jet
is manifested in the detector; this was outlined in the previous section.
Generally speaking, $q\anti q \gl\gl$ events will have 4 jets
and missing momentum. As noted in the previous section, 
the most crucial kinematical aspect of the $\gl$-jets
is their distribution as a function of $\beta$.
The number of $\gl$-jets as a function of
$\beta$ is presented in Fig.~\ref{lepbetadist} for $\mgl=5\gev$ and $25\gev$. 
We see that a light gluino with 
$\mgl\lsim 5\gev$ has a $\beta$ distribution that peaks at 
$\beta\sim 0.98$ while a heavier gluino
with $\mgl\sim 25\gev$ has a broad $\beta$ peak centered about $\beta\sim 0.6$,
with the most probable $\beta$ values lying between $0.5$ and $0.7$.
The implications of these $\beta$ ranges at these two masses were
already indicated in the previous section.
The reason that we will not be able to obtain limits from LEP
data for very small $\mgl$ values is that as
the gluino bound state mass decreases below $5\gev$, the initial
$\beta$ of the $\gl$ increases. As a result, the energy
loss in the first few hadronic collisions increases significantly.
For a mass of $\lsim 1\gev$,
the energy loss is essentially complete (that is the calorimeters will
contain the hadron).

The most relevant LEP experimental analyses currently available are those
related to the search for pair production of neutralinos, $Z\to \cnone\cntwo$,
with $\cntwo\to q\anti q\cnone$.
The OPAL \cite{opal} and L3 \cite{l312} analyses have the
highest statistics and place limits on
$\cnone\cntwo$ production in the ${\rm jets}+\ptmiss$ channel that are
potentially relevant for the $q\anti q\gl\gl$ final state.
However, the L3 analysis is restricted entirely
to $2j+\ptmiss$ final states. Only the OPAL analysis is relevant 
to any $nj+\ptmiss$ final state with $n\geq 2$. Typically, $q\anti q\gl\gl$
events give $n=2$, 3, or 4, depending upon
the amount of energy deposition by the $\gl$-jets.

The OPAL analysis is based on dividing the event into two hemispheres
as defined by the thrust direction of the visible jets. We have implemented
their procedures in a parton-level
Monte Carlo and computed the efficiency for the $Z\to q\anti q\gl\gl$
events to pass their cuts as a function of $\mgl$ for
various choices of the charged fragmentation probability $P$.
Our precise procedures are as follows.
In the OPAL analysis of multi-jet events, 
each event is divided into two hemispheres by the plane 
normal to the thrust axis, where the thrust $T$ is defined as
\begin{equation}
T = {\rm max}_{\hat n} \frac{\sum_i |\vec p_i \cdot \hat n|}
                   {\sum_i |\vec p_i|}
\label{tform}
\end{equation}
and the thrust axis is the $\hat n$ that leads to
the maximum.  In the OPAL analysis, the $\vec p_i$ are assigned to
calorimeter clusters and associated tracks as described 
in the previous section. Associated energies are computed as if
the track/cluster composites have very small mass.
The sum of the (visible) four-momenta in a given
hemisphere defines the four-momentum of the `jet' 
associated with that hemisphere; note that the `jet'
need not have zero invariant mass.
OPAL then separates events into 
mono- or di-`jet' events, where a mono-`jet' event is one having
a `jet' in only one hemisphere. Mono-`jet' events are
discarded. The following cuts are then applied to the di-`jet' events:
\begin{eqnarray}
&\half(M_{\rm vis}^{\rm hem.~1}+M_{\rm vis}^{\rm hem.~2})< 20\gev \;,
\quad
{M_{\rm vis}}/{E_{\rm cm}} > 0.27 \;, &
\nonumber\\
&p_T > 10\; {\rm GeV} \;,
\quad
p_z < 20\;{\rm GeV}  \;, &
\nonumber\\
&T > 0.7\;,
\quad
{\rm min}[T_{\rm hem. 1},T_{\rm hem. 2}]>0.7 \;,&
\nonumber\\
&\cos \theta_{\rm acol} < 0.98  \;,
\quad
|\cos \theta_{\rm miss}| < 0.94 \;, &
\nonumber\\
&\cos \theta_{\rm acol}  <0.95\;, \cos \theta_{\rm acop} < 0.98 
\quad
    \mbox{if both `jet's are in $|\cos\theta|<0.71$ } \;,&
\nonumber\\
&\cos \theta_{\rm acol}  <0.90\;, \cos \theta_{\rm acop} < 0.95 
\quad
    \mbox{if either `jet' is in $|\cos\theta| \ge 0.71$ } \;,&
\nonumber
\end{eqnarray}
where $(\pi-\theta_{\rm acol})$ is the three-dimensional angle between the two 
`jet's, $(\pi-\theta_{\rm acop})$ 
is the angle between the two `jet's in the $x$-$y$
plane, $\theta_{\rm miss}$ is the polar angle of the missing momentum, 
$M_{\rm vis}$ is the visible mass,
and $\vec p$ (used to compute $p_z$ and $p_T$) 
is the vector sum of all (visible) three momenta.
In the above,
$M_{\rm vis}^2$ is computed by summing all the visible four-momenta (as
defined earlier) in the event and taking the square.
The square of $M_{\rm vis}^{\rm hem.}$ for each hemisphere is computed
by summing the visible four-momenta in the hemisphere and squaring.
The thrust, $T_{\rm hem.}$, for each hemisphere is defined by
going to the center-of-mass 
for that hemisphere (defined by the sum of all visible three-momenta
in the hemisphere being zero) and computing the thrust as in Eq.~(\ref{tform})
using only the three-momenta of that hemisphere.

In applying the above procedures to the Monte Carlo events, it is necessary
to adopt an algorithm for including the effects of detector resolution.
In our computations, 
all cluster/track momenta and energies are smeared using
the stated OPAL hadronic calorimeter
energy resolution of $\Delta E/E=120\%/\sqrt{E(\gev)}$. 
We note that energy smearing is important in that it generally
increases the OPAL acceptance efficiencies by virtue
of the fact that, on average, jet-energy mismeasurement tends to enhance
the amount of missing momentum. This enhancement is especially important
for $\mgl$ and $P$ choices (\eg\ $\mgl=25\gev$ and $P=1$)
such that the missing momentum before smearing is small.
Another important ingredient is properly accounting for the
fact that the $R$-hadron does not take
the entire momentum of the $\gl$. We have employed
the standard Peterson \cite{peterson} form for the fragmentation
function of $\gl\to R$:
\beq
D_{\gl}^R=Cz^{-1}\left[1-{1\over z}-{\eps_{\gl}\over 1-z}\right]^{-2}\,,
\label{pform}
\eeq
where we will take $\eps_{\gl}=(0.3\gev/\mgl)^2$. Here, the $R$-hadron
carries a fraction $z$ of the momentum of the $\gl$ and a normal 
(light quark or gluon) jet
carries the remainder. The $R$-hadron is then treated in the calorimeter
as we have described in the previous section.
The energy of the remainder (effectively zero-mass)
jet is taken equal to its momentum 
and is assumed to be entirely deposited in the calorimeter.
Typically, fragmentation does not have a large influence
on the efficiency with which events are retained,
especially in cases for which the $\gl$-jet energy is
measured to be a large fraction of the $\gl$'s initial kinetic energy.

The OPAL data corresponds to $N_{\rm had}=4.4\times 10^6$ hadronic $Z$
decays. The expected number of $q\anti q\gl\gl$ events after cuts is then
\beq
N={N_{\rm had}\br(Z\to q\anti q\gl\gl)\times\mbox{efficiency}\over
\br(Z\to {\rm hadrons})}\,,
\label{evtnumz}
\eeq
where we use the efficiency as computed via the Monte Carlo.
After cuts, OPAL observes 2 events with an expected background of $B=2.3$ 
events.
The 95\% upper limit on a possible new physics signal is then $S=4$ events,
corresponding to $\br(Z\to q\anti q\gl\gl)\times\mbox{efficiency}
\sim 6.4\times 10^{-7}$. 
How low a value of $\mgl$ can be eliminated depends
upon the efficiency at low $\mgl$. Because of the very high raw event
rate at low $\mgl$ values, quite small efficiency can be tolerated.
We will see that we can exclude gluino masses above $3-4\gev$.

\begin{figure}[p]
\leavevmode
\begin{center}
\epsfxsize=4.in
\hspace{0in}\epsffile{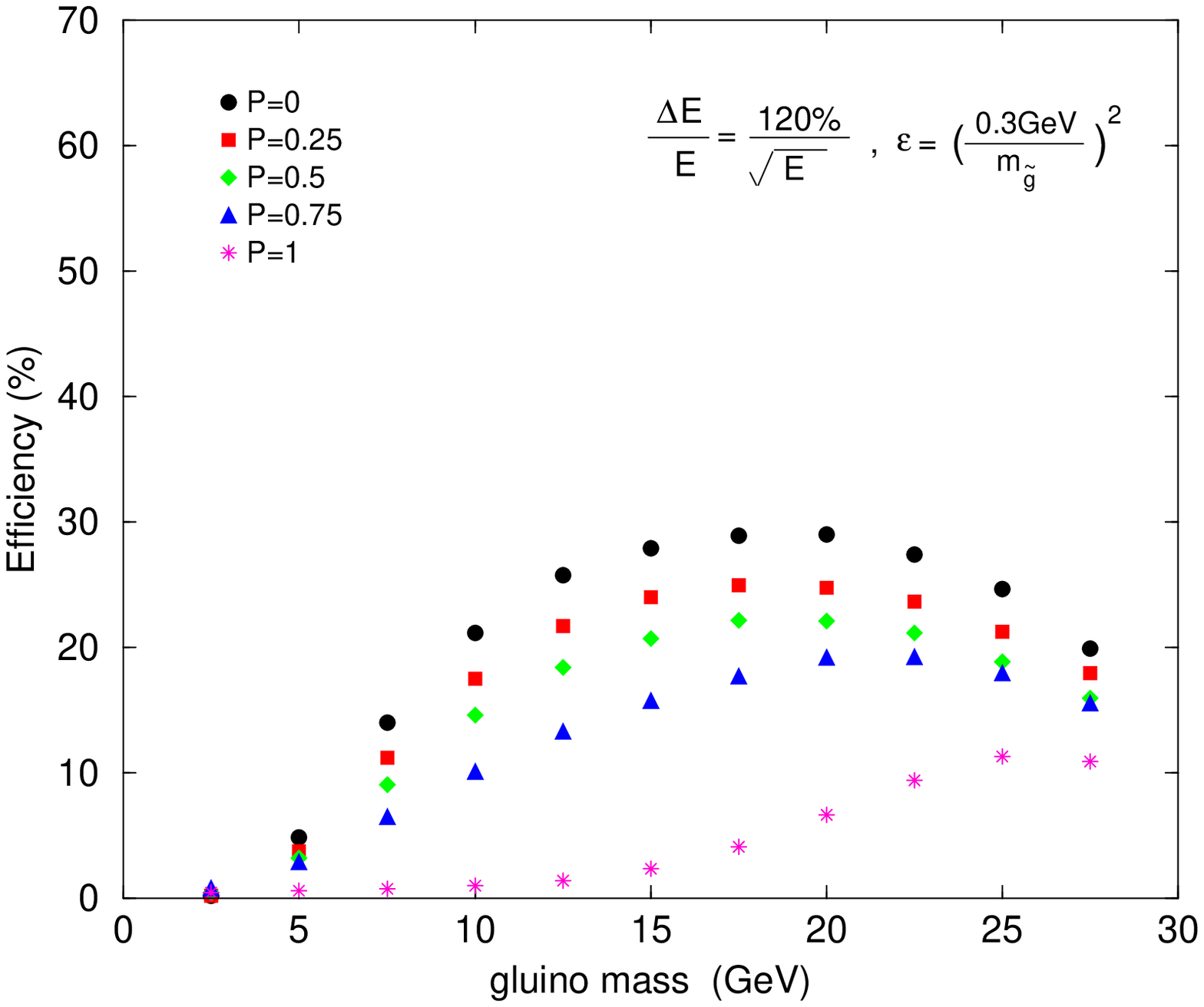}
\medskip
\epsfxsize=4.in
\hspace{0in}\epsffile{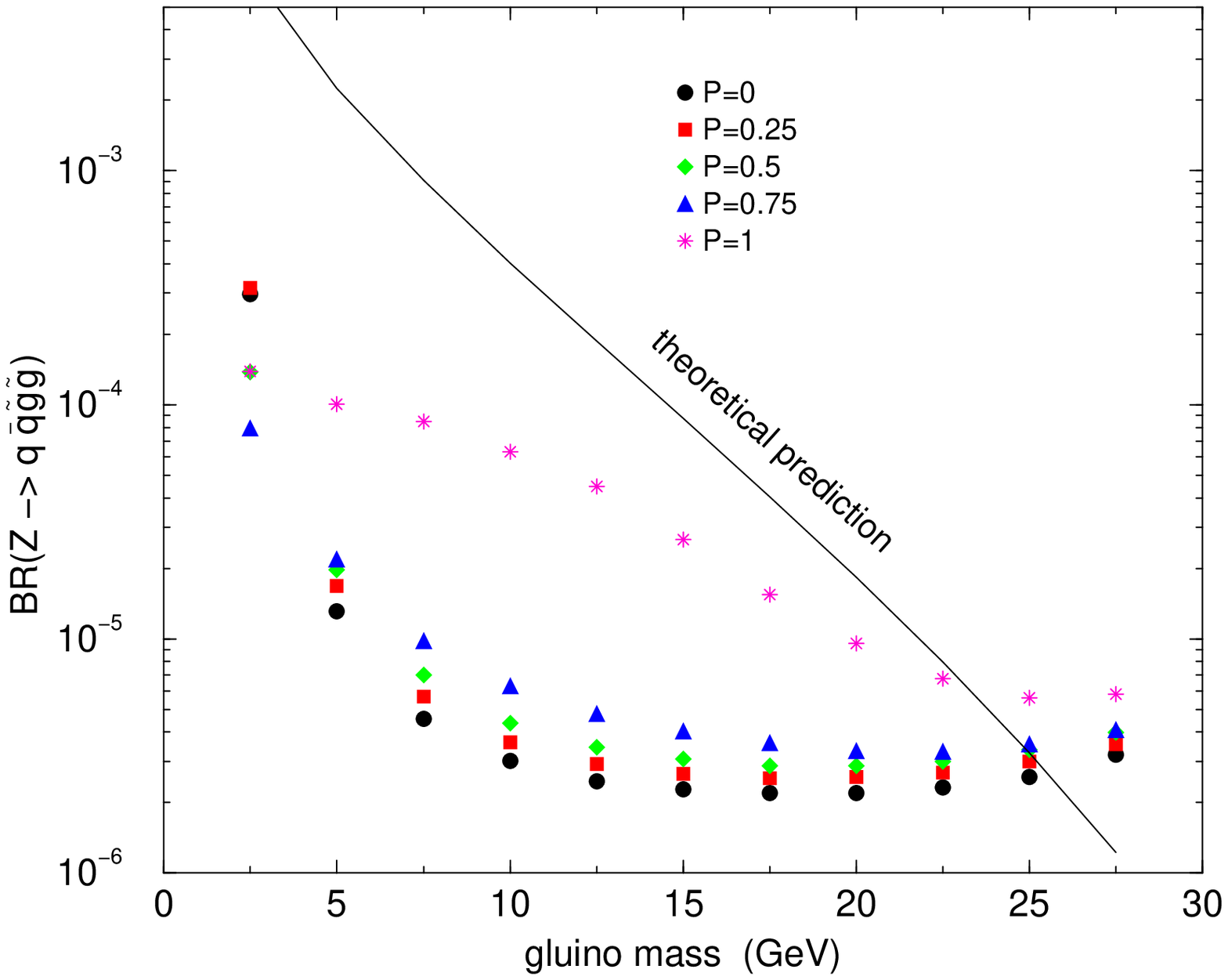}
\end{center}
\caption[]{In the upper window, we plot the OPAL $q\anti q \gl\gl$ 
event efficiency (after all cuts) in the $P=0,1/4,1/2,3/4,1$ cases,
as computed using event-by-event determination of 
$\ejet$ [using Eq.~(\ref{muonjet})] for each $\gl$.
For $P\neq 0,1$, changes of the $R$-hadron
charge as it passes through the detector are randomly implemented.
Both smearing and fragmentation effects are included.
The lower window gives, as a function of $\mgl$,
the corresponding 95\% CL upper limits compared to the
theoretical prediction for $\br(Z\to q\anti q \gl\gl)$. Results are
for the SC1 choices of $\lam_T=19$~cm and $\vev{\Delta E}$ case (1).}
\label{pscannormal} 
\end{figure}

As described in the previous section,
to obtain a reliable result for the range of $\mgl$ that the OPAL
analysis excludes, we have computed the efficiency for
$q\anti q\gl\gl$ events to pass the full set of cuts when
Eq.~(\ref{muonjet}) is employed for each $\gl$ on an event-by-event basis,
including (for $P\neq 0,1$) random changes (with
probability determined by $P$) of the $R$-hadron
charge at each of the hadronic interactions it experiences
as it passes through the detector. We have considered
the three scenarios --- SC1, SC2, and SC3 --- 
for choices of $\lam_T$ and the $\vev{\Delta E}$
case that were outlined at the end of the previous section.
In Fig.~\ref{pscannormal}, we plot the resulting 
OPAL efficiency for $q\anti q \gl\gl$
events after all cuts as a function of $\mgl$ for $P=0,1/4,1/2,3/4,1$
for the SC1 choices, including calorimeter energy smearing and 
fragmentation effects. Also shown are 
the resulting 95\% CL upper limits on $\br(Z\to q\anti q\gl\gl)$.
We see that for any $P$ not near 1, the entire
range from low $\mgl\sim 3\gev$ to high $\mgl\sim 25\gev$ is 
unambiguously excluded. For $P\sim 1$, the
largest value of $\mgl$ that can be excluded is about $23\gev$.
[The $\mgl\gsim 23\gev$ limit for $P=1$ is similar to,
but somewhat higher than, the limit obtained by
searching for heavily-ionizing tracks at OPAL (discussed later in
section 6).] 

\begin{figure}[h]
\leavevmode
\begin{center}
\epsfxsize=4.in
\hspace{0in}\epsffile{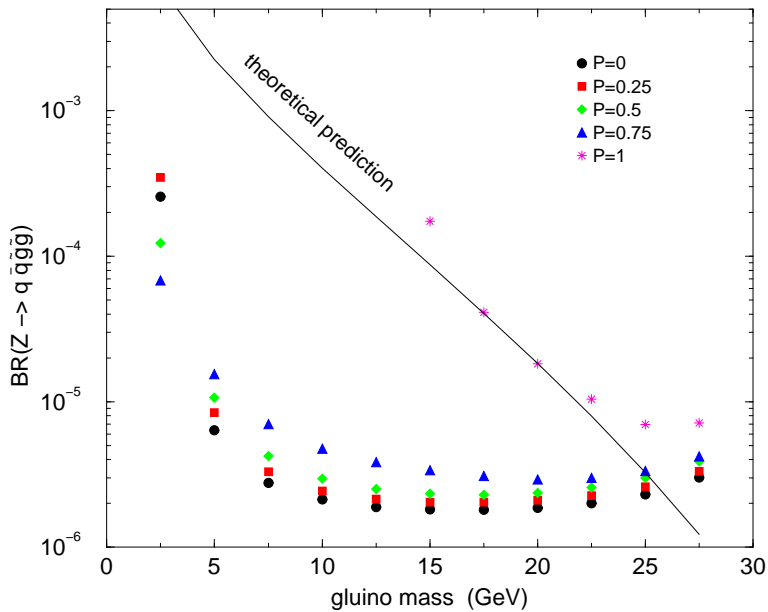}
\end{center}
\caption[]{95\% CL upper limits as in Fig.~\ref{pscannormal} except that
we do not include the effects of energy smearing or fragmentation.}
\label{pscannosmear} 
\end{figure}

In Fig.~\ref{pscannosmear} we present 
the 95\% CL limits obtained without including either energy smearing
or Peterson fragmentation. This figure shows that the limits are little
altered except for $P\sim 1$, in which case the OPAL analysis 
does not exclude any significant range of $\mgl$. It is energy smearing
that is the dominant factor in obtaining a significant efficiency for
event acceptance when $P\sim 1$.  Even though $P\sim 1$ leads to 
$\ejet\sim p_{\rm true}$ at the parton level [for the $\beta$
values typical for the $\mgl=5-25\gev$ mass range (see Fig.~\ref{pvisibleopal})]
and thus small missing momentum at the parton level,
energy smearing produces large event-by-event fluctuations in the 
measured energy of each $\gl$ jet which lead to substantial missing
momentum for many events.

\epsfxsize=4.in


\begin{figure}[p]
\leavevmode
\begin{center}
\epsfxsize=4.in
\hspace{0in}\epsffile{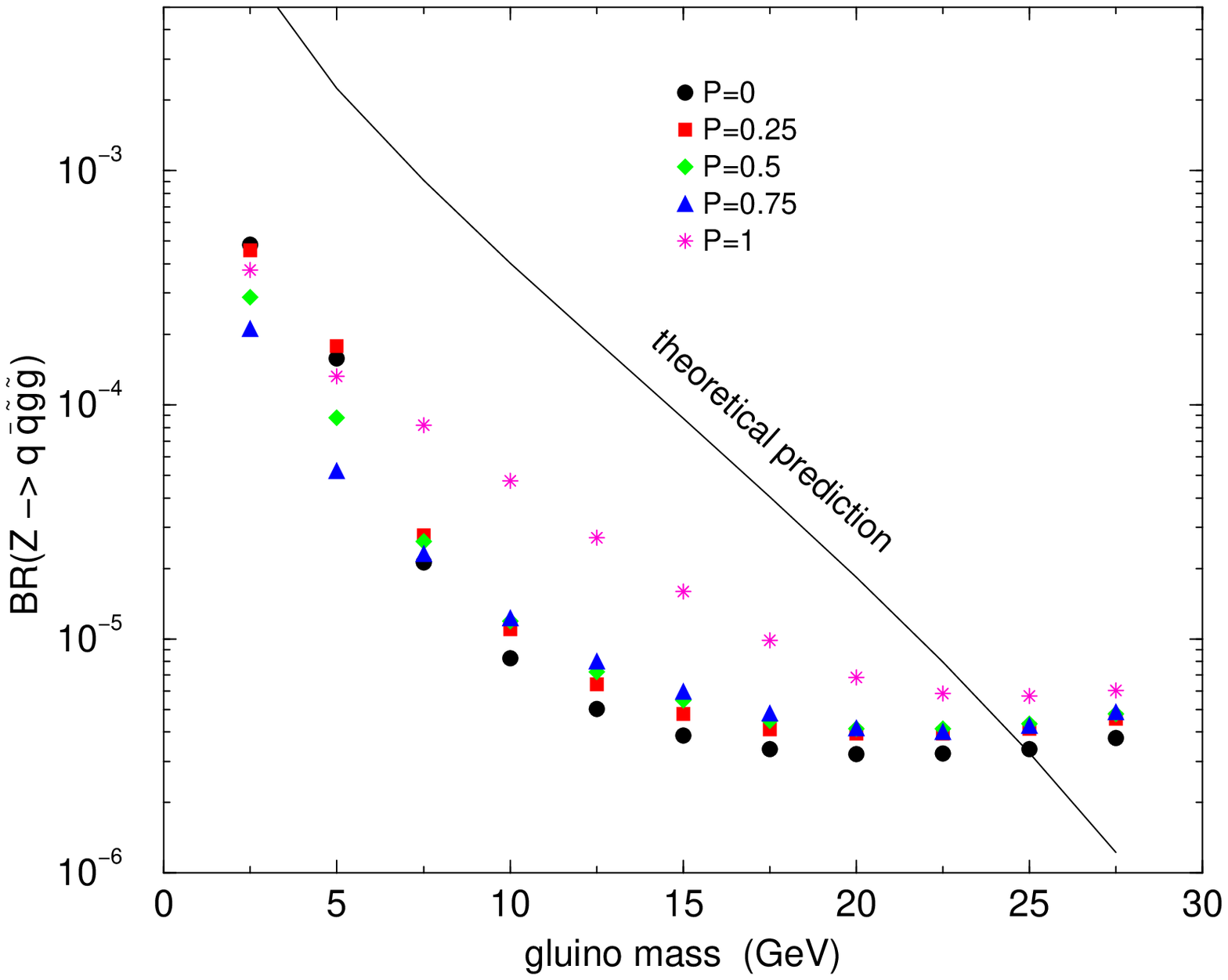}
\medskip
\epsfxsize=4.in
\hspace{0in}\epsffile{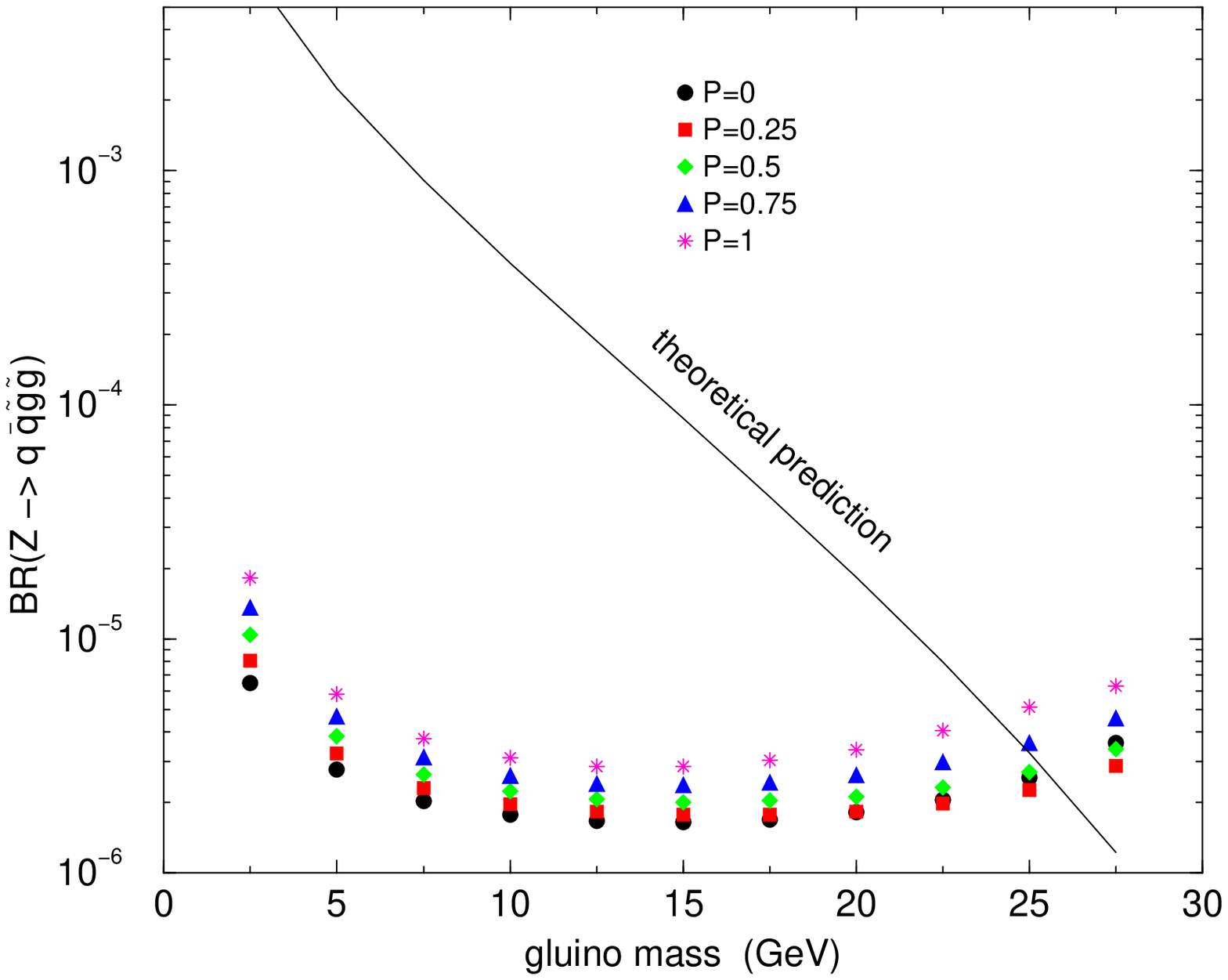}
\end{center}
\caption[]{95\% CL upper limits as in Fig.~\ref{pscannormal} except that
we use the SC2 choices of $\lam_T=9.5$~cm and $\vev{\Delta E}$
case (1) in the upper window and
the SC3 choices of $\lam_T=38$~cm and $\vev{\Delta E}$
case (2) in the lower window.}
\label{pscanwcs} 
\end{figure}

Results analogous to those obtained for the SC1 choices 
of $\lam_T=19$~cm and $\vev{\Delta E}$ case (1), and presented in 
Fig.~\ref{pscannormal}, are presented for the SC2 and SC3 choices
[SC2: $\lam_T=9.5$~cm, $\vev{\Delta E}$ case (1). SC3: $\lam_T=38$~cm,
$\vev{\Delta E}$ case (2)]  in Fig.~\ref{pscanwcs}.
In fact, these possible extremes always give higher
efficiencies and a slightly larger range of $\mgl$ exclusion than found
in the SC1 case.

We expect that re-analysis of the LEP data sets using cuts more
appropriate to the $q\anti q\gl\gl$ final state for given 
values of $P$ and $\mgl$ will yield only a small
improvement over the results obtained using the existing $\cnone\cntwo$
analysis cuts. At large $\mgl$, the event rates are falling so rapidly 
that the 95\% CL upper limit is not likely to be increased
by more than a few GeV. Ruling out $\mgl$ values significantly below $3-5\gev$ 
will be difficult since for such $\mgl$
the gluino looks so much like a normal jet that only
the still controversial analyses of Ref.~\cite{nogluinos} are likely to
prove relevant. Still, we would recommend attempting to make use
of the threshold in the mass recoiling against the two energetic
jets of the the event present at $\mrecoil\sim 2\mgl$.
Perhaps the background
could be reduced to zero by an appropriate set of cuts including
one requiring $\mrecoil\gsim 2\mgl$.

It is also worth nothing that 
the jet energy as computed using the OPAL procedure of Eq.~(\ref{muonjet})
is often larger than the actual $\gl$ energy for large $P$.
This may be interesting at LEP, since there it is possible to
compare the total measured or `visible' energy associated with an
event to the total center of mass energy. 
By summing the assigned energies of all jets,
one would find events in which the total
energy exceeds the center of mass energy when $P$ is near 1.
Indeed, the above Monte Carlo generates a significant
number of such events when $\mgl$ is small.
To our knowledge, the LEP experimental groups
have not analyzed their events in
a manner that would be sensitive to such a discrepancy.

Finally, we briefly discuss $\epem\to\gl\gl$ production via quark-squark loops.
Again, only the existing $Z$-pole data might possibly yield a useful 
constraint. As discussed in Refs.~\cite{krol,css}, even if the squarks
are all completely degenerate, the $Z\to\gl\gl$ branching ratio can be
non-zero by virtue of the top mass being much greater than the bottom mass.
However, Ref.~\cite{css} finds $\br(Z\to\gl\gl)<2\times 10^{-4}$ for all $\mgl$
if the common squark mass ($\wtil m$) is above $\sim 200\gev$.
The typical event would contain two back-to-back jets. 
But these would not generally have equal energy due to
the fact that fluctuations would be
substantial, especially if $P$ is in a range such that there would
sometimes, and sometimes not, be a charged track identified as a muon
contained in one or both of the jets. For small 
deposited energy per $\gl$-jet, as typical for small $P$, the
net apparent energy of the typical event would be below $\mz$, possibly
causing such events to be confused with the two-photon background.
For large enough $P$ and smaller $\mgl$, many of the events
would be anomalous in that the sum of their apparent energies
would exceed $\mz$. We are uncertain if any of the LEP analyses
would have been sensitive to such events appearing at a level
corresponding to $\br(Z\to\gl\gl)\sim 1-2\times 10^{-4}$.
In any case, the $\gl\gl$ event rate can 
be suppressed to an unobservable level simply
by taking $\wtil m$ sufficiently large. (Roughly, $\br(Z\to\gl\gl)$ falls
as $1/\wtil m^2$.) Thus, no model-independent $\mgl$ limits from the $\gl\gl$
final state are possible.

\section{Present and Future Tevatron Constraints from jets + $\ptmiss$}

In the \glsp\ scenario, with all other SUSY particles taken to be much
heavier, the only standard hadron-collider SUSY signal is 
${\rm jets}+\ptmiss$. Current mSUGRA
analyses of this channel do not apply since the $\gl$ does not
cascade decay ($\gl\to q\anti q \cnone,\ldots$) to additional jets.
In the \glsp\ scenario, for a given value of $\mgl$, fewer hard jets
are expected and the amount of missing momentum is typically
smaller. Consequently, the limits that can be placed
on $\mgl$ from Tevatron data will be weaker.\footnote{The 
situation being considered is not dissimilar to the O-II model 
case where the gluino, $\cnone$ and $\cpmone$
are all nearly degenerate with one another. The RunI Tevatron limits
for this latter scenario were determined in Ref.~\cite{guniondrees2}.}
Still, we will find that 
substantial constraints can be placed on the \glsp\ scenario
using existing Tevatron data, and that even stronger constraints
will arise from RunII data.

In assessing the ability of the Tevatron to discover or exclude
a heavy \glsp, we have employed cuts that mimic those employed
by CDF in analyzing RunI data in  the ${\rm jets}+\ptmiss$ channel.
CDF cuts \cite{cdfcuts,cdffinal} are employed rather than D0
cuts \cite{d0cuts} since the CDF 
jet-energy and $\ptmiss$ requirements are weaker
than required in the D0 cuts. For the same integrated
luminosity, weak cuts allow greater sensitivity
to the heavy \glsp\ situation in which the most energetic jets come
from gluons radiated from the initial state colliding partons.
The precise CDF cuts used are those employed
in Ref.~\cite{guniondrees2}; they are designed to 
duplicate the experimental procedures of Ref.~\cite{cdfcuts} 
to the extent possible in the context of a Monte Carlo simulation. 
\bit
\item LI: No (isolated) leptons with $E_T>10\gev$.
\item MPT: $\ptmiss>60\gev$.
\item NJ: There are $n(jets)\geq 3$ with $|\eta_{\rm jet}|<2$ and $E_T>15\gev$,
using a coalescence cone size of $\Delta R=0.5$.
\item Azimuthal separation requirements as follows:
\bit
\item J1MPT: $\Delta\phi(\ptmiss,j_1)<160^\circ$;
\item JMPT: $\Delta\phi(\ptmiss,j(E_T>20\gev))>30^\circ$. 
\eit These are designed, in
particular, to reduce QCD jet mismeasurement background.
\eit
Events were generated using ISAJET-7.37 \cite{isajet}.
Each event was passed through a toy calorimeter
with cells of size $\Delta\eta\times\Delta\phi=0.1\times 0.1$ extending
out to $|\eta|=4$. Electromagnetic and hadronic resolutions
of  $15\%/\sqrt E$ and $70\%/\sqrt E$, respectively, 
were chosen to approximate those of CDF.
The most important cut is the missing transverse momentum (MPT) cut. This is
especially true at low $\mgl$. Typically 
only a small fraction of the events are retained after the MPT cut. 
The next most important cut is the jet-number (NJ) cut.
Typically, for $P$ and $\mgl$ choices that give larger MPT cut acceptance,
the NJ cut acceptance is smaller.
At the higher $\mgl=140\gev$ mass, the cuts retain a larger
fraction of events than at lower mass. (But, of course, the cross section
is smaller at high mass.) 

\begin{figure}[h]
\leavevmode
\begin{center}
\epsfxsize=4.15in
\hspace{0in}\epsffile{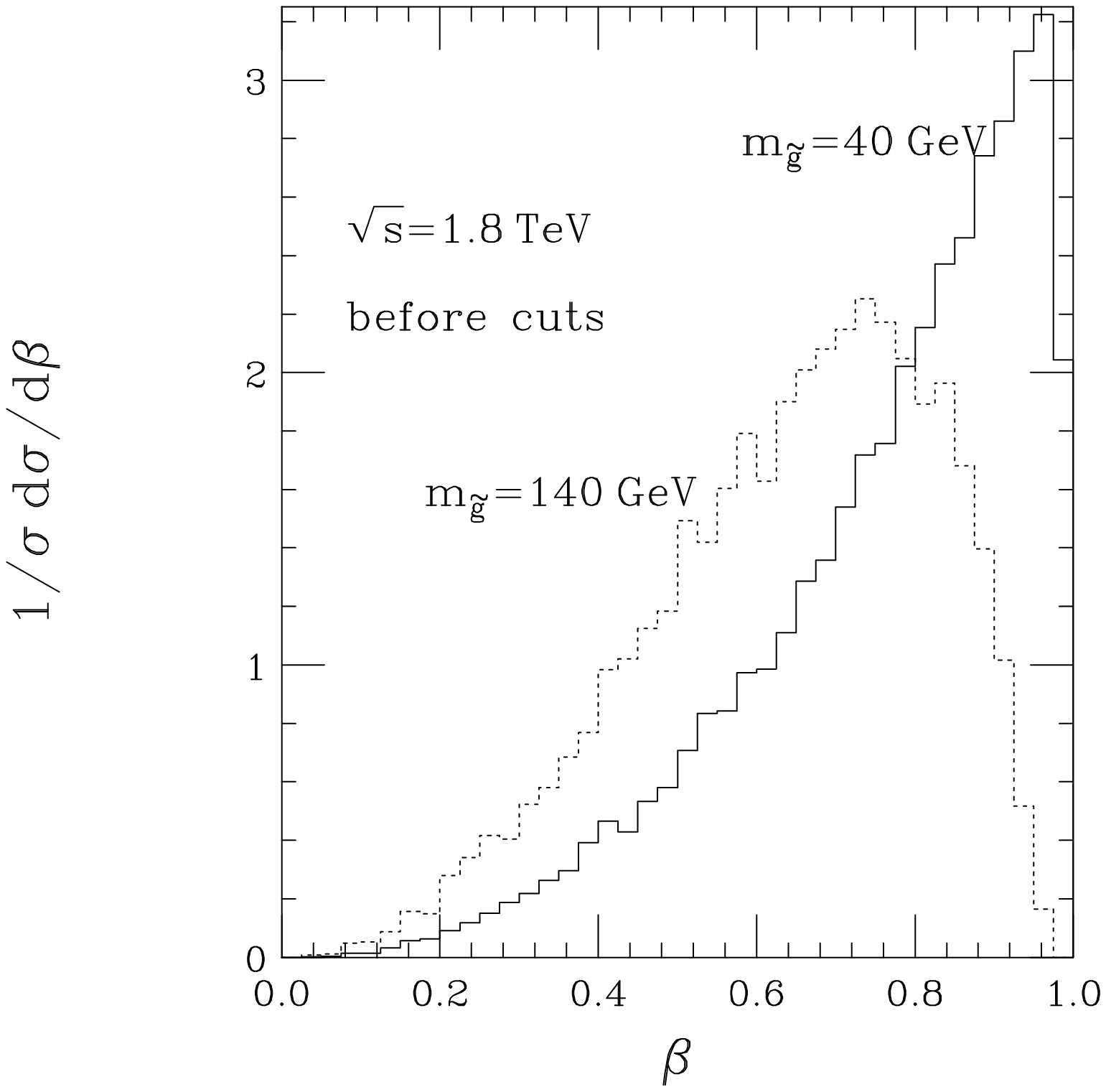}
\end{center}
\caption[]{The $\beta$ distributions of the $\gl$'s
produced in $p\anti p \to \gl\gl$, before
cuts, for $\mgl=40\gev$ and $\mgl=140\gev$, taking $\protect\rts=1.8\tev$.}
\label{betaplots} 
\end{figure} 

\begin{figure}[p]
\leavevmode
\begin{center}
\epsfxsize=3.8in
\hspace{0in}\epsffile{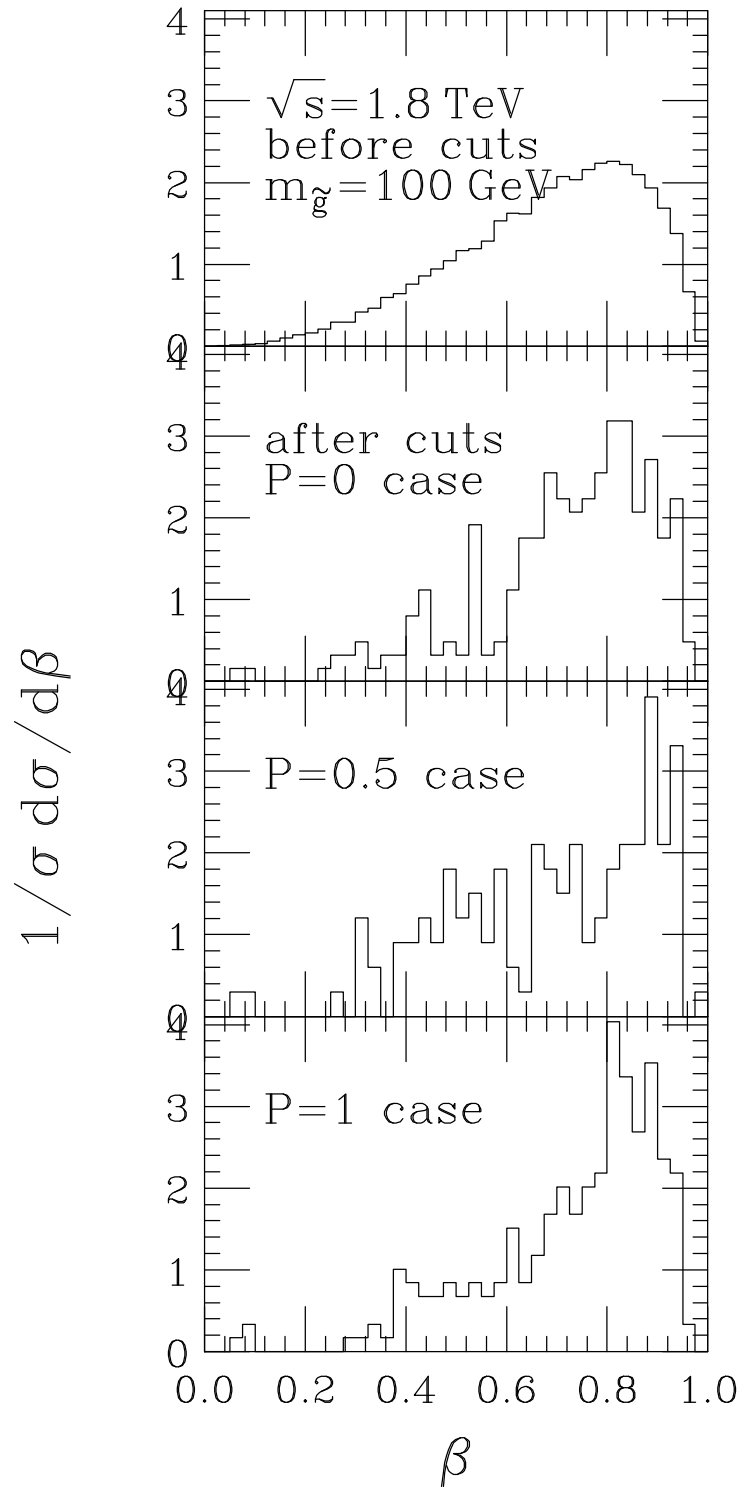}
\end{center}
\caption[]{In the top window, the $\beta$ distribution of the $\gl$'s
produced in $p\anti p \to \gl\gl$, before
cuts, for $\mgl=100\gev$, taking $\protect\rts=1.8\tev$.
In the lower three windows, distributions in $\beta$ after cuts
are compared for $P=0$, 1/2 and 1.}
\label{beta100plot} 
\end{figure}

In order to relate the Tevatron situation to the discussion of section 3,
it is useful to present the $\beta$ distribution of the $\gl$
for several $\mgl$ values. In Fig.~\ref{betaplots},
we present the $\beta$ distributions, before cuts,
for $\mgl=40\gev$ and $140\gev$, \ie\ values  
near the upper and lower ends of the interesting mass range. 
For $\mgl\leq 40$, $\beta$ is typically $\geq 0.95$; for $\mgl\sim 140\gev$,
the $\beta$ distribution peaks near $\beta\sim 0.75$, with most
events having $0.5\leq \beta\leq 0.9$. 
The $\beta$ distributions, both before and after cuts (taking
$P=0$, 1/2 and 1), are given for $\mgl=100\gev$ in Fig.~\ref{beta100plot}.
Referring back to Fig.~\ref{pvisiblecdf} and related comments, we
see that in all cases the most probable $\beta$ values are such that
the measured $\ejet$ of most
$\gl$-jets will be much smaller than the true momentum,
thereby leading to large missing momentum as defined in the analysis.

\begin{figure}[h]
\leavevmode
\begin{center}
\epsfxsize=4.15in
\hspace{0in}\epsffile{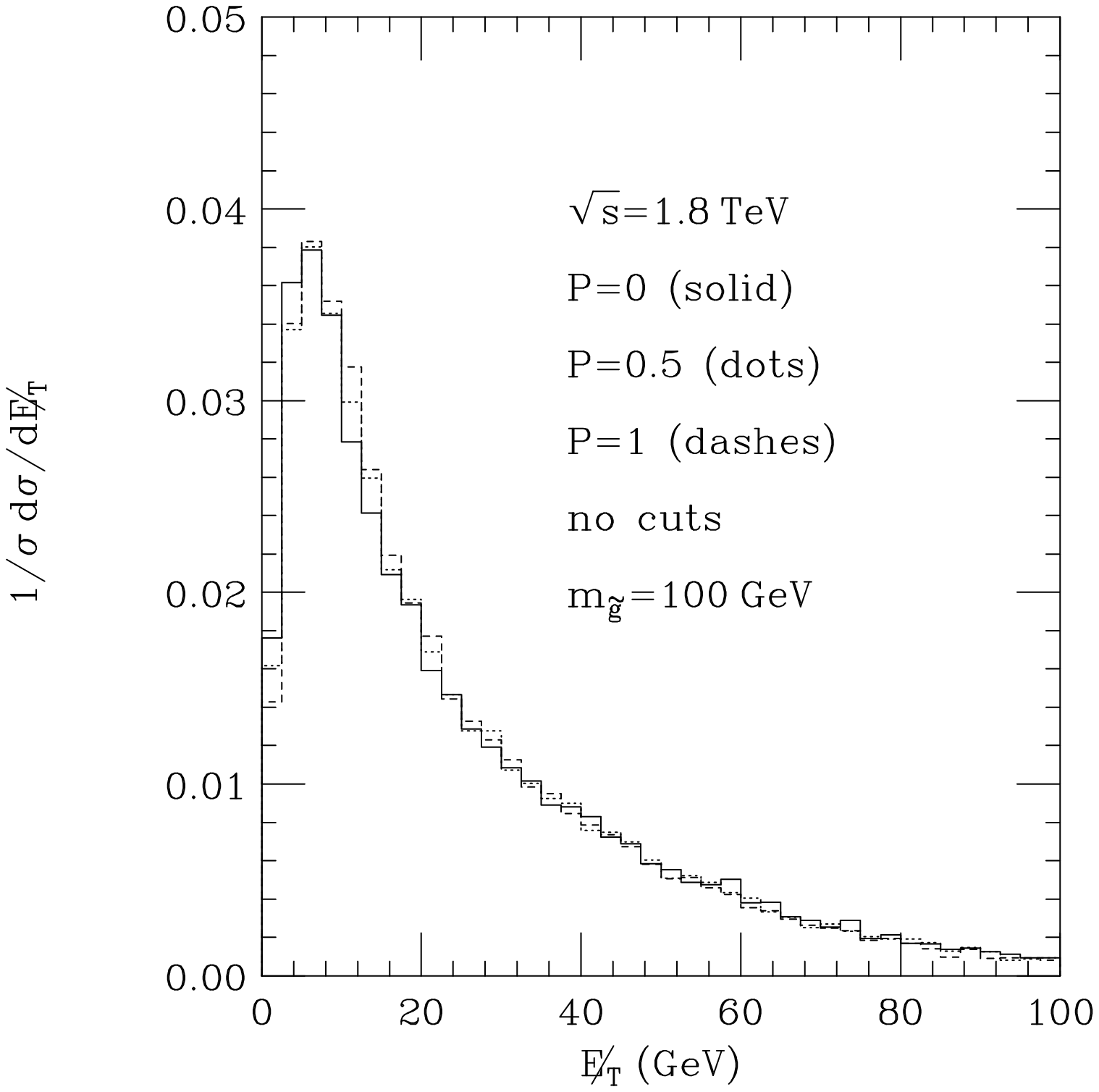}
\end{center}
\caption[]{The $\ptmiss$  distribution (before cuts) for $p\anti p\to \gl\gl$
events at $\protect\rts=1.8\tev$ 
is illustrated for $\mgl=100\gev$ and $P=0$, 1/2 and 1.}
\label{ptmiss100plot} 
\end{figure} 

The distribution in $\ptmiss$ that results is illustrated for $\mgl=100\gev$
in Fig.~\ref{ptmiss100plot}.  There, we see a substantial tail
with $\ptmiss>60\gev$ that is essentially independent of the choice
of $P$.  This independence of $P$ is
due to the small dependence of the $\beta$ distribution on
$P$ (as illustrated in Fig.~\ref{beta100plot}) and to the CDF procedure
in which events where one of the $\gl$-jets looks muonic are discarded
and no correction is applied to the calorimetric energy
measurement for a retained $\gl$-jet that contains a penetrating track.

Let us now turn to determining the limits on a \glsp\ from the CDF data.
To do so, we compare the 
cross section for $\gl\gl$ pair production after cuts
to the SM background expected by CDF.
For the above CDF cuts and $\rts=1.8\tev$, Ref.~\cite{cdfcuts}
quotes a background rate of 28.7 events for $L=19\pbi$, corresponding
to $\sigma_B=1.51\pb$.  (A background rate of 33 events
is quoted for the very slightly different $\geq 3$ jet
cuts of the final published CDF analysis, Ref.~\cite{cdffinal};
we prefer to stick to the cuts of Ref.~\cite{cdfcuts}.)
The 95\% CL lower limit on $\mgl$ is obtained
when the signal rate declines below the $1.96 \sigma$ level,
corresponding to $\sigma_S\sim 553\fb$ (after cuts). We note
that this is about the same as the $\sigma_S\sim 614\fb$ required
for a $5\sigma$ signal at $L=0.1\fbi$. This latter cross
section level will be indicated on our figures. In RunII,
systematic uncertainties in the background will very probably determine the
limit of sensitivity. Indeed, the 95\% CL and $5\sigma$ 
levels for $\sigma_S$ are much lower for $L\geq 2\fbi$ than the $\sigma_S$
sensitivity limit defined by $S/B>0.2$ (\ie\ $\sigma_S>302\fb$).
For instance,
the 95\% CL cross section upper limits would be $53.9\fb$ ($15.2\fb$)
for $L=2\fbi$ ($25\fbi$), respectively. If systematics
can be understood at a better than 20\% level, then the limits
that could be obtained from RunII using RunI cuts
would improve substantially as compared to the $S/B>0.2$ level limits.
Correspondingly, a \glsp\ signal with $S/B\sim 0.2$ would have
a very high nominal $S/\sqrt B$. Clearly, optimization of the cuts
and procedures can be expected to improve upon these first estimates
of sensitivity at RunII.

\begin{figure}[p]
\leavevmode
\begin{center}
\epsfxsize=4.15in
\hspace{0in}\epsffile{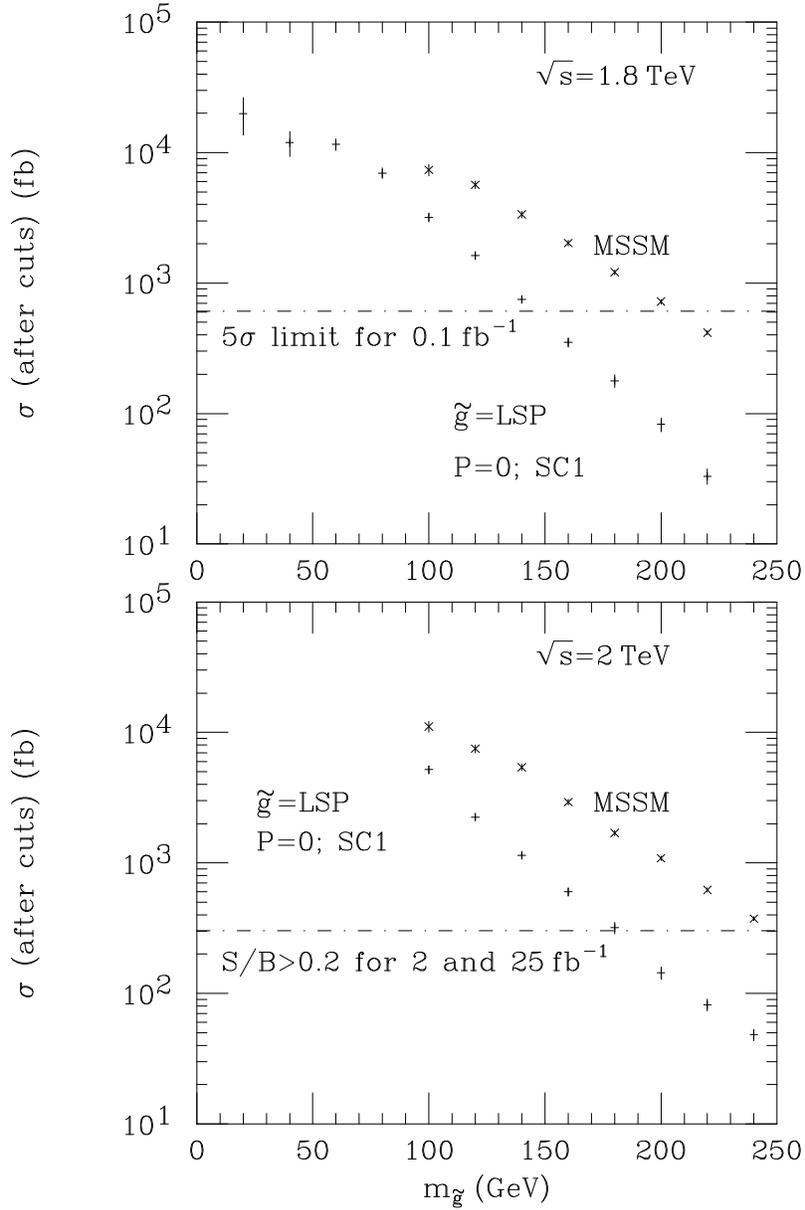}
\end{center}
\caption[]{The cross section (after cuts) in the ${\rm jets}+\ptmiss$ channel 
is compared to (a) the $5\sigma$ level for $L=0.1\fbi$
(also roughly the 95\% CL upper limit for $L=19\pbi$) at $\rts=1.8\tev$
and (b) the $S/B=0.2$ level at RunII ($L\geq 2\fbi$, $\rts=2\tev$)
as a function of $\mgl$ for $P=0$. SC1 choices
of $\lam_T=19$~cm and $\vev{\Delta E}$ case (1) are employed.}
\label{p0cdf} 
\end{figure} 

\begin{figure}[p]
\leavevmode
\begin{center}
\epsfxsize=4.15in
\hspace{0in}\epsffile{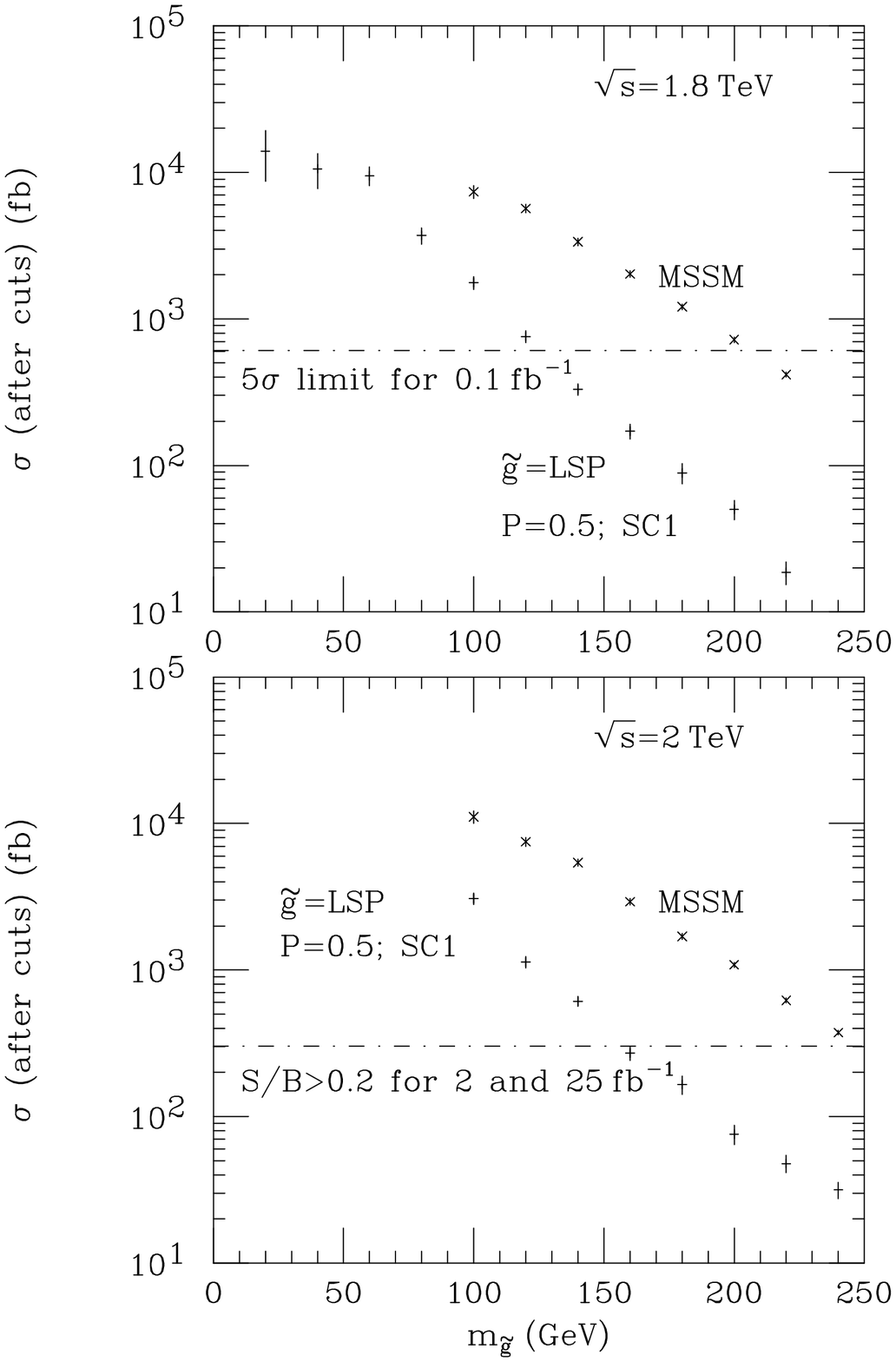}
\end{center}
\caption[]{The cross section (after cuts) in the ${\rm jets}+\ptmiss$ channel 
is compared to (a) the $5\sigma$ level for $L=0.1\fbi$
(also roughly the 95\% CL upper limit for $L=19\pbi$) at $\rts=1.8\tev$
and (b) the $S/B=0.2$ level at RunII ($L\geq 2\fbi$, $\rts=2\tev$)
as a function of $\mgl$ for $P=1/2$, using event-by-event
determination of the momentum (=energy) of each $\gl$-jet 
(including the probabilistic treatment of
charge-exchanges at each hadronic collision)
in events such that neither $\gl$-jet is ``muonic'' (see text). SC1 choices
of $\lam_T=19$~cm and $\vev{\Delta E}$ case (1) are employed.}
\label{phalfcdf} 
\end{figure} 

\begin{figure}[p]
\leavevmode
\begin{center}
\epsfxsize=4.15in
\hspace{0in}\epsffile{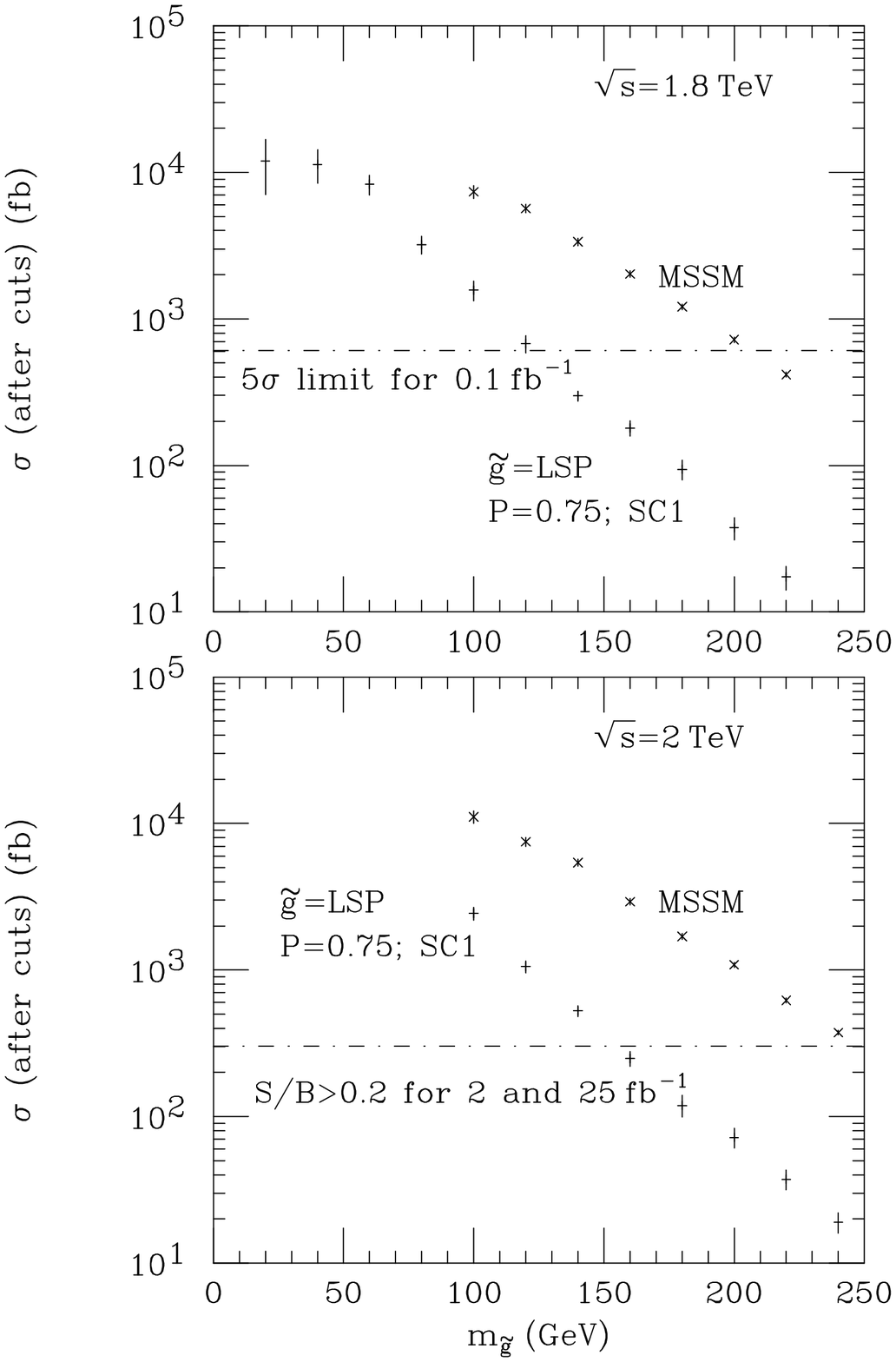}
\end{center}
\caption[]{The cross section (after cuts) in the ${\rm jets}+\ptmiss$ channel 
is compared to (a) the $5\sigma$ level for $L=0.1\fbi$
(also roughly the 95\% CL upper limit for $L=19\pbi$) at $\rts=1.8\tev$
and (b) the $S/B=0.2$ level at RunII ($L\geq 2\fbi$, $\rts=2\tev$)
as a function of $\mgl$ for $P=3/4$, using event-by-event
determination of the momentum (=energy) of each $\gl$-jet 
(including the probabilistic treatment of
charge-exchanges at each hadronic collision)
in events such that neither $\gl$-jet is ``muonic'' (see text). SC1 choices
of $\lam_T=19$~cm and $\vev{\Delta E}$ case (1) are employed.}
\label{p34cdf} 
\end{figure} 

\begin{figure}[p]
\leavevmode
\begin{center}
\epsfxsize=4.15in
\hspace{0in}\epsffile{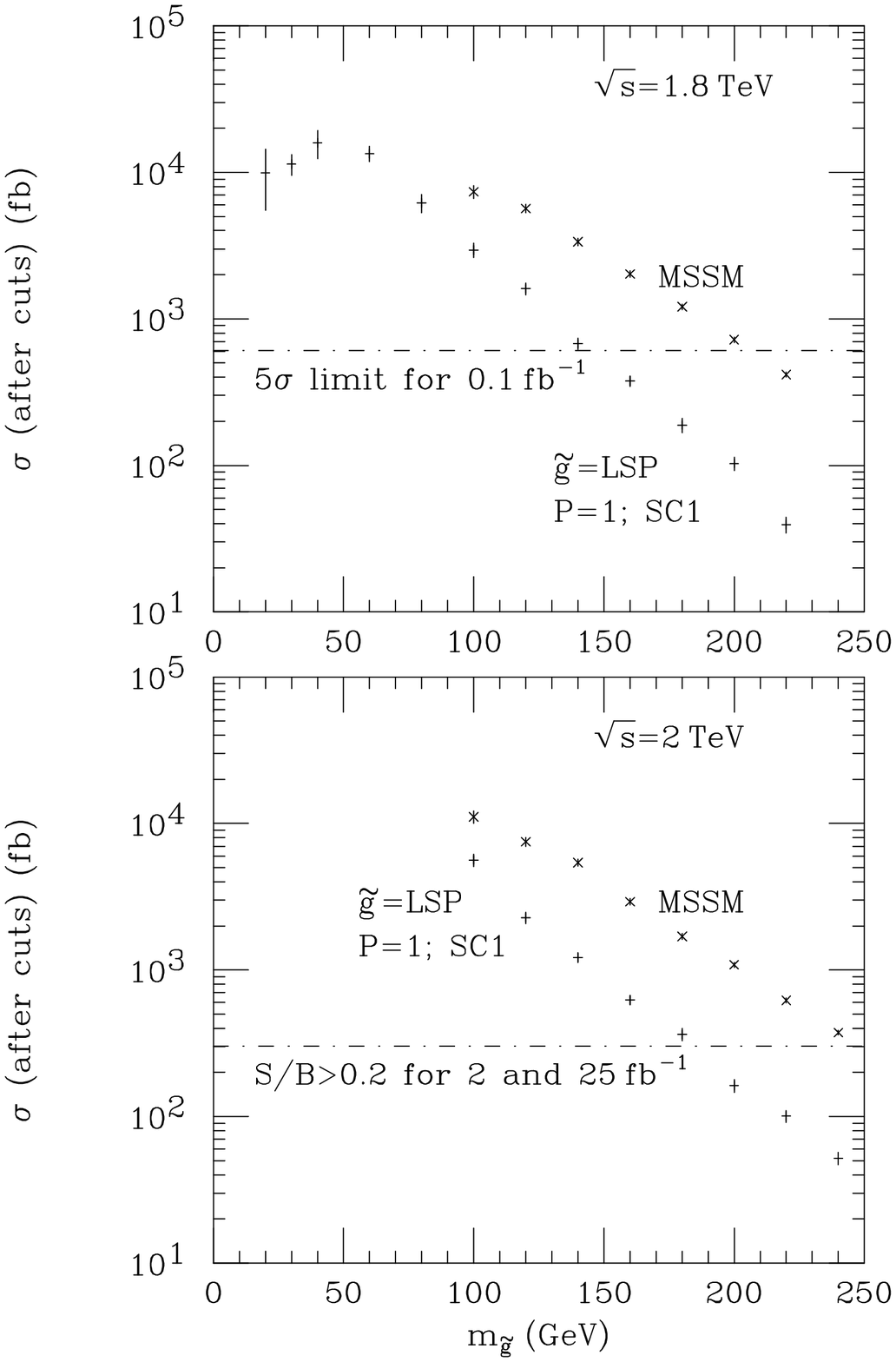}
\end{center}
\caption[]{The cross section (after cuts) in the ${\rm jets}+\ptmiss$ channel 
is compared to (a) the $5\sigma$ level for $L=0.1\fbi$
(also roughly the 95\% CL upper limit for $L=19\pbi$) at $\rts=1.8\tev$
and (b) the $S/B=0.2$ level at RunII ($L\geq 2\fbi$, $\rts=2\tev$)
as a function of $\mgl$ for $P=1$, using event-by-event
determination of the momentum (=energy) of each $\gl$-jet 
in events such that neither $\gl$-jet is ``muonic'' (see text). SC1 choices
of $\lam_T=19$~cm and $\vev{\Delta E}$ case (1) are employed.}
\label{p1cdf} 
\end{figure}

In Figs.~\ref{p0cdf}, \ref{phalfcdf}, \ref{p34cdf} and \ref{p1cdf},
we plot the cross section, $\sigma_S$, after cuts, as a function of $\mgl$
for $P=0$, $1/2$, $3/4$ and $1$,
for the SC1 choices of $\lam_T=19$~cm and $\vev{\Delta E}$ case (1).
Also shown on these plots is the $L=0.1\fbi$
$S/\sqrt B=5$ cross section level (which, as discussed above,
is about the same as the 95\% CL lower limit for $L=0.19\fbi$). 
We see that, at 95\% CL, 
current CDF analyses \cite{cdfcuts,cdffinal} of the $L=19\fbi$ data set
require $\mgl\gsim 150,130,130,140\gev$ for $P=0,1/2,3/4,1$, respectively,
and that, for all $P$, $\mgl$ values are excluded
from the upper limit all the way down to $\leq 20\gev$ at a very high CL.
Note that the $130-150\gev$ lower limit on $\mgl$ obtained is
substantially below the lower limit that RunI data
places on $\mgl$ in a typical MSSM model.
For easy comparison, Figs.~\ref{p0cdf}, \ref{phalfcdf}, \ref{p34cdf}
and \ref{p1cdf} all show the cross
section (after cuts) resulting from gluino pair production
in the MSSM model considered in Ref.~\cite{cdfcuts} with $\msq=1000\gev$,
$\mu=-400\gev$ and $\tanb=4$; one sees that RunI data
yields a 95\% CL limit of roughly $\mgl\gsim 210\gev$.

We re-emphasize that in the Monte Carlo
we have treated each $\gl$-jet on an event-by-event basis. 
In this way, the decision as to whether a given $\gl$-jet 
is ``muonic'' is made event-by-event, including
(for $P<1$) the possibility of charge changes (allowed
for in random fashion on an event-by-event basis according to the chosen $P$)
at each hadronic interaction as the $\gl$ traverses the detector.

\begin{figure}[p]
\leavevmode
\begin{center}
\epsfxsize=4.15in
\hspace{0in}\epsffile{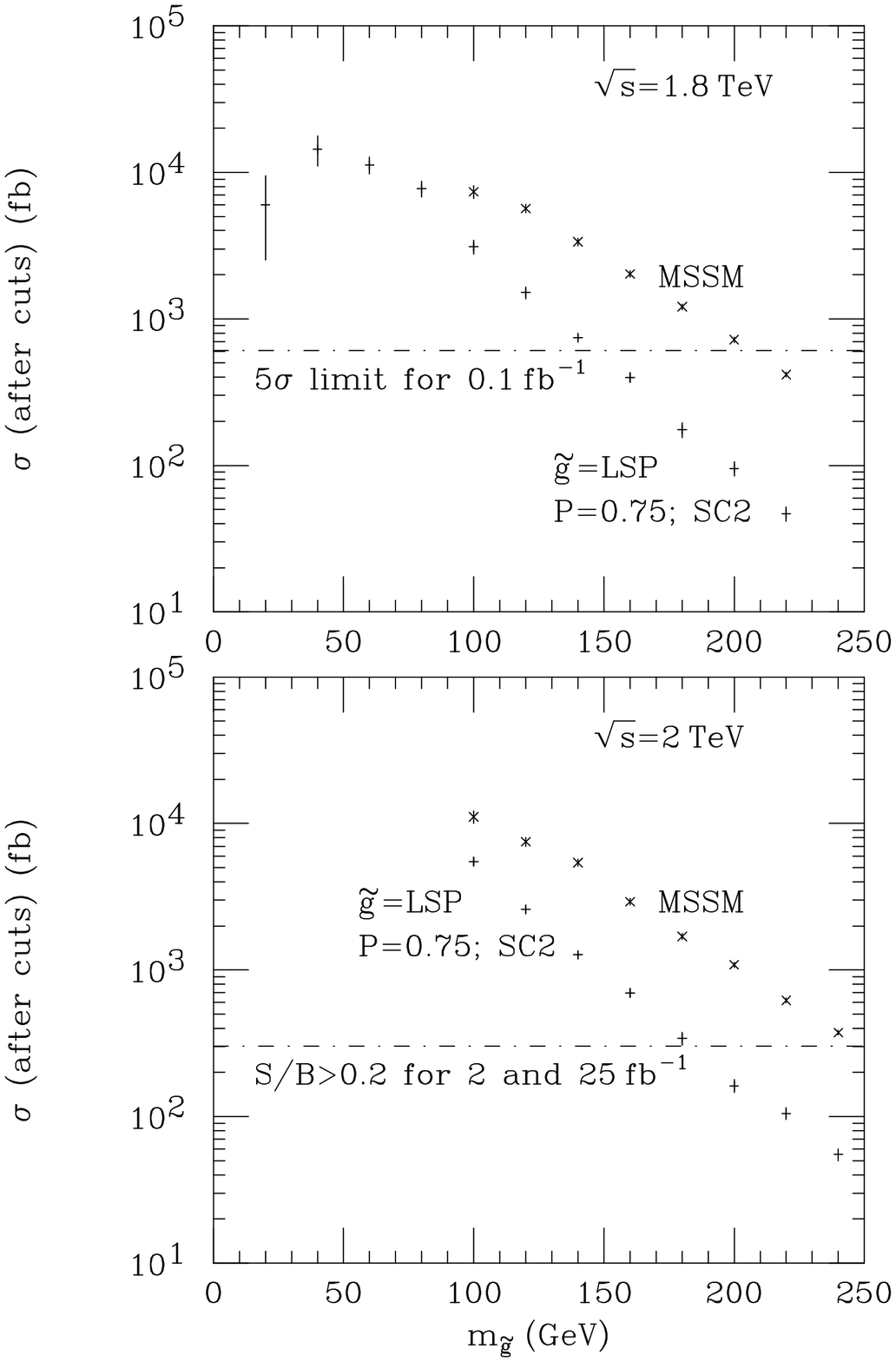}
\end{center}
\caption[]{As in Fig.~\ref{p34cdf}, except that SC2 choices
of $\lam_T=9.5$~cm and $\vev{\Delta E}$ case (1) are employed.}
\label{p34sc2cdf} 
\end{figure} 

\begin{figure}[p]
\leavevmode
\begin{center}
\epsfxsize=4.15in
\hspace{0in}\epsffile{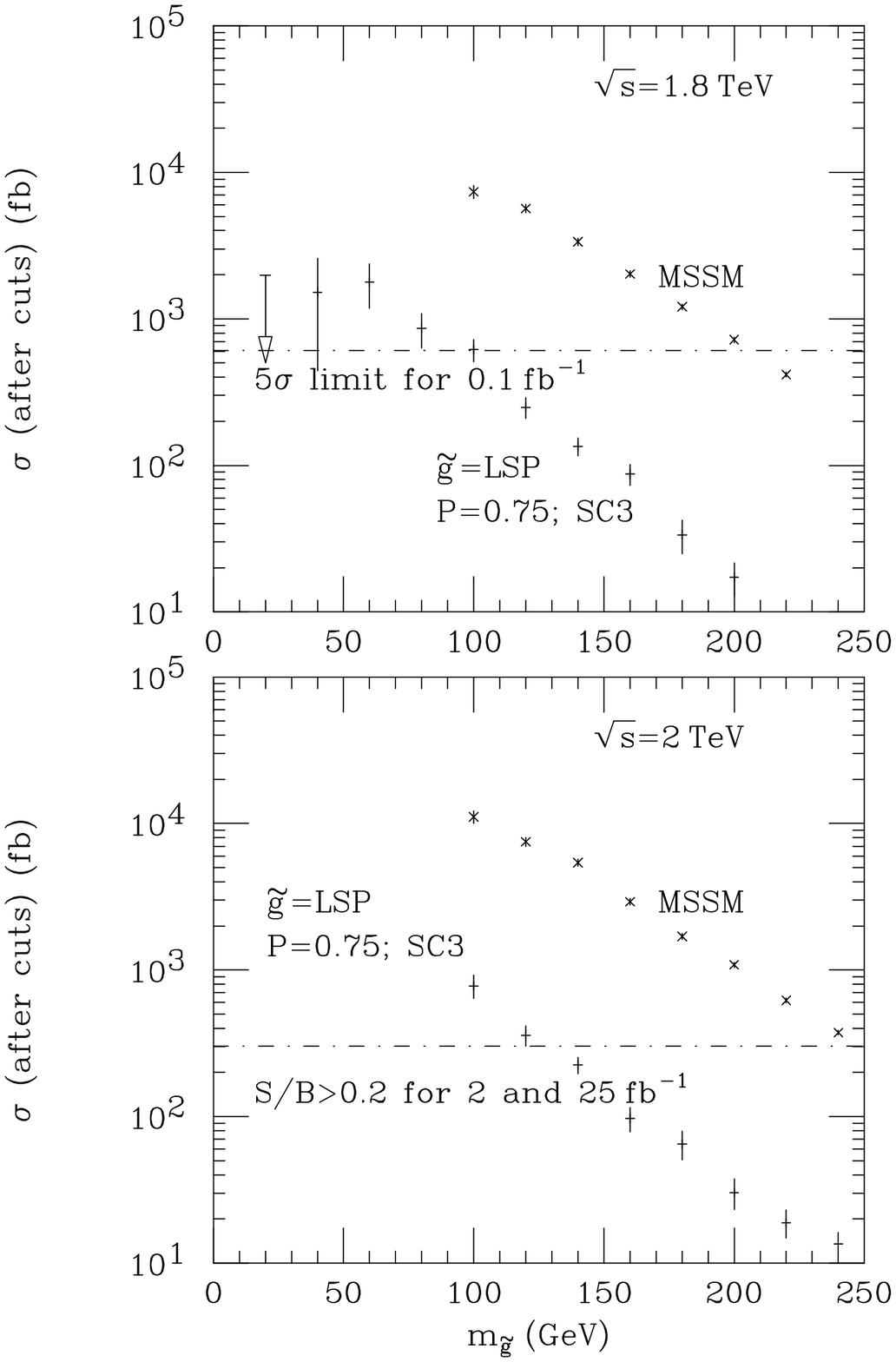}
\end{center}
\caption[]{As in Fig.~\ref{p34cdf}, except that SC3 choices
of $\lam_T=38$~cm and $\vev{\Delta E}$ case (2) are employed.}
\label{p34sc3cdf} 
\end{figure} 

\begin{figure}[p]
\leavevmode
\begin{center}
\epsfxsize=4.15in
\hspace{0in}\epsffile{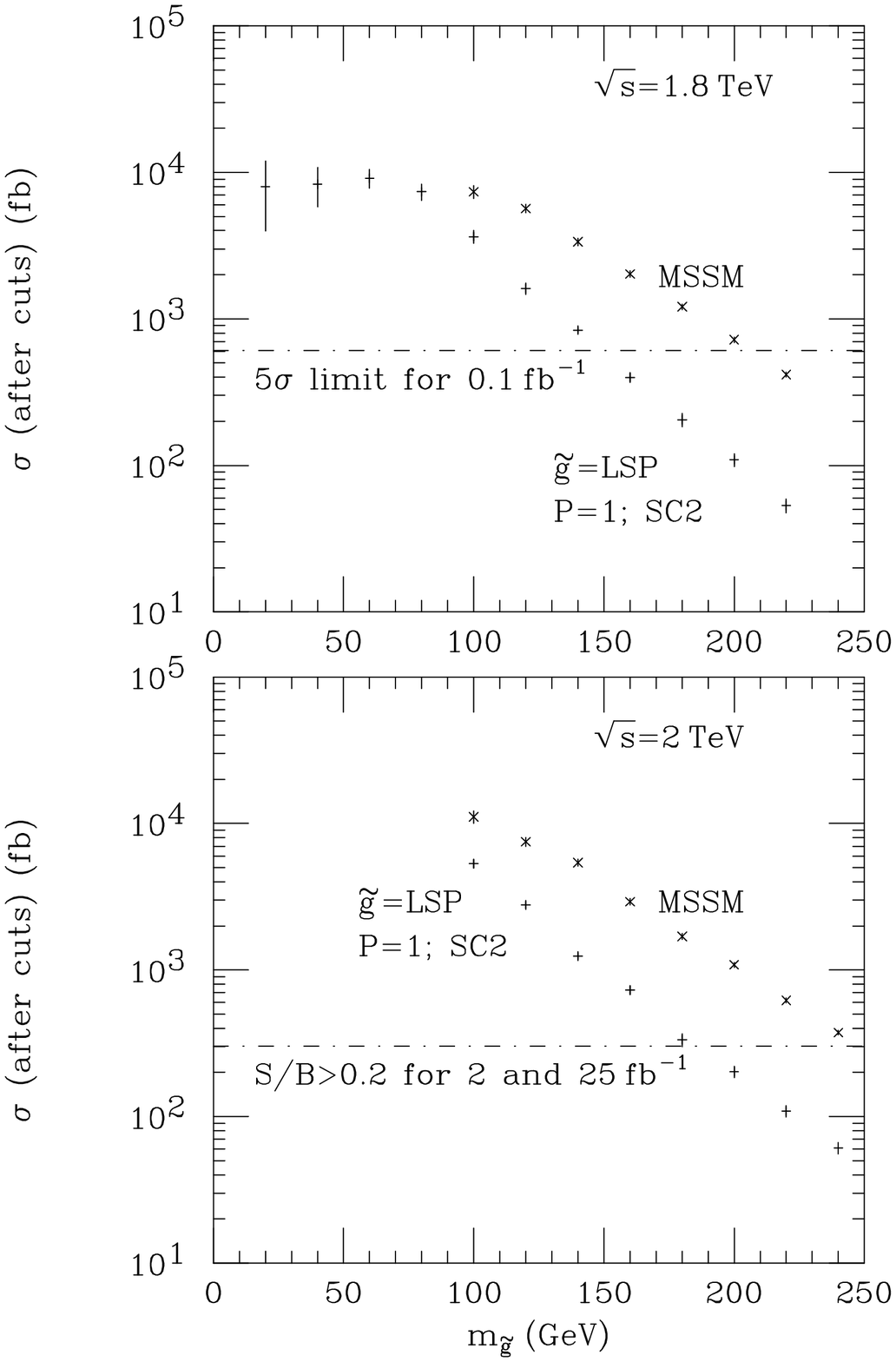}
\end{center}
\caption[]{As in Fig.~\ref{p1cdf}, except that SC2 choices
of $\lam_T=9.5$~cm and $\vev{\Delta E}$ case (1) are employed.}
\label{p1sc2cdf} 
\end{figure} 

\begin{figure}[p]
\leavevmode
\begin{center}
\epsfxsize=4.15in
\hspace{0in}\epsffile{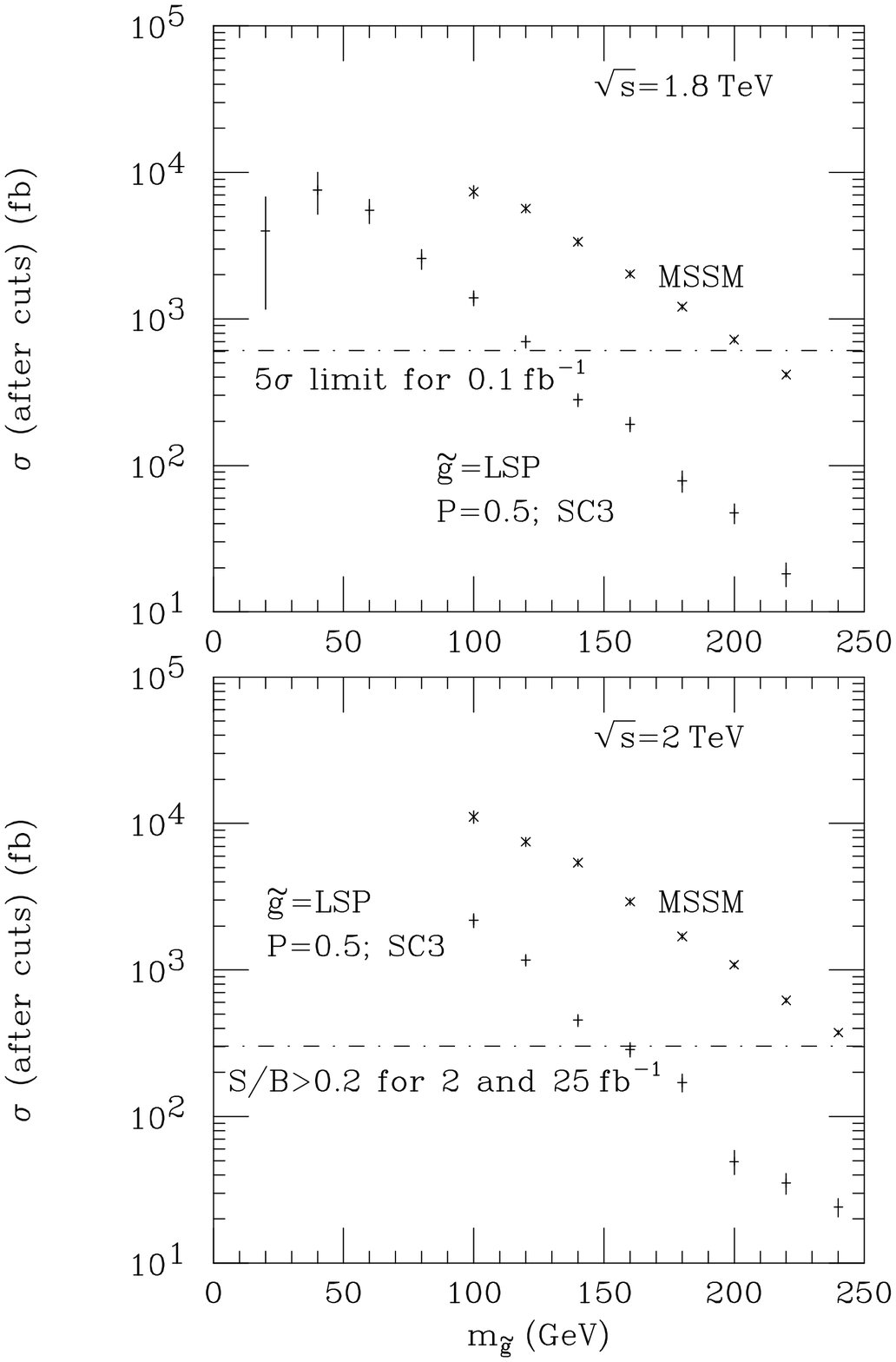}
\end{center}
\caption[]{As in Fig.~\ref{phalfcdf}, except that SC3 choices
of $\lam_T=38$~cm and $\vev{\Delta E}$ case (2) are employed.}
\label{phalfsc3cdf} 
\end{figure}

As for the OPAL analysis, we wish to assess sensitivity of our CDF results
to the choices of $\lam_T$ and $\vev{\Delta E}$ case.  In order to
do so we present several results for the
extreme choices defined earlier in section 3, and denoted by 
scenario labels SC2 and SC3.
First, in Figs.~\ref{p34sc2cdf} and \ref{p34sc3cdf}, we present
$P=3/4$ results for the SC2 and SC3 choices, respectively.  We observe that
when $P$ is large SC2 (SC3) choices result 
in stronger (much weaker) limits from the CDF analysis.
The poor SC3 results are easily understood as follows. For the SC3 choices,
significantly less energy is deposited by a $\gl$-jet. (The
hadronic energy losses are smaller for the longer
$\lam_T$ and smaller case (2) $\vev{\Delta E}$'s,
and the ionization energy losses are smaller because the $\gl$
does not slow down as much due to the smaller hadronic energy losses.)
As a result, when $P$ is large the $\gl$-jet is much more likely
to be declared to be ``muonic'', both because it is highly probable
that it will make it to either the inner or outer muon chamber,
and be charged therein, 
and also because the total energy deposit will not exceed the CDF cutoff
and thereby prevent its being
declared to be a ``muonic'' jet. Thus, many more events are discarded.
As $P$ increases above 3/4, the cross section obtained for
a given $\mgl$ after cuts decreases further.
For example, for $P=1$ current CDF jets + missing momentum data and 
analysis procedures provide no constraints on $\mgl$ for the SC3 choices,
whereas Fig.~\ref{p1sc2cdf} shows that strong constraints are
provided for the SC2 choices.
Finally, in Fig.~\ref{phalfsc3cdf}, we show that, for $P=1/2$
(and smaller), even if we make the SC3 choices the limits on $\mgl$ are
nearly as strong as for the SC1 choices of Fig.~\ref{phalfcdf}.
For SC2 choices, the corresponding plot would show
even stronger limits than for the SC1 choices.

Thus, the jets + missing momentum
data and analysis of CDF only allows a $\gl$ with $\mgl\leq 130\gev$
if the $\gl$ has a high charged-fragmentation probability {\it and}
rather weak hadronic interactions. Fortunately,
the CDF heavily-ionizing-track analysis discussed later
provides strong constraints for large $P$ that exclude this
possibility for $\mgl\geq 50\gev$ (which should be extendable to lower $\mgl$
values). As we have repeatedly noted, the lack of sensitivity of
the RunI CDF jets + missing momentum analysis would disappear if the data
is re-analyzed without eliminating events containing a muonic jet.
We urge the CDF collaboration to perform this re-analysis.

As one possible backup at low $\mgl$, we looked at whether or not
UA1 \cite{ua1data} and UA2 \cite{ua2data} data could be used to exclude
$\mgl$ in the $\mgl\sim 30\gev$ region. We find, however, that
no limits on $\mgl$ in this (or any other mass region)
are possible from the UA1 and UA2 data.
Another backup at low $\mgl$ could be an
analysis of pre-scaled data (\ie\ data not taken at
the full trigger rate) accumulated using lower $p_T$ cuts on the jets.
For example, CDF took about $1\pbi$ of data using a low-$E_T$ four-jet trigger
\cite{stuart}.  Such data might be useful since at lower $\mgl$
the standard CDF cuts employed above
tend to yield a rather small efficiency for accepting signal
events. We have not examined this data in detail.

Let us now consider RunII. Returning to Figs.~\ref{p0cdf}, \ref{phalfcdf},
\ref{p34cdf} and \ref{p1cdf},
we see that the limits based on $S/B>0.2$ will rise
to $\mgl\geq 180,160,160,180\gev$ for $P=0,1/2,3/4,1$, respectively,
for RunII (with $L>0.5\fbi$).
If systematics could be controlled so
that a signal with $S/B\lsim 10\%$ becomes reliable, each
of these lower limits would be increased by about $30\gev$.
All these potential lower bounds are, of course, still substantially
lower than the $\mgl$ lower bound 
that can be achieved in the reference MSSM model
for the same $S/B$ criterion (\eg\ $250\gev$ for $S/B>0.2$). 
It is worth noting that RunII limits will be much less sensitive
to $\lam_T$ and $\vev{\Delta E}$. As shown in Fig.~\ref{p34sc3cdf},
even the SC3 choices will allow exclusion of all $\mgl\lsim 130\gev$.

We end by noting that if the squarks are not much heavier than the $\gl$, then
the $\gl\gl$ cross section at the Tevatron will be reduced due to
negative interference effects in the $q\anti q\to\gl\gl$
amplitude from squark exchanges.
However, the $gg\to\gl\gl$ amplitude is unaffected. Further,
additional very prominent signals will emerge from squark
production channels that will more than compensate.  Thus, the approach
of taking all other SUSY particles to be much heavier than the $\gl$
can be expected to yield the most conservative limits for the \glsp\ models.

\section{The OPAL signal for a charged gluino hadron}

OPAL has searched \cite{opalrpm} for $\epem\to q\anti q \gl\gl$ events
in which the $\gl$'s fragment to a charged $\rpm$
that traverses their two-meter radius tracking chamber. They
look for events with an anomalous value for the ionization $dE/dx$
as compared to the momentum $|\vec p|$.
Both quantities are measured in the tracking chamber. As a result,
penetration of the track to the muon detectors is {\it not} required.
After appropriate kinematical cuts and cuts on the region of
the $dE/dx-|\vec p|$ plane accepted, there is only one candidate event.
They convert this into a 95\% CL limit on the number of signal events.
To interpret this limit they compute the expected number of gluinos
produced and accepted and multiply by the probability $P$ for $\gl\to R^{\pm}$
fragmentation.\footnote{This is not quite the correct procedure
in cases where both gluinos are accepted; the appropriate multiplication
factor {\it per gluino} in that case is $P-P^2/2$.}
They place 95\% CL upper limits on $P$ as given in Table~\ref{opalplim}.

\begin{table}[h]
\centering
\begin{tabular}{|c|c c c c c c c|}
\hline
$\mgl$ & 1.5 & 2.3 & 3.0 & 5.0 & 10.0 & 15.0 & 20.0 \\
\hline
$P^{\,\rm max}_{95\%~CL}$ & 0.37 & 0.20 & 0.14 & 0.06 & 0.13 & 0.33 & 1.03 \\
\hline
\end{tabular}
\caption{The OPAL 95\% CL upper limit on the probability $P$
for $\gl\to \rpm$ fragmentation as a function of $\mgl$.}
\label{opalplim}
\end{table}
\bigskip

As always, it is important to keep in mind that 
if the $\rpm$ decays to a neutral state of any kind
with a lifetime shorter than $\sim 10^{-7}~\mbox{sec}$, then $P$
is effectively zero since the $\rpm$ will decay before traversing
the tracker.
Assuming a sufficiently long lifetime for the $\rpm$, the limits
of Table~\ref{opalplim} can be interpreted in the context of the model
for $P$ described earlier.  For $P=1$, $1/2$ and 1/4, one excludes 
$\mgl=1-20\gev$, $1.2-16.6\gev$ and $1.9-13.6$, respectively. We have
already seen that the OPAL jets plus missing momentum
analysis excludes $\mgl=3-25\gev$ for any $P$ value not
too close to 1; for $P\sim 1$, the upper limit declines to $\sim 23\gev$. 
Thus, the limits
from our analysis of the OPAL jets plus missing momentum channel are nicely
complementary to the OPAL heavily-ionizing track limits; they confirm
one another for a substantial range of $\mgl$.

\section{The CDF signal for a penetrating charged gluino hadron}

The strength of this signal depends on the model used for gluino
interactions  and upon details of the detector. 
CDF's central muon system consists of two muon detection scintillators
separated by iron.  To be identified as a penetrating charged particle,
a particle must (a) penetrate the iron, (b) be charged at the scintillator
layer just before it enters the iron and 
(c) be charged at the exit detection layer.
To be identified as a heavily-ionizing particle, the particle must
also be charged as it exits from the primary interaction and its ionization
must be clearly larger than minimal.

Let us recall the picture we shall
employ for the gluino as it traverses the detector.
As in the OPAL analysis, the primary produced $\gl$ 
is assumed to have some probability $P$ to fragment (immediately)
to a charged $R^{\pm}$-hadron. 
The ionization of the $R^{\pm}$ will be measured
shortly after emerging from the interaction vertex.  The $R^{\pm}$ then
undergoes a certain number of hadronic interactions as it passes
through the calorimeters before arriving at the inner muon detection layer 
preceding the iron. As described earlier, we imagine
that at each hadronic interaction the light quark's and/or gluons
are stripped from the $R$-hadron (whether neutral or charged
at the time) leaving the bare gluino which then
has the same probability $P$ to again become charged. Thus,
the probability that the $R$-hadron is charged 
just before entering the muon iron is again $P$.
As it traverses the iron it 
will undergo several more hadronic interactions 
and so the probability that it 
exits as a charged $R$-hadron is once again $P$. Altogether, we must reduce
the cross section (after cuts to be discussed below) by $P^3$.
Once again, this assumes that all the possible charged $R$-hadron
states are effectively stable
as they travel through the detector.
If they decay rapidly to the $\rzero$ or another neutral
state, then this must be taken into
account by an appropriate reduction of $P$.

\begin{figure}[h]
\leavevmode
\begin{center}
\epsfxsize=4.15in
\hspace{0in}\epsffile{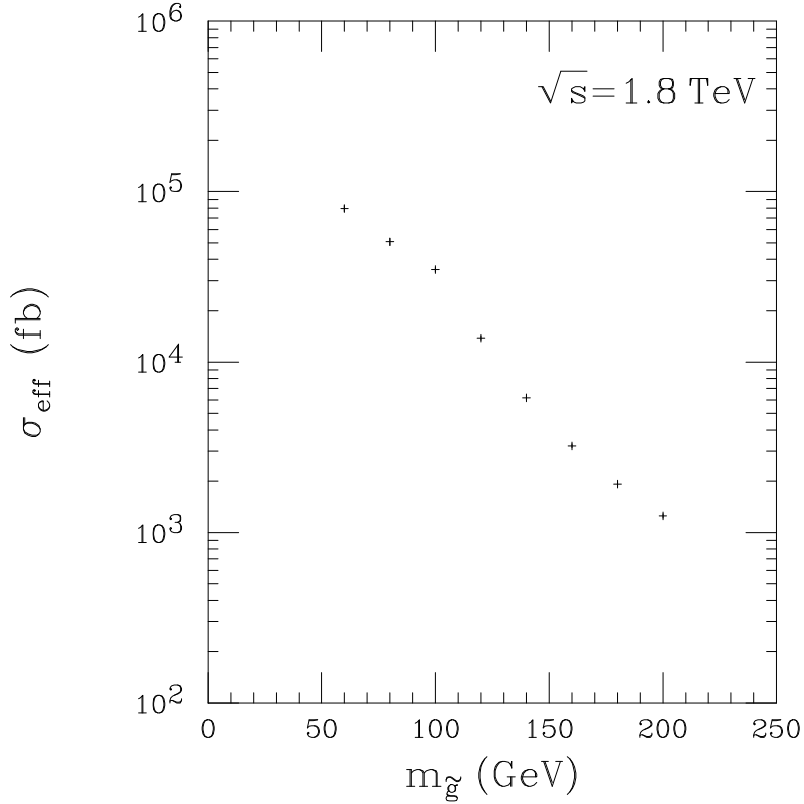}
\end{center}
\caption[]{The effective cross section $\sigma_{\rm eff}$
for one or more $\gl$ to pass the heavily-ionizing penetrating particle
cuts of Eqs.~(\ref{trigger}) and (\ref{recon}), including
the efficiencies quoted in the text.}
\label{stablesig} 
\end{figure}

Whatever the value of $P$, we compute the event acceptance efficiency
as follows \cite{stuart2}.  For a given $\mgl$, we generate events 
using ISAJET. We impose the triggering
requirement that at least one of the $\gl$'s has
\beq 
|\eta|<0.6~\mbox{and}~ p_T>15\gev\,.
\label{trigger}
\eeq
An efficiency of 0.8 is included for triggering on such a $\gl$.
We next demand that at least one of the $\gl$'s satisfy the following
heavily-ionizing, stable charged particle `reconstruction' requirements:
\bea
&|\eta|<1.0\,,\quad|\vec p|>35\gev\,,\beta>\beta_{\rm min}&\nonumber\\
&\beta\gamma<0.85~\mbox{for $\mgl>100\gev$ or}~\beta\gamma<0.7~\mbox{for
$\mgl<100\gev$}\,.&
\label{recon}
\eea
We note that the $\beta\gamma<0.7$ requirement we impose
for $\mgl<100\gev$ is such that only events
in which ionization is at least three times minimal (as
compared to twice minimal if only $\beta\gam<0.85$ is required)
are accepted.  This cut
is stronger than that of the actual analysis \cite{stuart2}.
We do this in the hope that the background will be even smaller
than the conservative number used later.
In the above, we use $\beta_{\rm min}$ as given by
the solid curve in Fig.~\ref{betamin}.
For $P$ substantially smaller than 1, this is quite conservative 
given that ionization energy loss will be much less than 
that employed in the figure, which is for $P=1$. Also, because
we use $\beta_{\rm min}$ for $P=1$ and
because typical $\beta$ values are substantially above
$\beta_{\rm min}$, this analysis is quite
insensitive to the choices of $\lam_T$ and $\vev{\Delta E}$ case.
Finally, an efficiency of 0.5 is included for the reconstruction. Note that
one $\gl$ could provide the trigger but fail the reconstruction
while the other $\gl$ could pass the reconstruction cuts.
In Fig.~\ref{stablesig} we plot the effective cross section $\sigma_{\rm eff}$
as a function of $\mgl$ after including the above cuts and efficiencies, 
but before including $P^3$. We note that no events pass the cuts
for $\mgl<50\gev$; the cuts would have to be weakened, which might
result in the introduction of substantial background.

\begin{figure}[h]
\leavevmode
\begin{center}
\epsfxsize=4.15in
\hspace{0in}\epsffile{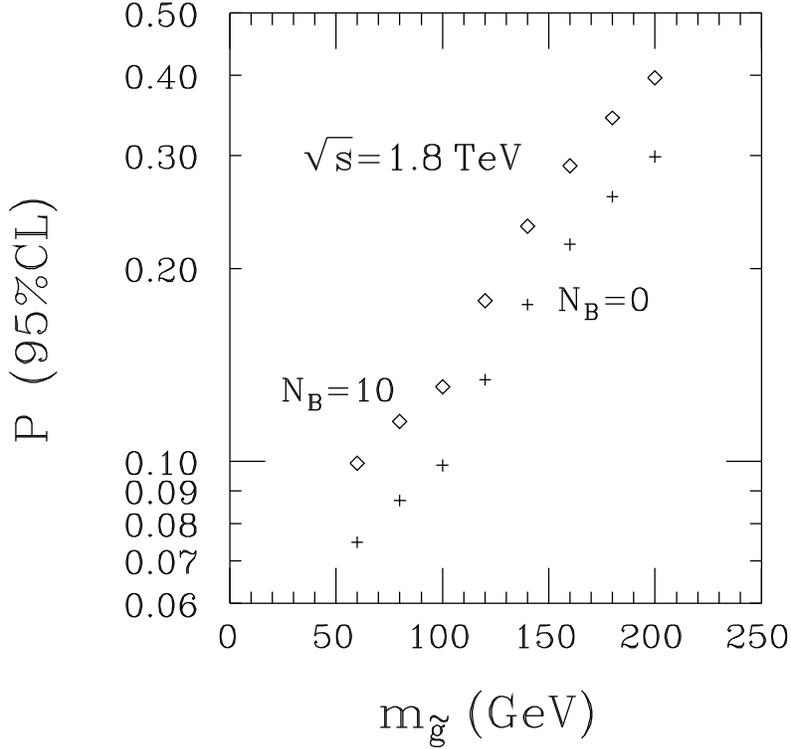}
\end{center}
\caption[]{The 95\%CL upper limit on the probability $P$
for a $\gl$ to fragment to a singly-charged $R^{\pm}$ hadron
after production and collision is given as a function of $\mgl$
for $N_B=0$ and 10 background events.}
\label{stableprob} 
\end{figure}

In Ref.~\cite{stable}, it is stated that there are zero background
events in $L=90\pbi$ of data after the mass~$>100\gev$ cuts.
The background level probably increases gradually 
as one lowers the $\mgl$ value considered down to $50\gev$.
(Current cuts do not allow sensitivity below this.)
However, even for the less stringent $\beta\gamma<0.85$ cut 
the background level is estimated at $<12$ events 
\cite{runiimeeting} for $\mgl=50\gev$.
To illustrate the situation,
let us consider the cases of $N_B=0$ and 10 background events.
At 95\% CL we require $LP^3\sigma_{\rm eff}<3$ ($N_B=0$)
or $<7$ ($N_B=10$). The resulting
95\% CL upper limits on $P$ are plotted as a function of $\mgl$
in Fig.~\ref{stableprob}. We see that the limits on $P$ are
significant.  In particular, for $\mgl\sim 50\gev$ and $N_B\sim 10$,
we find that $P>0.09$ is excluded. For $\mgl\sim 100\gev$ and $N_B=0$,
$P>0.1$ is ruled out, rising to $P>0.2$ for $\mgl\sim 150\gev$.
For $\mgl\geq 50\gev$, this result confirms the RunI jets+$\ptmiss$
analyses that exclude values of $\mgl$
below $130-150\gev$ down to $<20\gev$ for any $P$ for SC1
$\lam_T$ and $\vev{\Delta E}$ case choices. The heavily-ionizing track
signal improves (though only slightly) for SC3 choices, and thus excludes
$\mgl\geq 50\gev$ (up to very big values) for $P\geq 1/2$ (\ie\
for $P$ values such that the jets+$\ptmiss$ signal fails for the SC3 choices).
We expect that, at large $P$, a CDF heavily-ionizing track
analysis with weakened cuts would probably be able to extend
the excluded $\mgl$ range down to the OPAL $\mgl\sim 22- 25\gev$ 
lower bound (that applies for any $P$) 
based on the OPAL jets+$\ptmiss$ analysis and probably also down
to the $\sim 20\gev$ bound (that applies for large $P$)
from the OPAL heavily-ionizing track search.
In any case, currently the only significant 
window for a \glsp\ in the $P$--$\mgl$ parameter space 
arises for SC3 choices and $P\gsim 3/4$. The window at $P\sim 3/4$
is $25\gev\leq\mgl\leq50\gev$,
widening to $23\gev\leq\mgl\leq 50\gev$ for $P\sim 1$.

\section{A gluino NLSP decaying to gluon plus gravitino}

For completeness, we consider the scenario in which the gluino
is not the LSP, but rather the NLSP (next-to-lightest supersymmetric
particle), with the gravitino ($\gtino$) being the (now invisible) LSP.  
Such a situation
can arise in GMSB models, including that of Ref.~\cite{raby}.
In this scenario, the gluino decays via $\gl\to g\gtino$.
Early universe/rare isotope limits are then irrelevant. 
Further, the decay will be prompt from the detector point of view
if $\mgtino$ is in the $\leq$ few eV region such that the $\gtino$
is guaranteed to have no impact on $\Omega h^2$
\cite{coslimit,murayamabigf}. (If
the scale of supersymmetry breaking is so large that the $\gl\to g\gtino$
decay lifetime is long enough that most $\gl$'s
exit the detector before decaying, then the results of previous sections
apply.) The first examination of this scenario at the Tevatron appears 
in Ref.~\cite{glnlsptev}. We are unaware of any studies
of this scenario for the $q\anti q\gl\gl$ final state
at LEP or LEP2. Here, we will give the 95\% CL excluded
mass domains based on the previously considered jets + missing momentum
analyses of OPAL \cite{opal} and CDF \cite{cdfcuts,cdffinal}.
In our analysis, we will assume that the branching
ratio of $\gl\to g\gtino$ is 100\% (as appropriate
if the $\gl$ is the NLSP), and that the decay is prompt.
We will also assume that
the $\gtino$ has negligible mass compared to $\mgl$, and that
other supersymmetric particles are much heavier than the gluino. 

\begin{figure}[p]
\leavevmode
\begin{center}
\epsfxsize=4.in
\hspace{0in}\epsffile{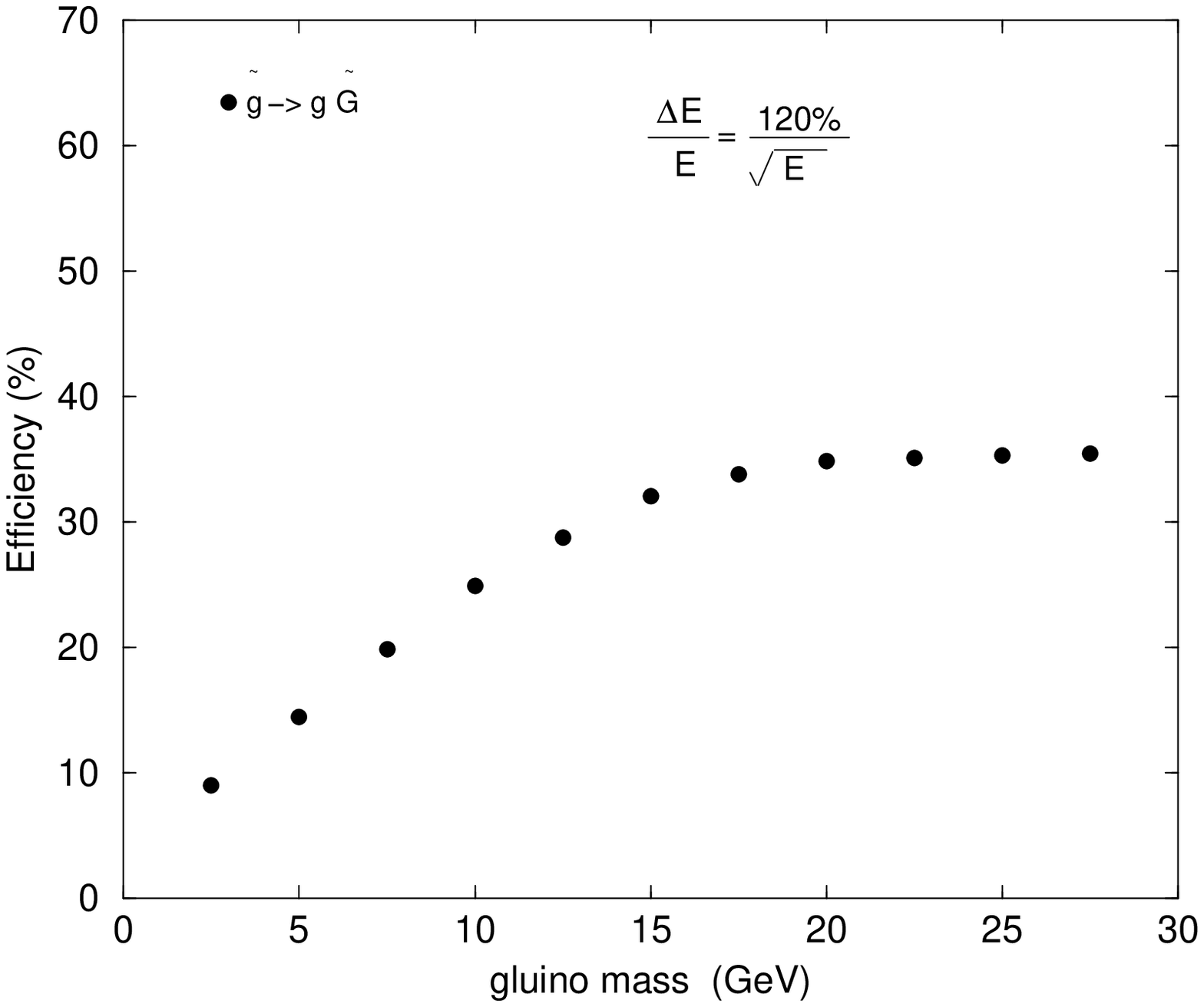}
\medskip
\epsfxsize=4.in
\hspace{0in}\epsffile{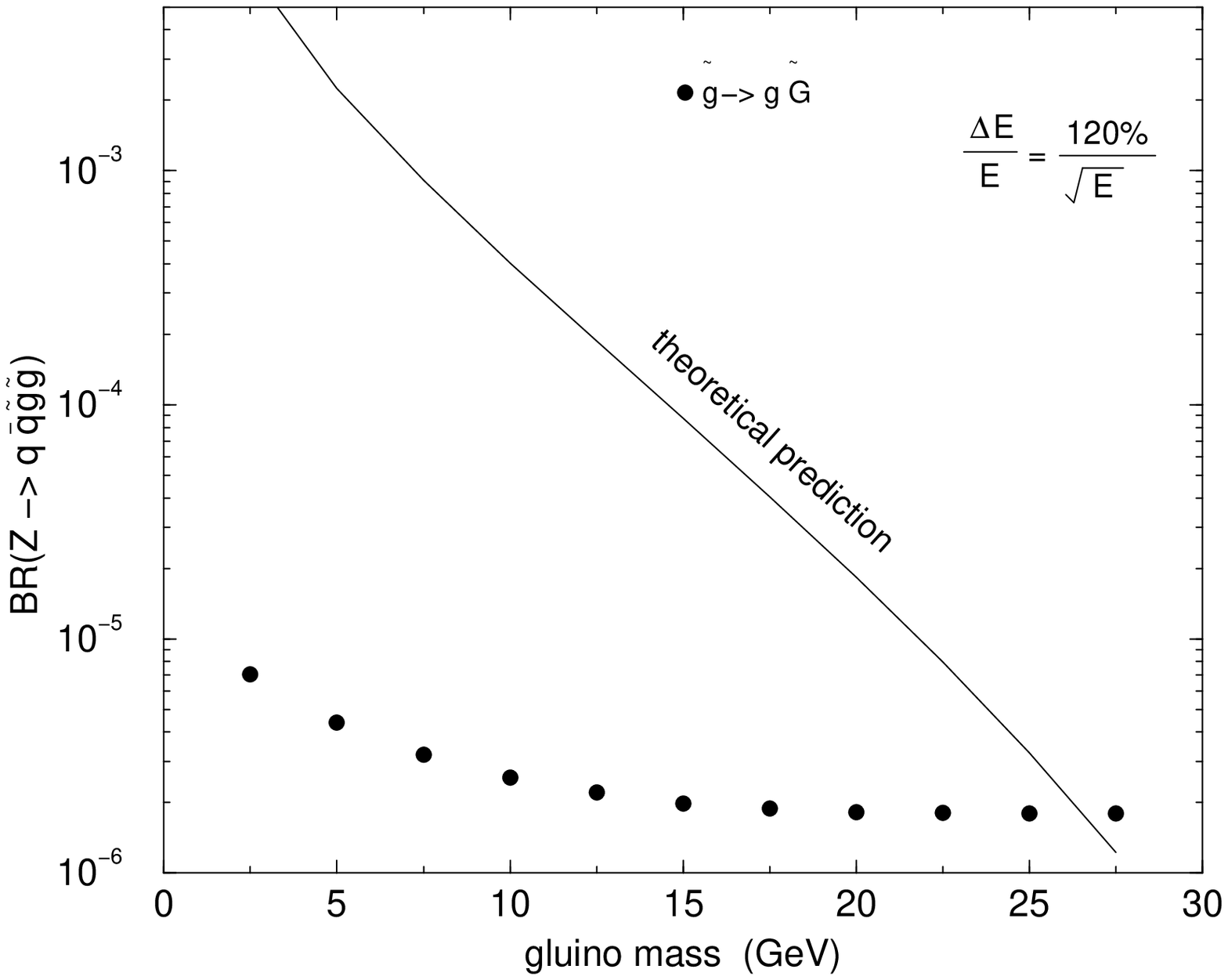}
\end{center}
\caption[]{In the upper window, we plot the OPAL $q\anti q \gl\gl$ 
event efficiency (after all cuts) in the $\gl\to g\gtino$
scenario. The lower window gives, as a function of $\mgl$,
the corresponding 95\% CL upper limits compared to the
theoretical prediction for $\br(Z\to q\anti q \gl\gl)$.}
\label{lepgravitino} 
\end{figure}

Consider first, the OPAL analysis. Using exactly the same procedures
and cuts as discussed earlier in section 4, but applied to $\epem\to Z\to
q\anti q\gl\gl\to q\anti q g g+\ptmiss$, we have determined the
efficiency for event acceptance and the resulting 95\% CL upper limit on 
$\br(Z\to q\anti q\gl\gl)$. These results appear in Fig.~\ref{lepgravitino}.
Gluino masses below about $26\gev$ are clearly excluded.

\begin{figure}[p]
\leavevmode
\begin{center}
\epsfxsize=4.15in
\hspace{0in}\epsffile{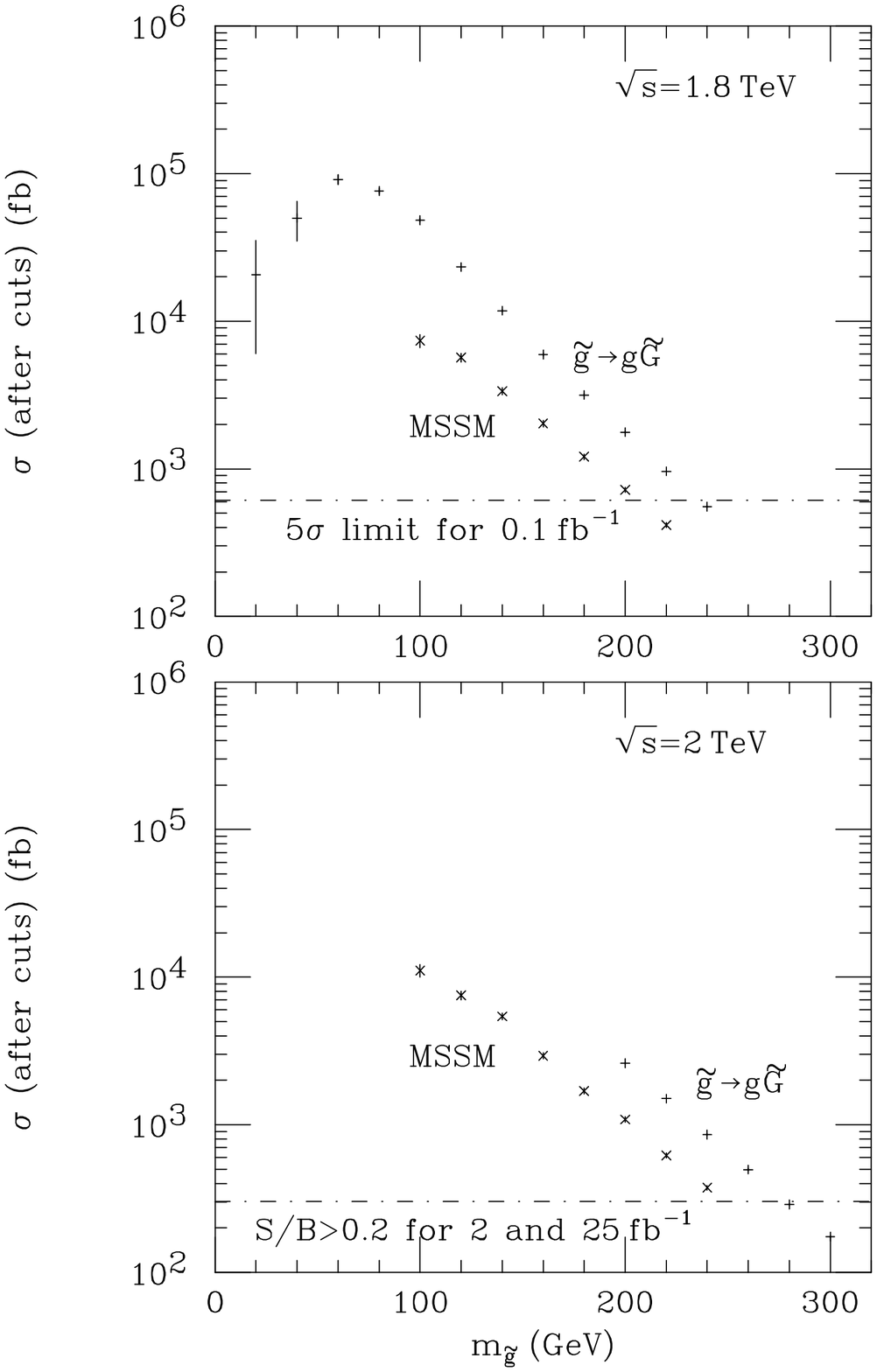}
\end{center}
\caption[]{The cross section (after cuts) in the ${\rm jets}+\ptmiss$ channel 
is compared to (a) the $5\sigma$ level for $L=0.1\fbi$
(also roughly the 95\% CL upper limit for $L=19\pbi$) at $\rts=1.8\tev$
and (b) the $S/B=0.2$ level at RunII ($L\geq 2\fbi$, $\rts=2\tev$)
as a function of $\mgl$ for the $\gl\to g\gtino$ case.}
\label{gnlspfig} 
\end{figure} 

For our CDF-based analysis of the $\gl\to g\gtino$ scenario
we employ the same procedures as in section 5. We compute
jets+$\ptmiss$ rates based on $p\anti p\to \gl\gl X$. 
The plots analogous to those
given earlier appear in Fig.~\ref{gnlspfig}. We observe that
the jets+$\ptmiss$ signal cross section (after cuts) for a $\gl$-NLSP is even
larger than in the reference MSSM model. All values of $\mgl\lsim 240\gev$
(down to very small values that clearly overlap
the OPAL exclusion region for this scenario) can be excluded at 95\% CL based
on the CDF $L=19\pbi$ data sample analysis.  This result
is stronger than the bound obtained in Ref.~\cite{glnlsptev}.
The same CDF analysis procedures
applied at RunII will be able to exclude $\mgl$ values up to about
$280\gev$.  Analyses optimized for such higher masses will presumably
be able to do even better.

Overall, it is clear that a gluino NLSP decaying to gluon plus light
gravitino can be excluded for essentially all $\mgl\lsim 240\gev$.

\section{Insights for other new physics analyses}

In this section, we wish to emphasize
a few interesting possibilities for other analyses for new physics
that can be extracted from the lessons learned in our specific studies.  

The primary point to note is that our results imply that the jets plus
missing momentum signal is immediately applicable for pair production of
any type of stable or semi-stable (\ie\ stable within the
detector) neutral or charged heavy particle that is produced
via the strong interactions. Examples of such particles abound in the
literature. 
\bit
\item Gauge-mediated supersymmetry breaking
models can contain colored messengers in the gauge-mediation sector
that are stable or semi-stable.
\item In models with extra generations, one or more of the heavy quarks
could be long-lived.
\item Semi-stable, strongly-interacting massive particles are
proposed as a source of ultra-high-energy cosmic ray events.
\eit
Pair production of a heavy stable particle 
produced via strong interactions gives rise to
a substantial missing momentum signal
due to the mismatch between the true momentum of 
each produced particle and the apparent energy of the jet 
associated with the particle
(as measured after including calorimeter response
and possible identification of any associated charged hadron
track as a muon track within the jet).
Further, the net missing momentum in a typical pair-production event
does not tend to be aligned with the visible energy of the jet associated
with any one of the heavy particles.
This is because in a realistic Monte Carlo 
the pair-production process initiated by quarks and/or gluons
in the colliding hadrons is accompanied by additional jets
with high transverse momentum coming from initial state `radiation'. 

In the case of pair production of
heavy stable quarks at a hadron collider,
the limits from the jets plus missing momentum
analysis would be very complementary to the heavily-ionizing,
penetrating track limits that rely more heavily on substantial modeling of
the charge exchange and fragmentation for a heavy quark
as it passes through the detector. As discussed earlier,
the rate for the latter signals scales roughly as $P^3$, where $P$
is the probability for the heavy quark to fragment to a charged (as
opposed to neutral) heavy hadron. For small enough $P$,
the missing momentum signal will be stronger than the penetrating track
signal. In addition, there is a very interesting hybrid
signal that should be analyzed.
A missing momentum trigger could be used to
isolate events in which to look for a heavily-ionizing track.\footnote{The
\glsp\ should also be searched for in the manner we describe.}
This could be more efficient than the present
CDF analysis which requires a penetrating
track in order to have a trigger rate such that all events
can be accepted.  The jets plus missing momentum trigger would eliminate
the need to require a penetrating track and one could just
search for a heavily-ionizing track in events accepted by the trigger.
The advantage would be that the probability for the heavily-ionizing
track (without requiring penetration) scales only as $P$ (rather than $P^3$).

It might be possible to take direct advantage of 
the mismatch between different ways of measuring the momentum
of a heavy particle that is contained in a charged state after
the initial interaction. The tracker would measure the
true momentum of the particle.  There are then two possibilities.
\bit
\item 
If the additional tracks are
not present that cause the track observed in the tracker
to be deemed as having penetrated to the muon detector, 
then this true momentum could
be directly compared to the momentum of the particle
as determined by the calorimeter response. We have seen
that there is generally
a very substantial difference. This situation would have
probability $\propto P(1-P^2)$ (including the probability
for the initial track in the tracker).
\item
Alternatively, if the track observed
in the tracker {\it is} deemed to have penetrated to the muon detector,
one could compare the true momentum to that computed for the jet
assuming the track belonged to a muon [see Eq.~(\ref{muonjet})].
The difference is substantial when the average $\beta$
of the produced particle is large.
\eit
In order to retain as many events as possible it would be best to
use a simple multi-jet trigger (without necessarily requiring missing
momentum). 
Of course, since we are looking for momentum discrepancies for
a single jet, it would be necessary 
to perform a very careful study of backgrounds,
such as that due to jets that are mismeasured and/or fragment to $K_L^0$'s.

\section{Summary and Conclusions}

We have examined constraints on any model in which
the gluino is the LSP.  In section 2, 
we considered the relic cosmological
density of a \glsp. We found that the relic density
depends very strongly on the presence and nature of non-perturbative
effects that could enter into the gluino and gluino-bound-state
annihilation cross sections. Assuming a completely perturbative
$\gl\gl$ annihilation cross section leads to a relic density
of $\Omega h^2\sim (\mgl/10\tev)^2$. For $\mgl\gsim 100\gev$, 
this level of relic density is probably inconsistent with bounds
from limits from heavy isotopes, underground detector
interaction rates and the like. However, we found that
non-perturbative effects can potentially decrease the relic density to
$\Omega h^2\sim 10^{-10}$ for all $\mgl\lsim 10\tev$,
a level that would be entirely consistent with all constraints.
Our conclusion is that, 
until the non-perturbative physics associated with gluino-gluino
annihilation can be clarified, no reliable limits on the \glsp\
can be obtained from constraints requiring knowledge of its relic density.
Thus, direct limits from accelerator experiments are of great interest.

In section 3, we studied the manner in which a (stable) \glsp\ 
is manifested in a typical detector. 
The critical issue for experimental analyses is the
average amount of visible momentum assigned to a gluino jet.
For a given detector, this
depends upon many ingredients, including the average
hadronic collision length of the $R$-hadron into which the $\gl$ fragments,
the average hadronic energy deposited in the various collisions experienced
by the $R$-hadron as it passes through the detector,
and the typical velocity and charge of the $R$-hadron.
The hadronic collision length was estimated using
the two-gluon model for total cross sections; one finds a collision
length that is somewhat longer
than for a typical light hadron. Collision lengths that 
are twice as large and one-half as large as our central
prediction were also considered.
Two cross section models were employed
for computing the average energy deposit (as a function of velocity)
in each hadronic collision. 
The (generally fluctuating) charge of the $R$-hadron as it passes
through the detector is also a crucial ingredient
and is characterized in terms of the probability $P$
for the $\gl$ to turn into a stable charged $\rpm$, such as $\gl u\anti d$, 
as opposed to a neutral state, such as the $\rzero=\gl g$,
after a hadronic collision. Simple quark counting models suggest
$P<1/2$ and probably much smaller if the $\gl g$ bound state is important.
For $P=0$, the energy (=momentum) assigned to a gluino
jet will be equal to the amount of the $\gl$ kinetic energy that is deposited
in the calorimeters due to hadronic collisions. For $P>0$,
the ionization energy deposits must be included and the possible
interpretation of an $\rpm$ track in the central
tracker as a muon within the $\gl$-jet must be taken into account.

In order to do this properly in a Monte Carlo
context, for any given value of $P$, the momentum
measured for each $\gl$ is computed on an event-by-event basis,
including (for $P\neq0,1$) random changes (according
to the value of $P$) of the charge of
the $R$-hadron at each hadronic collision as the $\gl$
passes through the detector. Procedures
are highly dependent upon the detector and specific analysis
in question. For example, in the LEP OPAL jets + missing momentum
analysis, if the $R$-hadron is an $\rpm$
in the tracker and penetrates as an $\rpm$
to the muon chamber, then the $\gl$-jet is declared to contain
a muon and a procedure for adding in the supposed
muon track momentum (and correcting for its presumed minimal
ionization energy deposit in the calorimeter) is followed. In contrast,
in the CDF jets + missing momentum analysis for Tevatron RunI,
if the $R$-hadron is an $\rpm$ in the tracker and appears
as an $\rpm$ in one of the muon chambers, and if the net
measured calorimeter energy is not too large, then the $\gl$-jet
is declared to be muonic and the event is discarded.

We studied the momentum typically              
assigned to the $\gl$-jet as a function of $P$, for 
the $\gl$ masses and velocities
of relevance, in the OPAL and CDF analyses. 
For all $P$ (for $P\leq1/2$), we found that the CDF (OPAL) procedure
implies that the momentum assigned to the $\gl$-jet
is (on average) only a small fraction of its actual momentum
unless $\mgl$ is smaller than a few GeV. This is true
even for the cross section choice that overestimates energy
deposits and even though, in the OPAL procedure,
we allow for the appropriate fraction of cases (determined
by $P$) in which the $\gl$ penetrates to the muon chamber and has an $\rpm$
track that is treated as a muon component of the jet
in reconstructing the jet energy.  Thus, when
the $\gl$ is the lightest supersymmetric particle, 
the jets plus missing momentum signature at colliders is, indeed, relevant.
In fact, this would be the dominant standard SUSY signal
if all other supersymmetric particles, in particular those with strong
production cross sections, are significantly heavier than the $\gl$. 

Section 3 ended with a discussion of the effects of incomplete containment
of a shower from a hadronic interaction that takes place near the
outer edge of the hadronic calorimeter (or outer edge of 
uninstrumented iron). Effects, on the OPAL and CDF
analyses summarized below, from the failure to include the
shower energy in the measured jet energy and from the extra tracks in
the subsequent muon-chamber(s) are outlined.

As noted, existing jets plus missing momentum analyses at both LEP and the
Tevatron are relevant to excluding a range of $\mgl$ values
in the \glsp\ scenario. In section 4, we demonstrated that
the OPAL LEP data analysis that has been performed in order to search
for $Z\to\cnone\cntwo$ (with $\cntwo\to q\anti q\cnone$) 
in the jets plus missing momentum channel can be applied
to $Z\to q\anti q\gl\gl$ events.  For $P=0,1/4,1/2,3/4$, 
we found that $\mgl$ values from $\sim 3\gev$ up to $\sim 25\gev$
are excluded at the 95\% CL, 
for all choices of path length $\lam_T$ and $\vev{\Delta E}$
energy loss (per hadronic collision) case considered.
For $P=1$, and after including energy smearing and fragmentation
effects, the upper limit of the excluded range declines
to $\mgl\sim 23\gev$ for our standard or ``SC1'' choices
of $\lam_T$ and $\vev{\Delta E}$ case. 
There is almost no change of the excluded range of $\mgl$ 
for possible extreme choices of $\lam_T$ and $\vev{\Delta E}$
(with scenario labels ``SC2'' and ``SC3'').
For the ``SC1'' choices, results for $P\sim 1$ are sensitive to
whether or not we include energy smearing and fragmentation
effects. If these effects are not included,
the fluctuations in measured jet energy are reduced and no
limit is possible for $P=1$ from OPAL jets + missing momentum data.
(But, as discussed below, much the same range of $3\lsim\mgl\lsim 20\gev$ 
is excluded by the heavily-ionizing track signal.)
In contrast, for $P\leq 3/4$, the excluded range of $\mgl$ is essentially
independent of whether or not energy smearing and fragmentation
are included.
Turning to LEP2, we noted that accumulated luminosities will
not be adequate to improve the LEP $Z$-pole limits. A next linear
collider operating at $\rts=500\gev$ would be able to
extend the LEP limits, but probably not beyond the limits
that are imposed by our Tevatron analysis.

In section 5, we analyzed constraints from the Tevatron, assuming
that all other SUSY particles are much heavier. We believe
the resulting limits on $\mgl$ to be conservative. 
We examined the jets plus missing momentum channel using cuts
and procedures based on the currently published CDF analysis of
$L=19\pbi$ of RunI data. The cross section limits
obtained by CDF translate to a range of excluded $\mgl$ values.
At 95\% CL, we exclude $\mgl$  up to $\sim 130-150\gev$ 
(the precise upper limit depending on $P$) down to at least $20\gev$
(at a very high CL),
for ``SC1'' or ``SC2'' choices of $\lam_T$ and $\vev{\Delta E}$ case.
For ``SC3'' choices (corresponding to long path length and small
hadronic energy deposits per collision for the $\gl$) 
the current CDF analysis can only exclude
the above range of $\mgl$ for $P\leq 1/2$. 
Thus, for all but ``SC3'' choices, the CDF RunI limit overlaps the OPAL limit
for any value of $P$, and all values of $\mgl$
in the $\sim 3\gev-130\gev$ range are excluded. 
For ``SC3'' $\lam_T$ and $\vev{\Delta E}$ case choices, these same CDF
limits apply only for $P\leq 1/2$. This lack of sensitivity of
the CDF analysis at large $P$ 
to long $\lam_T$ and/or small $\vev{\Delta E}$ could
be eliminated by a re-analysis of the data that retains muonic jets.

RunII Tevatron data in the jets plus missing momentum 
channel can be expected to 
extend the exclusion region to higher masses; depending upon $P$,
we found that roughly
$\mgl\lsim 160-180\gev$ will be excluded for ``SC1'' or ``SC2''
choices of $\lam_T$ and $\vev{\Delta E}$ case. 
For ``SC3'' choices and high $P$, only $\mgl\lsim 130\gev$
would be excluded. Such sensitivity 
is substantially worse than that found for the MSSM with mSUGRA
boundary conditions, for which one can probe out to roughly
$\mgl\lsim 250\gev$. Possibly the RunII reach in the \glsp\ scenario could
be extended if systematic errors are smaller than anticipated.
The above limitation assumes that $S/B>0.2$ is required for a detectable
signal. Alternative cuts, with smaller $B$ at high $\mgl$, might also
yield a larger reach. Although we have not specifically
performed the analysis, the Tevatron results suggest
that the LHC can be expected to rule out a 
\glsp\ with $\mgl$ up to at least 1 TeV.

We also explored limits on a \glsp\ deriving from the non-observation 
of a pseudo-stable charged track which is heavily-ionizing. 
The strength of such signals depends on $P$. 
In section 6, we reviewed the OPAL results. OPAL performed a direct
search for such states using cuts in the $dE/dx-|\vec p|$ plane, concluding
that for $P\sim 1/2$ ($P\sim 1$) one can exclude $\mgl$ in the $\sim 1-17\gev$
($\sim 1-20\gev$) mass range. For heavy-ionization signals
at higher masses we must turn to the Tevatron.
CDF looks for events containing a pseudo-stable penetrating charged
track which is heavily-ionizing. In section 7,
we demonstrated that, depending upon $P$,
$\gl$-pair production can lead to a significant cross section
(after imposing the CDF cut, penetration
and ionization requirements for identifying such events
with small background).  We have estimated the upper limit
from RunI data on the probability $P$
of charged fragmentation of the $\gl$. 
The upper limit can be roughly parameterized
as $P\sim 0.3(\mgl/200\gev)$ for $100\lsim\mgl\lsim 250\gev$.
For $\mgl<140\gev$, this means that $P<0.18$ is required.
Meanwhile: the jets plus missing momentum limits based on OPAL and CDF
analyses exclude $3\gev\lsim \mgl\lsim130-150\gev$ for $P\leq 1/2$;
the OPAL jets plus missing momentum 
analysis excludes $\sim 3\gev\lsim \mgl\lsim 25\gev$ for any $P$ not
too near 1 ($\sim 3\gev\lsim\mgl\lsim 23\gev$ for $P=1$);
and, the CDF jets plus missing momentum
analysis excludes $\mgl$ from $\sim 20\gev$ to $\sim 130\gev$ 
for $P=3/4$ and $P=1$ for all but ``SC3'' choices of 
$\lam_T$ and $\vev{\Delta E}$ case. For $P\geq 1/2$ (independent
of $\lam_T$ and $\vev{\Delta E}$), the CDF
heavily-ionizing track analysis excludes $50\leq\mgl\leq 200\gev$. 
This leaves only the possibility that ``SC3' choices apply, 
that $P$ lies in the (less likely) $P\gsim 3/4$ range, and that $\mgl$ lies
in the $\sim 23-50\gev$ window.
Very probably, an extension of the  CDF heavily-ionizing
penetrating particle analysis with weakened cuts appropriate to
these lower masses could exclude this window.

For completeness, in section 8 we also considered the scenario
where the gluino is the NLSP (next-to-lightest supersymmetric particle)
and the gravitino is the LSP. Such a situation 
is quite possible in models with gauge-mediated supersymmetry breaking.
In this scenario, the gluino decays via $\gl\to g\gtino$ and the $\gtino$
is invisible.  There is then a strong jets+$\ptmiss$ signal at 
both LEP and the Tevatron. We repeated the LEP OPAL-based analysis
and the $L=19\pbi$ CDF-based analysis for 
this case and found that $\mgl\lsim 240\gev$
can be excluded at 95\% CL. RunII should be able to extend the
excluded region to at least $\mgl\sim 280\gev$.

Finally, we urge our experimental colleagues to take note of 
our remarks in section 9 regarding the
applicability of our procedures in the jets plus missing momentum 
channel, or hybrid procedures such as combining a jets plus missing
momentum trigger with a heavily-ionizing track requirement, 
to placing limits on other exotic particles, such as a heavy stable quark.
We also note that a search for heavily-ionizing tracks in events
with jets plus missing momentum should prove very valuable for
excluding $P>1/2$ \glsp\ scenarios.

\section{Acknowledgements}
This work was supported in part by the DOE under contracts 
No. DE-FG03-91ER40674 and No. DE-FG05-87ER40319,
and in part by the Davis Institute for High Energy Physics.
The hospitality of the Aspen Center for Physics, where this work
was begun, is gratefully acknowledged.
We would like to thank S. Brodsky, 
M. Drees, J. Kiskis, S. Mani, S. Raby and K. Tobe for helpful discussions.
We are especially grateful to M. Albrow, H. Frisch, 
T. LeCompte, J. Hauser, S. Mani,  D. Stuart and R. Van Kooten for
extremely valuable discussions regarding the D0, CDF and OPAL detectors
and algorithms, 
and for critical examination of our procedures for computing energy losses.
%
%

\clearpage
%
\end{document}

